\documentclass[11pt,a4paper]{article}

\def\bea#1\eea{\begin{eqnarray}#1\end{eqnarray}}
\def\be#1\ee{\begin{equation}#1\end{equation}}
\def\ba#1\ea{\begin{align}#1\end{align}}
\def\la{\label}
\def\nl{\nonumber\\} 
\def\non{\nonumber}

\def\yz#1\yz {{\color{blue} YZ: #1}}
\def\sh#1\sh {{\color{blue} SH: #1}}

\usepackage[utf8]{inputenc}
\usepackage{jheppub}
\usepackage{amsmath}
\usepackage{multicol}
\usepackage{bbm}
\usepackage{enumerate}
\usepackage[inline]{enumitem}
\usepackage{bibentry}
\usepackage{comment}
\usepackage{amsthm}
\usepackage{mathrsfs}
\usepackage{upgreek}
\usepackage{amssymb}
\usepackage{bm}
\usepackage{setspace}
\usepackage{array,multirow,arydshln}
\usepackage{bigdelim}
\usepackage{scalerel}
\usepackage{diagbox}
\usepackage{caption}
\usepackage{tabularx}

\usepackage{tikz}

\usetikzlibrary{shapes.geometric,arrows,arrows.meta,decorations.pathmorphing,decorations.markings,patterns}

\def\<{\langle}
\def\>{\rangle}
\def\a{\alpha}

\def\c{\cdot}
\def\C{\!\cdot\!}
\def\e{\epsilon}

\def\er{\eqref}

\def\PT{\text {PT}}

\font\tenshuffle=shuffle10 \font\sevenshuffle=shuffle7 \font\fiveshuffle=shuffle7 at 5pt
\def\shuffle{{%
\def\Dshuffle{\mathbin{\hbox{\tenshuffle\char'001}}}%
\def\Sshuffle{\mathbin{\hbox{\sevenshuffle\char'001}}}%
\def\SSshuffle{\mathbin{\hbox{\fiveshuffle\char'001}}}%
\mathchoice{\Dshuffle}{\Dshuffle}{\Sshuffle}{\SSshuffle}}}    

\newcommand*{\halfway}{0.5*\pgfdecoratedpathlength+4.2pt}
\newcommand*{\halfwayb}{0.5*\pgfdecoratedpathlength+2.4pt}

\newcommand{\patha}[3]{\tikz[baseline={([yshift=-1ex]current bounding box.center)},every node/.style={font=\footnotesize},dir/.style={decoration={markings, mark=at position \halfwayb with {\arrow{Latex[scale=0.6]}}},postaction={decorate}}]{\filldraw (0,0) circle (1pt)  node[above=-1.5pt]{$#1$} (0.5,0) circle (1pt)  node[above=-1.5pt]{$#2$} (1,0) circle (1pt) node[above=-1.5pt]{$#3$};\draw[thick,dir] (0,0) -- (0.5,0);\draw[thick,dir] (0.5,0) -- (1,0);}}

\newcommand{\pathb}[2]{\tikz[baseline={([yshift=-1ex]current bounding box.center)},every node/.style={font=\footnotesize},dir/.style={decoration={markings, mark=at position \halfwayb with {\arrow{Latex[scale=0.6]}}},postaction={decorate}}]{\filldraw (0,0) circle (1pt)  node[above=-1.5pt]{$#1$} (0.5,0) circle (1pt)  node[above=-1.5pt]{$#2$};\draw[thick,dir] (0,0) -- (0.5,0);}}

\newcommand{\pathc}[4]{\tikz[baseline={([yshift=-1ex]current bounding box.center)},every node/.style={font=\footnotesize},dir/.style={decoration={markings, mark=at position \halfwayb with {\arrow{Latex[scale=0.6]}}},postaction={decorate}}]{\filldraw (0,0) circle (1pt)  node[above=-1.5pt]{$#1$} (0.5,0) circle (1pt)  node[above=-1.5pt]{$#2$} (1,0) circle (1pt) node[above=-1.5pt]{$#3$} (1.5,0) circle (1pt) node[above=-1.5pt]{$#4$};\draw[thick,dir] (0,0) -- (0.5,0);\draw[thick,dir] (0.5,0) -- (1,0);\draw[thick,dir] (1,0) -- (1.5,0);}}

\preprint{UUITP-25/19}

\title{String Correlators: Recursive Expansion, Integration-by-Parts and Scattering Equations}

\author[a,b]{Song He,}

\author[c]{Fei Teng}

\author[a,b]{and Yong Zhang}
\affiliation[a]{CAS Key Laboratory of Theoretical Physics, Institute of Theoretical Physics, Chinese Academy of Sciences, Beijing 100190, China}
\affiliation[b]{School of Physical Sciences, University of Chinese Academy of Sciences, No.19A Yuquan Road, Beijing 100049, China}
\affiliation[c]{Department of Physics and Astronomy, Uppsala University, 75108 Uppsala, Sweden}
\emailAdd{songhe@itp.ac.cn}
\emailAdd{fei.teng@physics.uu.se}
\emailAdd{yongzhang@itp.ac.cn}
\date{\today}

\abstract{We further elaborate on the general construction proposed in~\cite{He:2018pol}, which connects, via tree-level double copy, massless string amplitudes with color-ordered QFT amplitudes that are given by Cachazo-He-Yuan formulas. The current paper serves as a detailed study of the integration-by-parts procedure for any tree-level massless string correlator outlined in the previous letter. We present two new results in the context of heterotic and (compactified) bosonic string theories. First, we find a new recursive expansion of any multitrace mixed correlator in these theories into a logarithmic part corresponding to the CHY integrand for Yang-Mills-scalar amplitudes, plus correlators with the total number of traces and gluons decreased. By iterating the expansion, we systematically reduce string correlators with any number of subcycles to linear combinations of Parke-Taylor factors and similarly for the case with gluons. Based on this, we then derive a CHY formula for the corresponding $(DF)^2 + {\rm YM} + \phi^3$ amplitudes. It is the first closed-form result for such multitrace amplitudes and thus greatly extends our result for the single-trace case. As a byproduct, it gives a new CHY formula for all Yang-Mills-scalar amplitudes. We also study consistency checks of the formula such as factorizations on massless poles.}

\begin{document}

\maketitle
\flushbottom


\section{Introduction and review}
Recently, the study of scattering amplitudes has uncovered new structures and symmetries in various quantum field theories (QFT), as well as surprising connections between them ({\it cf.}~\cite{ArkaniHamed:2012nw,Henn:2014yza,Elvang:2015rqa}). The double-copy construction provides a notable example, which describes gravitational scattering amplitudes as ``squares" of gauge-theory ones. At tree level, such relations can be derived as the field-theory limit of the celebrated Kawai-Lewellen-Tye (KLT) relations~\cite{Kawai:1985xq} between tree amplitudes in open and closed string theory~\cite{Bern:1998sv}. Based on a remarkable duality between color and kinematics due to Bern, Carrasco and Johansson (BCJ)~\cite{Bern:2008qj}, double copy has been extended to quantum regime and become the state-of-the-art method for multiloop calculations in supergravity theories~\cite{Bern:2010ue,Bern:2013uka,Bern:2014sna,Johansson:2017bfl,Bern:2018jmv}. 

The Cachazo-He-Yuan (CHY) formulation~\cite{Cachazo:2013hca, Cachazo:2013iea} has provided a new way to manifest and extend the double copy. Based on the universal scattering equations connecting kinematics of massless particles to moduli space of punctured Riemann spheres~\cite{Cachazo:2013gna}, the CHY formula expresses tree amplitudes in a large class of theories as integrals over moduli space localized to the solutions of scattering equations. Together with loop-level generalizations~\cite{Adamo:2013tsa,Geyer:2015bja,Cachazo:2015aol,Geyer:2016wjx,Geyer:2018xwu}, they have led to new double-copy realization of various theories~\cite{Cachazo:2014xea}, and one-loop extensions of KLT relations and amplitude relations~\cite{He:2016mzd,He:2017spx}. What underpins both tree and loop-level CHY formulas are worldsheet models known as ambitwistor string theory~\cite{Mason:2013sva,Casali:2015vta}, where CHY integrands can be obtained as correlators therein. There has been significant progress~\cite{Siegel:2015axg,Casali:2016atr,Azevedo:2017yjy,Mizera:2017rqa,Bjerrum-Bohr:2014qwa} for connecting ambitwistor string theory to the usual string theory, but a complete understanding is still lacking. 

String theory has played a crucial role in these developments since the discovery of KLT relations. In particular, amplitude representations that respect color-kinematics duality at tree and loop level have both been realized by string-theory based methods~\cite{Mafra:2011kj,Mafra:2011nv,Mafra:2011nw,Mafra:2014gja, He:2015wgf}. Amplitude relations in gauge theory, {\it e.g.} BCJ relations~\cite{Bern:2008qj}, and those in Einstein-Yang-Mills (EYM)~\cite{Chiodaroli:2014xia}, can also find origin in string theory~\cite{BjerrumBohr:2009rd,Stieberger:2009hq,Stieberger:2016lng,Schlotterer:2016cxa}. More interestingly, it has been realized that tree-level superstring amplitudes themselves can be obtained via a double copy~\cite{Broedel:2013tta}. The first example is the discovery that one can decompose disk amplitudes for massless states of type-I theory into field-theory KLT products of universal basis of disk integrals, later called $Z$ integrals, and super-Yang-Mills (SYM) amplitudes: ``$\text{type-I} = Z \otimes \text{SYM}$''~\cite{Mafra:2011nw,Mafra:2011nv}. The key point is that all nontrivial $\alpha'$-dependence of string amplitudes is encoded in the $Z$ integrals, which can also be interpreted as amplitudes in an effective field theory of biadjoint scalars dubbed as $Z$ theory~\cite{Carrasco:2016ldy,Mafra:2016mcc,Carrasco:2016ygv}. 

It has been realized in~\cite{Huang:2016tag,Azevedo:2018dgo} that such a double copy for string amplitudes is general, since it also applies to cases for bosonic and heterotic strings. For (compatified) bosonic open string amplitudes, the same double copy works where the field-theory amplitudes now contain tachyon poles, and they were shown~\cite{Azevedo:2018dgo} to come from the $(DF)^2 + {\rm YM} + \phi^3$ Lagrangian~\cite{Johansson:2017srf}, with $\alpha'$ related to its mass parameter: 
\begin{align*}
\text{(compactified) bosonic open string}= Z\otimes\big[(DF)^2 + {\rm YM} + \phi^3 \big]\,. 
\end{align*}
Furthermore, by replacing $Z$ integrals by certain sphere integrals, one can generalize the double copy structure to the massless amplitudes of closed and heterotic strings. As conjectured in~\cite{Schlotterer:2012ny,Stieberger:2013wea,Stieberger:2014hba} and proven in~\cite{Schlotterer:2018zce,Brown:2018omk}, the latter can be obtained as the single-valued (sv) projection~\cite{Schnetz:2013hqa, Brown:2013gia} of open-string amplitudes: 
\begin{align*}
\text{type-II} &= \text{sv(type-I)} \otimes \text{SYM}\,,\\
\text{heterotic} &= \text{sv(type-I)}\otimes\big[(DF)^2+\text{YM}+\phi^3\big] \,.
\end{align*}

In a recent letter~\cite{He:2018pol}, we have initiated a systematic study on the double-copy of tree-level massless string amplitude in terms of field-theory amplitudes defined by CHY formulas. The nontrivial part of such CHY formulas can be directly obtained from the original string correlator via an integration-by-parts (IBP) process~\cite{He:2018pol}, which we review here. A generic massless open-string tree amplitude is given by a disk integral:
\be\label{string}
{\cal M}_n^{\rm string}(\rho)=\int_{\rho}\underbrace{\frac{d^n z}{{\rm vol}\,\text{SL}(2, \mathbb{R})} \prod_{i<j} |z_{i j}|^{s_{ij}}}_{:= d\mu_n^{\rm string}}
~{\cal I}^{\rm string}_n(z)\,,\qquad {\rm KN}:=\prod_{i<j} |z_{ij}|^{s_{ij}}\,,
\ee
where $z_{ij}:=z_i-z_j$ and $s_{ij}:= \alpha' k_i\!\cdot\! k_j$ are the Mandelstam variables. The color ordering $\rho \in S_n/\mathbb{Z}_n$ is realized by the integration domain $z_{\rho(i)}<z_{\rho(i{+}1)}$. We denote the Koba-Nielsen factor as ${\rm KN}$ and the integral measure including it as $d\mu_n^{\rm string}$. One can fix three punctures, {\it e.g.} $(z_1, z_{n-1}, z_n)=(0, 1, \infty)$, using the SL$(2, \mathbb{R})$ redundancy, and the product in the Koba-Nielsen factor is over $1\leqslant i<j\leqslant n{-}1$ with this fixing. The (reduced) {\it string correlator} ${\cal I}^{\rm string}_n$ is a rational function of $z$'s, and we only require it to have correct SL$(2)$ weight: ${\cal I}^{\text{string}}_n \to \prod_{a=1}^n (\gamma+\delta z_a)^2 {\cal I}^{\text{string}}_n$ under $z_a \to -\frac{\alpha +\beta z_a}{\gamma+\delta z_a}$ with $\alpha\delta - \beta\gamma = 1$. As shown in~\cite{He:2018pol}, using IBP relations, one can write any such integral as a double-copy of field-theory color-ordered amplitudes and the $Z$ integrals \be
{\cal M}_n^{\rm string}(\rho)={\cal M}_n^{\rm FT} \otimes Z_\rho:=\sum_{\alpha, \beta\in S_{n{-}3}} {\cal M}^{\rm FT}_n(\alpha) S[\alpha|\beta] Z_\rho (\beta)\,,
\ee
where $\alpha, \beta$ are color orderings in a minimal basis, and the KLT double copy is defined using $(n{-}3)!$-dimensional matrix $S[\alpha|\beta]$ known as the field-theory momentum kernel~\cite{Bern:1998sv,BjerrumBohr:2010yc}. The $Z$ integral is a disk integral over a Parke-Taylor (PT) factor of~\cite{Broedel:2013tta}:
\be\label{ztheory}
Z_\rho(\pi):=\!\int_\rho d\mu_n^{\rm string}\,{\rm PT}(\pi)\,,\qquad\text{PT}(\pi):=\frac{1}{z_{\pi_1\pi_2}z_{\pi_2\pi_3}\cdots z_{\pi_n\pi_1}}\,.
\ee
Each color-ordered, field-theory amplitude ${\cal M}_n^{\rm FT}$ is defined by a CHY formula, whose color ordering is given by a PT factor $\text{PT}(\rho)$; the nontrivial part is a half-integrand ${\cal I}^{\text{CHY}}_n$ that is obtained from the original ${\cal I}_n^{\rm string}$ by IBP, which specifies the theory and external states:
\be\label{field}
{\cal M}_n^{\rm FT} (\rho):=\!\int\!\!\underbrace{\frac{d^n z}{{\rm vol}\,\text{SL}(2, \mathbb{C})}\prod_i{}' \delta\big(\sum_{j\neq i} \frac{s_{i\,j}}{z_{i\,j}}\big)}_{:=d\mu_n^{\rm CHY}} {\rm PT}(\rho)\,{\cal I}^{\text{CHY}}_n (z)\,.
\ee 
Here the integrals are localized by the $n{-}3$ delta functions imposing scattering equations~\cite{Cachazo:2013gna,Cachazo:2013hca}. 

Before moving on, let us pause and talk about an equivalent way of expressing string amplitudes as double copy. Note that the KLT double copy of $Z$ integrals with $\mathcal{M}(\rho)$ only concerns the Parke-Taylor factor ${\rm PT}(\rho)$ of the latter, and leaves ${\cal I}^{\text{CHY}}_n$ intact (which is independent of the ordering $\rho$). Thus it is natural to put the double-copy inside the CHY integral and write the string amplitude as a CHY formula:
\be
{\cal M}_n^{\rm string}(\rho)=\int d\mu_n^{\rm CHY}~{\cal Z}_\rho (z)~{\cal I}^{\text{CHY}}_n (z)\,,
\ee
where we have defined a {\it universal} CHY half-integrand ${\cal Z}_\rho(z):=Z_\rho \otimes {\rm PT} (z)$ for any open-string amplitude.\footnote{Such half-integrands have been studied earlier: it was called string-deformed Parke-Taylor factor in~\cite{Mizera:2017sen} and also implicitly studied for the higher-energy limit in~\cite{Cachazo:2013gna}.} This part is present regardless of type-I, bosonic or other possible theories, and the difference between these theories is represented by ${\cal I}_n^{\text{CHY}}$ only. For the closed string case, we simply replace the $Z$ integral in the definition of ${\cal Z}$ by single-valued projection of open-string amplitudes. While this rewriting has been known for a while, we emphasize that here the nontrivial, theory-dependent part in such CHY formulas, ${\cal I}^{\text{CHY}}_n$, can be obtained from the string correlator ${\cal I}^{\rm string}_n$ as follows. 

Using the technique developed in~\cite{He:2018pol}, one in fact obtains an {\it equivalence class} of CHY half-integrands from the string correlator ${\cal I}^{\rm string}_n$ through the following two steps: 
\begin{itemize}
	\item First we algorithmically reduce the string correlator ${\cal I}^{\text{string}}_n$, via IBP relations, to an equivalent class of logarithmic functions ${\cal I}_n$, which can be used as a CHY half-integrand in Eq.~\eqref{field}.
	\item Next we use scattering equations (SE) to obtain equivalent half-integrands, ${\cal I}_n^{\rm CHY}$, which are no longer logarithmic but usually take a more compact form and make some useful properties more manifest.
\end{itemize}
Logarithmic functions are defined to have only logarithmic singularities, {\it i.e.} simple poles, on boundaries of the moduli space of $n$-punctured Riemann spheres. Equivalently, it can be written as a linear combination of PT factors~\cite{Mizera:2017cqs,Arkani-Hamed:2017tmz,Arkani-Hamed:2017mur}. Note that we have an equivalence class of logarithmic functions ${\cal I}_n \overset{\rm IBP}\cong {\cal I}^{\rm string}_n$: any ${\cal I}_n$ gives the same string integral as that of ${\cal I}^{\rm string}_n$. Since they are also equivalence by SE, any ${\cal I}_n$ gives the same ${\cal M}_n^{\rm FT}$ as well. One can usually use SE to simplify ${\cal I}_n$ greatly and obtain ${\cal I}_n^{\rm CHY}$: while being non-logarithmic, usually it allows an all-multiplicity expression!

In this paper, we obtain two new results, corresponding to the two steps above,  for the scalar-gluon correlators of compactified open bosonic strings, or equivalently the holomorphic part of heterotic strings. Recall that a general mixed string correlator for $r$ gluons and $m{+}1$ scalar traces reads~\cite{Schlotterer:2016cxa}
\begin{align}\label{eq:strInt}
\mathcal{I}^{\rm string}_n(z)=R(i_1,i_2,\ldots,i_r)\prod_{t=1}^{m+1}\text{PT}(W_t)\,,
\end{align}
where the PT factors follow the definition in Eq.~\eqref{ztheory}. The $R(i_1,i_2,\ldots,i_r)$ correlator, containing the gluon polarization vectors, is given by a cycle expansion:
\begin{align}\la{rrrde}
R(i_1,i_2,\cdots,i_r) =\sum_{(I)(J)\cdots (K) \in S_r} {\cal R}_{(I)}{\cal R}_{(J)}\cdots {\cal R}_{(K)} \,.
\end{align}
Here, we sum over all the permutations of $\{i_1,\ldots,i_r\}$, and write them as products of cycles $(I),(J),\ldots,(K)$. For length-one and two cycles, we have
\begin{align}\la{calr}
{\cal R}_{(i)}=C_{i} := \sum_{j\neq i}C_{i,j}=\sum_{j\neq i}\frac{\e_i\cdot k_j}{z_{i\,j}} \,, \qquad  {\cal R}_{(ij)}=\frac{\epsilon_{i} \C \e_{j}}{\alpha' z_{i\,j}^2}\,,
\end{align}
while $\mathcal{R}_{(I)}=0$ for longer cycles. In $C_i$, the summation is over all the particle labels that are different from $i$. We note that Eq.~\eqref{eq:strInt} also appears in the heterotic string correlator $\mathcal{I}^{\text{string}}_n(z)\mathcal{K}(\bar z)$ for $r$ gravitons and $m{+}1$ gluons traces, where $\mathcal{K}(\bar z)$ is an antiholomorphic type-I superstring correlator.

For step 1, we will propose a systematic method for performing IBP to reduce any multitrace mixed correlator to logarithmic functions. This is based on a new recursive expansion we discover for such string correlators, as we show in~\eqref{ntracea} for pure-scalar case and~\eqref{stringExpansion} for mixed case. The correlator can be expanded into two parts: the first part is a logarithmic function involving a set of {\it labeled trees} ${\cal T}$, which has appeared in previous studies of Yang-Mills-scalar CHY integrand~\cite{Teng:2017tbo,Du:2017gnh}, and corresponds to ${\cal I}_n$ from IBP reducing (compactified) superstring correlators; the second part contains terms with the total number of traces and gluons decreased. By iterating the expansion, any multitrace mixed correlator can be reduced to a logarithmic function. In our arXiv submission, we provide an ancillary Mathematica file which implements the expansion and does IBP reduction for any number of traces and gluons. 

For step 2, we further use SE to rewrite the logarithmic function ${\cal I}_n$ to a closed-form CHY half-integrand, ${\cal I}_n^{\rm CHY}$ for generic multitrace $(DF)^2+\text{YM}+\phi^3$ amplitudes. In~\cite{He:2018pol} we have presented a simple formula for all single-trace results, which we checked to high multiplicities but we had not found a proof then. In this paper we extend our construction to any number of traces, and write ${\cal I}_n^{\rm CHY}$  in a relatively simple form once a basic operation called {\it fusion} is defined. The result is expressed as a sum over {\it total partitions}, and it turns out that even for the single-trace case, we get an equivalent but distinct formula than that in~\cite{He:2018pol}. The outline of the paper is as follows.

We first present the general CHY-half integrand, ${\cal I}_n^{\rm CHY}$ in section~\ref{cquheflh}. As a byproduct, in the $\alpha'\to 0$ limit, our formula also gives a new formula for all multitrace Yang-Mills-scalar amplitudes, which are different from the original one in~\cite{Cachazo:2014xea}. In section \ref{lkhfqj}, we present the recursive expansion of the string correlator into a manifestly logarithmic part (which will be reviewed in Appendix~\ref{sec:trees}), and additional terms for which we can use the expansion again. 
We derive the expansion for pure-scalar case and outline the derivation for mixed cases, with some details left in the Appendix~\ref{sec:multibranch} and~\ref{sec:appb}. In section \ref{fuqiwepfh}, we will illustrate how to use the result from IBP to obtain the CHY formula summarized in section \ref{cquheflh}. In section \ref{fq3uofi}, we present an important check of the result which is factorization on massless poles.

\section{The formula for all multitrace amplitudes}\la{cquheflh}
In this section, we present the complete CHY half-integrands for multitrace amplitudes in $(DF)^2+{\rm YM}+\phi^3$ theory. The allowed external particles are massless gluons $A_\mu^a$ and bi-adjoint scalars $\phi^{a\tilde{a}}$, where $a$ is in the adjoint of a gauge group $U(N)$ and $\tilde{a}$ is in the adjoint of a global symmetry group $U(\tilde{N})$. We consider tree-level amplitudes with a fixed color ordering in $a$, represented by $\text{PT}(\rho)$ in Eq.~\eqref{field}. By ``multitrace'', we mean that the global adjoint indices $\tilde{a}$ of the scalars have the structure of $m{+}1$ traces. 

In the limit $\alpha'\rightarrow 0$, the half-integrand reduces to the usual Yang-Mills-scalar one, where the gluons $\{i_1,i_2,\ldots,i_r\}$ are packed into a reduced Pfaffian $\text{Pf}\,'(\Pi)$, and the scalar multitrace structure is described by $m{+}1$ PT factors $\text{PT}(W_i)$, see section 3 of~\cite{Cachazo:2014xea}. Alternatively, we can single out one trace, say $\text{PT}(W_{m+1})$, but treat the rest on the same footing as gluons. 
As we will see, the result is given by ``fusions" of all possible partitions of the set $\{W_1,\ldots,W_m,i_1,\ldots,i_r\}$. At finite $\alpha'$, we need to introduce some $\alpha'$-deformation of the fusion and consider a generalization of the partition: it turns out that we need the so-called {\it total partitions}. 

\subsection{Fusion of traces and gluons}\label{sec:fusion}
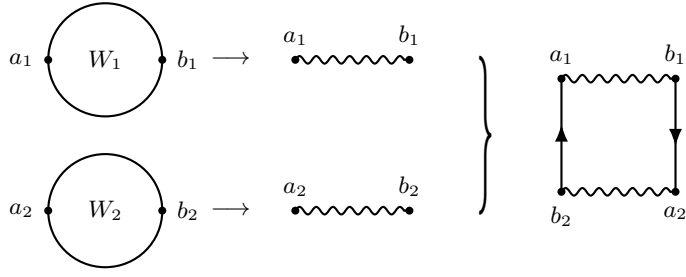
\begin{figure}[t]
	\centering
	\begin{tikzpicture}[every node/.style={font=\footnotesize,},vertex/.style={inner sep=0,minimum size=3pt,circle,fill},wavy/.style={decorate,decoration={coil,aspect=0, segment length=2.2mm, amplitude=0.5mm}},dir/.style={decoration={markings, mark=at position \halfway with {\arrow{Latex}}},postaction={decorate}},scale=1]
	\node at (0,0) {$W_1$};
	\node at (-0.75,0) [label={left:{$a_1$}},vertex] {};
	\node at (0.75,0) [label={right:{$b_1$}},vertex] {};
	\draw[thick] (0,0) circle (0.75cm);
	\node at (1.65,0) {$\longrightarrow$};
	\begin{scope}[xshift=2.5cm]
	\node at (-0,0) [label={above:{${a_1}$}},vertex] {};
	\node at (1.5,0) [label={above:{${b_1}$}},vertex] {};
	\draw[thick,wavy] (0,0) -- (1.5,0);
	\end{scope}
	\begin{scope}[yshift=-2cm]
	\node at (0,0) {$W_2$};
	\node at (-0.75,0) [label={left:{$a_2$}},vertex] {};
	\node at (0.75,0) [label={right:{$b_2$}},vertex] {};
	\draw[thick] (0,0) circle (0.75cm);
	\node at (1.65,0) {$\longrightarrow$};
	\end{scope}
	\begin{scope}[yshift=-2cm,xshift=2.5cm]
	\node at (0,0) [label={above:{${a_2}$}},vertex] {};
	\node at (1.5,0) [label={above:{${b_2}$}},vertex] {};
	\draw[thick,wavy] (0,0) -- (1.5,0);
	\end{scope}
	\begin{scope}[yshift=-0.25cm,xshift=6cm]
	\node at (0,0) [label={above:{$a_1$}},vertex] {};
	\node at (1.5,0) [label={above:{$b_1$}},vertex] {};
	\node at (0,-1.5) [label={below:{$b_2$}},vertex] {};
	\node at (1.5,-1.5) [label={below:{$a_2$}},vertex] {};
	\draw[thick,wavy] (0,0) -- (1.5,0); 
	\draw[thick,wavy] (0,-1.5) -- (1.5,-1.5); 
	\draw[thick,dir] (1.5,0) -- (1.5,-1.5);
	\draw[thick,dir] (0,-1.5) -- (0,0);
	\end{scope}
	\begin{scope}[yshift=-1cm,xshift=5cm]
	\node at (0,0) {$\stretchto[700]{\}}{60pt}$};
	\end{scope}
	\end{tikzpicture}
	\caption{Gluing two cycles $W_1$ and $W_2$ into a single one. A directed edge from node $i$ to $j$ represents a factor of $\frac{1}{z_{ij}}$ while a wavy edge is defined by Eq.~\eqref{eq:wavy}.}
	\label{fig:fusion2w}
\end{figure}

The first operation we introduce is the \emph{(weighted) fusion} between two traces $W_1$ and $W_2$ represented by the PT factors $\text{PT}(W_1)$ and $\text{PT}(W_2)$:
\begin{align}
\label{eq:fusion2}
\langle W_1,W_2\rangle:=\frac{1}{2}\sum_{\substack{a_1,b_1\in W_1 \\ a_2,b_2\in W_2}}s_{b_1a_2}s_{b_2a_1}\frac{z_{b_1a_1}z_{b_2a_2}}{z_{b_1a_2}z_{b_2a_1}}\,\text{PT}\,(W_1)\,\text{PT}\,(W_2)\,,
\end{align}
The cross ratio on the right hand side glues $\text{PT}(W_1)$ and $\text{PT}(W_2)$ into a single PT factor:
\begin{align}
\label{eq:sum2}
\frac{z_{b_1a_1}z_{b_2a_2}}{z_{b_1a_2}z_{b_2a_1}}\,\text{PT}\,(W_1)\,\text{PT}\,(W_2)=(-1)^{|B_1|+|B_2|}\!\sum_{\substack{\sigma_1\in A_1\shuffle B_1^T \\ \sigma_2\in A_2\shuffle B_2^T}}\!\text{PT}(a_1,\sigma_1,b_1,a_2,\sigma_2,b_2)\,.
\end{align}
The sets $A_i$ and $B_i$ are determined as follows for $i=1,2$.
For each choice of $a_i,b_i\in W_i$, we write $\text{PT}(W_i)=\text{PT}(a_i,A_i,b_i,B_i)$, using the cyclicity of PT factors. We then break $W_i$ into words $(a_i,\sigma_i,b_i)$, sum over all the $\sigma_i$'s in $A_i\shuffle B_i^T$ (the shuffle of $A_i$ and the reverse of $B_i$), and then glue the two words $(a_i,\sigma_i,b_i)$ into $\text{PT}(a_1,\sigma_1,b_1,a_2,\sigma_2,b_2)$. This process is shown schematically in figure~\ref{fig:fusion2w}. The sum over shuffle can be represented by a wavy line:
\begin{align}\label{eq:wavy}
\begin{tikzpicture}[baseline={([yshift=-1.5ex]current bounding box.center)},every node/.style={font=\footnotesize,},vertex/.style={inner sep=0,minimum size=3pt,circle,fill},wavy/.style={decorate,decoration={coil,aspect=0, segment length=2.2mm, amplitude=0.5mm}},dir/.style={decoration={markings, mark=at position \halfway with {\arrow{Latex}}},postaction={decorate}}]
\node at (-0,0) [label={above:{${a}$}},vertex] {};
\node at (1.5,0) [label={above:{${b}$}},vertex] {};
\draw[thick,wavy] (0,0) -- (1.5,0);
\end{tikzpicture}:=\text{PT}(W)z_{ba}=\text{PT}(a,A,b,B)z_{ba}=\sum_{\sigma\in A\shuffle B^T}\frac{(-1)^{|B|}}{z_{a\sigma_1}z_{\sigma_1\sigma_2}\cdots z_{\sigma_{|\sigma|b}}}\,,
\end{align} 
where $b$ is chosen as the end connected to an edge pointing away from $W$ (see figure~\ref{fig:fusion2w}). The fusion $\langle W_1,W_2\rangle$ merges two color traces into a single trace since it is a linear combination of $\text{PT}(\rho)$ with $\rho\in\text{perm}(W_1\cup W_2)$.
As a concrete example, we consider the fusion of $\text{PT}(12)$ and $\text{PT}(345)$:
\begin{align}
\langle (12),(345)\rangle&=\text{PT}(12)\text{PT}(345)\Big[s_{23}s_{51}\frac{z_{21}z_{53}}{z_{23}z_{51}}+s_{23}s_{41}\frac{z_{21}z_{43}}{z_{23}z_{41}}+s_{24}s_{51}\frac{z_{21}z_{54}}{z_{24}z_{51}}+(1\leftrightarrow 2)\Big]\nonumber\\
&=s_{23}s_{51}\text{PT}(12345)-s_{23}s_{41}\text{PT}(12354)+s_{24}s_{51}\text{PT}(12435)+(1\leftrightarrow 2)\,.
\end{align}
The generalization to the fusion of $r$ cycles is straightforward (with $a_{r+1}:=a_1$):
\begin{align}\label{eq:fusionr}
\langle W_1,W_2,\ldots,W_r\rangle :\!\!&=\frac{1}{2}\Bigg[\prod_{i=1}^r\sum_{a_i,b_i\in W_i}\frac{s_{b_ia_{i+1}}z_{b_{i}a_{i}}}{z_{b_ia_{i+1}}}\,\text{PT}(W_i)\Bigg]\nonumber\\
&=\frac{1}{2}\Bigg[\prod_{i=1}^{r}\sum_{a_i,b_i\in W_i}s_{b_ia_{i+1}}\Bigg]\begin{tikzpicture}[baseline={([yshift=-.5ex]current bounding box.center)},every node/.style={font=\footnotesize,},vertex/.style={inner sep=0,minimum size=3pt,circle,fill},wavy/.style={decorate,decoration={coil,aspect=0, segment length=2.2mm, amplitude=0.5mm}},dir/.style={decoration={markings, mark=at position \halfway with {\arrow{Latex}}},postaction={decorate}}]
\pgfmathsetmacro{\spa}{1};
\pgfmathsetmacro{\spb}{0.75};
\node (a1) at (0,0) [label={above:{$a_1$}},vertex] {};
\node (b1) at (\spa,0) [label={above:{$b_1$}},vertex] {};
\node (a2) at (\spa+\spb,0) [label={above:{$a_2$}},vertex] {};
\node (b2) at (2*\spa+\spb,0) [label={above:{$b_2$}},vertex] {};
\node (br) at (0,-\spb) [label={below:{$b_r$}},vertex] {};
\node (ar) at (\spa,-\spb) [label={below:{$a_r$}},vertex] {};
\node (a3) at (2*\spa+\spb,-\spb) [vertex] {};
\node (b3) at (\spa+\spb,-\spb) [vertex] {};
\draw[thick,wavy] (a1.center) -- (b1.center);
\draw[thick,wavy] (a2.center) -- (b2.center);
\draw[thick,wavy] (ar.center) -- (br.center);
\draw[thick,dir] (b1.center) -- (a2.center);
\draw[thick,dir] (br.center) -- (a1.center);
\draw[thick,dir] (b2.center) -- (a3.center) node [pos=0.36,right=0pt]{$\vdots$};
\draw[thick,wavy] (a3.center) -- (b3.center) node [pos=0.5,below=0pt]{$\cdots$};
\draw[thick,dir] (b3.center) -- (ar.center);
\end{tikzpicture}\,.
\end{align}
Each wavy line represents a summation over $A_i\shuffle B_i^T$ with $i$ inferred by the ends points. This also defines the fusion for $r=1$ case: \eqref{eq:fusionr} reduces to $\langle W\rangle=s_W\text{PT}(W)$, where $s_W:=\sum_{i<j \in W}s_{ij}$. The definition~\eqref{eq:fusionr} is clearly cyclic, and the factor $\frac{1}{2}$ cancels the double counting due to the reflection symmetry $\langle W_1,W_2,\ldots,W_r\rangle=\langle W_r,\ldots,W_2,W_1\rangle$.

Next, we include gluons into the fusion of traces. We find it convenient to view them as length-one words carrying polarizations, and consider them on the same footing as traces. For a single gluon, we define the fusion to be
\begin{align}
\langle i\rangle:=-C_i =-\sum_{j\neq i}\frac{\epsilon_i\!\cdot\! k_j}{z_{ij}}\,.
\end{align}
The fusion between two or more gluons $\langle i_1,i_2,\ldots,i_r\rangle$ is given as follows. For each length-one word $i$, we assign a field strength $f_i^{\mu\nu}=k_i^\mu\epsilon_i^\nu-k_i^\nu\epsilon_i^\mu$. We then glue all the $i$'s in order and Lorentz contract the field strengths in the same way:
\begin{align}\la{fqoe}
\langle i_1,i_2,\ldots,i_r\rangle &:=\frac{1}{2}\text{tr}(f_1f_2\cdots f_r)\,\text{PT}(i_1,i_2,\ldots,i_r):=\frac{1}{2}\text{tr}(f_1f_2\cdots f_r)
\begin{tikzpicture}[baseline={([yshift=-.5ex]current bounding box.center)},every node/.style={font=\footnotesize,},vertex/.style={inner sep=0,minimum size=3pt,circle,fill},wavy/.style={decorate,decoration={coil,aspect=0, segment length=2.2mm, amplitude=0.5mm}},dir/.style={decoration={markings, mark=at position \halfway with {\arrow{Latex}}},postaction={decorate}}]
\node (i1) at (0,1) [label={above:{$i_1$}},vertex] {};
\node (i2) at (1,1) [label={above:{$i_2$}},vertex] {};
\node (i3) at (1,0) [vertex] {};
\node (ir) at (0,0) [label={below:{$i_r$}},vertex] {};
\draw[thick,dir] (i1.center) -- (i2.center);
\draw[thick,dir] (i2.center) -- (i3.center);
\draw[thick,dir] (i3.center) -- (ir.center);
\draw[thick,dir] (ir.center) -- (i1.center);
\node at (0.55,0) [above=0pt]{$\cdots$};
\end{tikzpicture}\,.
\end{align}
These objects have already appeared in the cycle expansion of gluon CHY integrands~\cite{Lam:2016tlk}.

If a trace $W$ is involved in the fusion with a gluon $i$, we break $W$ as in figure~\ref{fig:fusion2w}, glue the length-one word $i$ to the end points, and then contract the momenta with the field strength $f_i$. Namely, we have
\begin{align}\label{eq:fusionWg}
\langle W,i\rangle :\!\!&=\frac{\alpha'}{2}\sum_{a,b\in W}(k_b\!\cdot\!f_i\!\cdot\!k_a)\frac{z_{ba}}{z_{bi}z_{ia}}\,\text{PT}(W)\nonumber\\
&=\frac{\alpha'}{2}\sum_{a,b\in W}(-1)^{|B|}(k_b\!\cdot\! f_i\!\cdot\! k_a)\!\sum_{\sigma\in A\shuffle B^T}\!\text{PT}(a,\sigma,b,i)
\end{align}
The generalization to arbitrary number of traces and gluons is straightforward:
\begin{align}\label{eq:fusionGen}
\langle W_1,\overbrace{i_1,\ldots,i_s}^{\mathsf{G}_1},\ldots,W_r,\overbrace{j_1,\ldots,j_\ell}^{\mathsf{G}_r}\rangle=\frac{\alpha'^r}{2}\Bigg[\prod_{i=1}^{r}\sum_{a_i,b_i\in W_i }\frac{(k_{b_i}\!\cdot\!f_{\mathsf{G}_i}\!\cdot\!k_{a_{i+1}})z_{b_ia_i}}{z_{b_i,\mathsf{G}_i,a_{i+1}}}\text{PT}(W_i)\Bigg]\,,
\end{align}
where, for example, $z_{b_1,\mathsf{G}_1,a_{2}}:=z_{b_1i_1}z_{i_1i_2}\cdots z_{i_sa_2}$ and $(f_{\mathsf{G}_1})_{\mu\nu}:=(f_{i_1}f_{i_2}\cdots f_{i_s})_{\mu\nu}$, etc.
Similarly, we also have a diagrammatic representation for the fusion. We illustrate with two traces and two sets of gluons in between:
\begin{align}\label{eq:fusionWiWj}
&\langle W_1,\overbrace{i_1,\ldots,i_s}^{\mathsf{G}_1},W_2,\overbrace{j_1,\ldots,j_\ell}^{\mathsf{G}_2}\rangle=\frac{\alpha'^2}{2}\!\sum_{\substack{a_1,b_1\in W_1 \\ a_2,b_2\in W_2}}\!(k_{b_1}\!\cdot\!f_{\mathsf{G}_1}\!\cdot\!k_{a_2})(k_{b_2}\!\cdot\!f_{\mathsf{G}_2}\!\cdot\!k_{a_1})\begin{tikzpicture}[baseline={([yshift=-.5ex]current bounding box.center)},every node/.style={font=\footnotesize,},vertex/.style={inner sep=0,minimum size=3pt,circle,fill},wavy/.style={decorate,decoration={coil,aspect=0, segment length=2.11mm, amplitude=0.5mm}},dir/.style={decoration={markings, mark=at position \halfway with {\arrow{Latex}}},postaction={decorate}},scale=0.84]
\node (a1) at (0,-0.6) [label={below:{$a_1$}},vertex] {};
\node (b1) at (0,0.6) [label={above:{$b_1$}},vertex] {};
\node (i1) at (0.75,0.6) [label={above:{$i_1$}},vertex] {};
\node (i2) at (1.5,0.6) [label={above:{$\cdots$}},vertex] {};
\node (ir) at (2.25,0.6) [label={above:{$i_r$}},vertex] {};
\node (b2) at (3,-0.6) [label={below:{$b_2$}},vertex] {};
\node (a2) at (3,0.6) [label={above:{$a_2$}},vertex] {};
\node (j1) at (0.75,-0.6) [label={below:{$j_\ell$}},vertex] {};
\node (j2) at (1.5,-0.6) [label={below:{$\cdots$}},vertex] {};
\node (jr) at (2.25,-0.6) [label={below:{$j_1$}},vertex] {};
\draw[thick,wavy] (a1.center) -- (b1.center); 
\draw[thick,wavy] (b2.center) -- (a2.center);
\draw[thick,dir] (b1.center) -- (i1.center);
\draw[thick,dir] (i1.center) -- (i2.center);
\draw[thick,dir] (i2.center) -- (ir.center);
\draw[thick,dir] (ir.center) -- (a2.center);
\draw[thick,dir] (b2.center) -- (jr.center);
\draw[thick,dir] (jr.center) -- (j2.center);
\draw[thick,dir] (j2.center) -- (j1.center);
\draw[thick,dir] (j1.center) -- (a1.center);
\end{tikzpicture}\,,
\end{align}
where the diagram stands for the sum
\begin{align}
\begin{tikzpicture}[baseline={([yshift=-.5ex]current bounding box.center)},every node/.style={font=\footnotesize,},vertex/.style={inner sep=0,minimum size=3pt,circle,fill},wavy/.style={decorate,decoration={coil,aspect=0, segment length=2.11mm, amplitude=0.5mm}},dir/.style={decoration={markings, mark=at position \halfway with {\arrow{Latex}}},postaction={decorate}},scale=0.85]
\node (a1) at (0,-0.6) [label={below:{$a_1$}},vertex] {};
\node (b1) at (0,0.6) [label={above:{$b_1$}},vertex] {};
\node (i1) at (0.75,0.6) [label={above:{$i_1$}},vertex] {};
\node (i2) at (1.5,0.6) [label={above:{$\cdots$}},vertex] {};
\node (ir) at (2.25,0.6) [label={above:{$i_r$}},vertex] {};
\node (b2) at (3,-0.6) [label={below:{$b_2$}},vertex] {};
\node (a2) at (3,0.6) [label={above:{$a_2$}},vertex] {};
\node (j1) at (0.75,-0.6) [label={below:{$j_\ell$}},vertex] {};
\node (j2) at (1.5,-0.6) [label={below:{$\cdots$}},vertex] {};
\node (jr) at (2.25,-0.6) [label={below:{$j_1$}},vertex] {};
\draw[thick,wavy] (a1.center) -- (b1.center); 
\draw[thick,wavy] (b2.center) -- (a2.center);
\draw[thick,dir] (b1.center) -- (i1.center);
\draw[thick,dir] (i1.center) -- (i2.center);
\draw[thick,dir] (i2.center) -- (ir.center);
\draw[thick,dir] (ir.center) -- (a2.center);
\draw[thick,dir] (b2.center) -- (jr.center);
\draw[thick,dir] (jr.center) -- (j2.center);
\draw[thick,dir] (j2.center) -- (j1.center);
\draw[thick,dir] (j1.center) -- (a1.center);
\end{tikzpicture}=(-1)^{|B_1|+|B_2|}\!\!\sum_{\substack{\sigma_1\in A_1\shuffle B_1^T \\ \sigma_2\in A_2\shuffle B_2^T}}\!\text{PT}(a_1,\sigma_1,b_1,i_1,\ldots,i_r,a_2,\sigma_2,b_2,j_1,\ldots,j_\ell)\,.
\end{align}
Loosely speaking, the gluons participating in fusions are turned into components of a color trace, while the polarization information appears as kinematic coefficients of the color traces.
To illustrate our result, let us write explicitly some low-multiplicity examples:
\begin{align}
\langle(12),3\rangle &=\alpha'(k_2\!\cdot\!f_3\!\cdot\!k_1)\text{PT}(123)\,,\nonumber\\
\langle(12),3,4\rangle &=\frac{\alpha'}{2}\big[(k_2\!\cdot\!f_3f_4\!\cdot\!k_1)\text{PT}(1234)+(k_1\!\cdot\!f_3f_4\!\cdot\!k_2)\text{PT}(2134)\big]\,,\nonumber\\
\langle(123),4\rangle&=\alpha'\big[(k_3\!\cdot\!f_4\!\cdot\!k_1)\text{PT}(1234)-(k_2\!\cdot\!f_4\!\cdot\!k_1)\text{PT}(1324)-(k_3\!\cdot\!f_4\!\cdot\!k_2)\text{PT}(2134)\big]\,,\nonumber\\
\langle(12),(34),5\rangle &=\frac{\alpha'}{2}\big[s_{23}(k_4\!\cdot\!f_5\!\cdot\!k_1)\text{PT}(12345)+s_{24}(k_3\!\cdot\!f_5\!\cdot\!k_1)\text{PT}(12435)+(1\leftrightarrow 2)\big]\,,\nonumber\\
\langle(12),5,(34),6\rangle &=\frac{\alpha'^2}{2}\big[(k_2\!\cdot\!f_5\!\cdot\!k_3)(k_4\!\cdot\!f_6\!\cdot\!k_1)\text{PT}(125346)+(k_2\!\cdot\!f_5\!\cdot\!k_4)(k_3\!\cdot\!f_6\!\cdot\!k_1)\text{PT}(125436)\nonumber\\
&\qquad\quad+(1\leftrightarrow 2)\big]\,.
\end{align}
Finally, as a technical convenience, we require the fusion operation be multilinear on traces, namely, the following relation should hold:
\begin{align}
\label{eq:linearity}
\langle\ldots,x\text{PT}(W_1)+y\text{PT}(W_2),\ldots\rangle:=x\langle\ldots,\text{PT}(W_1),\ldots\rangle+y\langle\ldots,\text{PT}(W_2),\ldots\rangle\,,
\end{align}
where $x$ and $y$ are independent of the worldsheet variables $z_i$. As a result, nested fusions like $\langle\langle W_1,i,W_2\rangle,W_3,j\rangle$ are well-defined. We note that fusions are not associative. For example, one can check that $\langle W_1,\langle W_2,W_3\rangle\rangle\neq \langle\langle W_1,W_2\rangle,W_3\rangle\neq\langle W_1,W_2,W_3\rangle$ by an explicit calculation.

\subsection{Partitions of set and symmetrized fusions}
A \emph{partition} of set ${A}$ is a family of nonempty subsets $\{\mathsf{A}_1,\mathsf{A}_2,\ldots,\mathsf{A}_m\}$ of ${A}$ that satisfies $\bigcup_{i=1}^{m}\mathsf{A}_i={A}$ and $\mathsf{A}_i\cap \mathsf{A}_j=\emptyset$ if $i\neq j$, where $1\leqslant m\leqslant |{A}|$ is the order of the partition. We refer $\mathsf{A}_i$ as a \emph{block} in the partition. 
For ${A}=\{\mathsf{a}_1,\mathsf{a}_2\}$ and $\{\mathsf{a}_1,\mathsf{a}_2,\mathsf{a}_3\}$, the collections of all partitions, denoted as $\mathbb{P}[{A}]$, are
\begin{align*}
\mathbb{P}[\{\mathsf{a}_1,\mathsf{a}_2\}]=\Big\{&\big\{\{\mathsf{a}_1,\mathsf{a}_2\}\big\},\big\{\mathsf{a}_1,\mathsf{a}_2\big\}\Big\}\,,\\
\mathbb{P}[\{\mathsf{a}_1,\mathsf{a}_2,\mathsf{a}_3\}]=\Big\{&\big\{\{\mathsf{a}_1,\mathsf{a}_2,\mathsf{a}_3\}\big\},\big\{\mathsf{a}_1,\mathsf{a}_2,\mathsf{a}_3\big\},\big\{\{\mathsf{a}_1,\mathsf{a}_2\},\mathsf{a}_3\big\},\big\{\{\mathsf{a}_2,\mathsf{a}_3\},\mathsf{a}_1\big\},\big\{\{\mathsf{a}_1,\mathsf{a}_3\},\mathsf{a}_2\big\}\Big\}\,.
\end{align*}
To avoid cluttered notations, we omit the curly bracket on singleton blocks when confusion is unlikely.\footnote{The total number of partitions for $n$ elements is known as the Bell number (\href{https://oeis.org/A000110}{https://oeis.org/A000110}), which equals $1$, $2$, $5$, $15$, $52$ for $n = 1$, $2$, $3$, $4$, $5$, etc.} For the partitions with $m\geqslant 2$, we can further partition each non-singleton $\mathsf{A}_i$ into two or more blocks, and continue the process until only singleton blocks remain. On the other hand, if we perform the same operation on the $m=1$ partition $\{A\}$, we get the same result but with an overall curly bracket~\cite{Stanley:2011}. Together they form the family of \emph{total partitions} of ${A}$, denoted as $\mathbb{T}[{A}]$.\footnote{The only exception is that $\mathbb{T}[\{\mathsf{a}_1\}]:=\{\{\mathsf{a}_1\}\}$. The number of total partitions under our definition equals the one in \href{https://oeis.org/A006351}{https://oeis.org/A006351}.} For example, we have
\begin{align*}
\mathbb{T}[\{\mathsf{a}_1,\mathsf{a}_2\}]=\Big\{&\big\{\mathsf{a}_1,\mathsf{a}_2\big\},\big\{\{\mathsf{a}_1,\mathsf{a}_2\}\big\}\Big\}\,, \\
\mathbb{T}[\{\mathsf{a}_1,\mathsf{a}_2,\mathsf{a}_3\}]=\Big\{&\big\{\mathsf{a}_1,\mathsf{a}_2,\mathsf{a}_3\big\},\big\{\{\mathsf{a}_1,\mathsf{a}_2\},\mathsf{a}_3\big\},\big\{\{\mathsf{a}_2,\mathsf{a}_3\},\mathsf{a}_1\big\},\big\{\{\mathsf{a}_1,\mathsf{a}_3\},\mathsf{a}_2\big\},\\
&\big\{\{\mathsf{a}_1,\mathsf{a}_2,\mathsf{a}_3\}\big\},\big\{\{\{\mathsf{a}_1,\mathsf{a}_2\},\mathsf{a}_3\}\big\},\big\{\{\{\mathsf{a}_2,\mathsf{a}_3\},\mathsf{a}_1\}\big\},\big\{\{\{\mathsf{a}_1,\mathsf{a}_3\},\mathsf{a}_2\}\big\}\Big\}\,.\nonumber
\end{align*}
For a total partition $\mathsf{A}=\{\mathsf{A}_1,\mathsf{A}_2\,\ldots,\mathsf{A}_m\}\in\mathbb{T}[{A}]$, each block $\mathsf{A}_i$ may contain nested curly brackets. In contrary, for $\mathsf{A}\in\mathbb{P}[{A}]$, the $\mathsf{A}_i$'s contain only singleton blocks. It is clear that by construction $\mathbb{P}[A]$ is always a subset of $\mathbb{T}[A]$.


We are interested in the case when the elements of ${A}$ are a collection of traces and gluons. Now we define the \emph{symmetrized fusion} with $\alpha'$ deformation that acts recursively on the block $\mathsf{A}_i$ of a total partition $\mathsf{A}\in\mathbb{T}[A]$:
\begin{itemize}
	\item If $\mathsf{A}_i=\{\mathsf{a}_1,\mathsf{a}_2,\ldots,\mathsf{a}_r\}$ contains only singleton blocks, we define
	\begin{align}
	\label{eq:Salpha1}
	\mathcal{S}_{\alpha'}(\mathsf{a}_1,\mathsf{a}_2,\ldots,\mathsf{a}_r):=\frac{1}{1-s_{\mathsf{a}_1\mathsf{a}_2\cdots\mathsf{a}_r}}\sum_{\pi\in S_r/\mathbb{Z}_r}\langle \mathsf{a}_{\pi(1)},\mathsf{a}_{\pi(2)},\ldots,\mathsf{a}_{\pi(r)}\rangle\,,
	\end{align}
	where $\mathsf{a}_j$ can either be a trace $W_j$ or a gluon $j$. For $r=1$, we use instead
	\begin{align}\label{eq:Salpha12}
	&\mathcal{S}_{\alpha'}(W):=\langle W\rangle=s_W\text{PT}(W)\,,& &\mathcal{S}_{\alpha'}(i):=\langle i\rangle=-C_i\,.
	\end{align}
	\item In the $\alpha'\rightarrow 0$ limit, we have
	\begin{align}
	\label{eq:S0}
	\mathcal{S}_0(\mathsf{a}_1,\mathsf{a}_2,\ldots,\mathsf{a}_r)=\sum_{\pi\in S_r/\mathbb{Z}_r}\langle \mathsf{a}_{\pi(1)},\mathsf{a}_{\pi(2)},\ldots,\mathsf{a}_{\pi(r)}\rangle\,,
	\end{align}
	but still $\mathcal{S}_0(W)=\langle W\rangle$ and $\mathcal{S}_0(i)=\langle i\rangle$ since they contribute to the leading $\alpha'$ order. 
	\item If $\mathsf{A}_i$ contains nested curly brackets, say $\mathsf{A}_i=\{\mathsf{A}'_1,\mathsf{A}'_2,\ldots,\mathsf{A}'_j,\mathsf{a}_{j+1},\ldots,\mathsf{a}_r\}$, we define
	\begin{align}
	\label{eq:Srecursion}
	\mathcal{S}_{\alpha'}(\mathsf{A}_i):=\mathcal{S}_{\alpha'}\big(\mathcal{S}_{\alpha'}(\mathsf{A}'_1),\mathcal{S}_{\alpha'}(\mathsf{A}'_2),\ldots,\mathcal{S}_{\alpha'}(\mathsf{A}'_j),\mathsf{a}_{j+1},\ldots,\mathsf{a}_r\big)\,.
	\end{align}
\end{itemize}
The symmetrization is an essential ingredient here since it restores the bosonic exchange symmetry of the gluons and color traces after the fusion.

Using the multilinearity~\eqref{eq:linearity} of the fusion, we can calculate generic $\mathcal{S}_{\alpha'}(\mathsf{A}_i)$ from the inner-most level. We give two examples to demonstrate the construction:
\begin{subequations}
	\begin{align}
	\mathcal{S}_{\alpha'}(\{(456),7\})&=\mathcal{S}_{\alpha'}\big((456),7\big)\nonumber\\
	&=\frac{\alpha'\big[(k_6\!\cdot\!f_7\!\cdot\!k_4)\text{PT}(4567)-(k_5\!\cdot\!f_7\!\cdot\!k_4)\text{PT}(4657)-(k_6\!\cdot\!f_7\!\cdot\!k_5)\text{PT}(5467)\big]}{1-s_{4567}}\,,\\
	\mathcal{S}_{\alpha'}(\{\{2,3\},4\})&=\mathcal{S}_{\alpha'}\big(\mathcal{S}_{\alpha'}(2,3),4\big)\nonumber\\
	&=\frac{\text{tr}(f_2f_3)}{1-s_{23}}\mathcal{S}_{\alpha'}\big((23),4\big)=\frac{\alpha'\text{tr}(f_2f_3)(k_3\!\cdot\!f_4\!\cdot\!k_2)}{(1-s_{234})(1-s_{23})}\text{PT}(234)\,.
	\end{align}
\end{subequations}
The symmetrized fusion is our basic building block for the multitrace CHY integrands, as we will show in the next subsection.


\subsection{Half-integrands for multitrace amplitudes}
Now we present the half-integrand for $m{+}1$ traces and $r$ gluons in the $(DF)^2+\text{YM}+\phi^3$ theory. It is given by an overall factor including $\text{PT}(W_{m+1})$ for the trace $W_{m+1}$ and tachyon poles for traces $\{W_1,\ldots,W_m\}$, times a sum over symmetrized fusions of all the total partitions $\mathsf{A}=\{\mathsf{A}_1,\mathsf{A}_2,\ldots,\mathsf{A}_{|\mathsf{A}|}\}$ of $\{W_1,\ldots,W_m,i_1,\ldots,i_r\}$:
\begin{align}
\label{eq:Imultitrace}
\mathcal{I}^{\text{CHY}}_n(W_1,\ldots,W_{m+1},i_1,\ldots,i_r)=\frac{\text{PT}(W_{m+1})}{\prod_{i=1}^{m}(1-s_{W_i})}\!\!\sum_{\mathsf{A}\in\mathbb{T}[W_1\ldots W_m,i_1\ldots i_r]}\!\!\!(-1)^{|\mathsf{A}|}\prod_{j=1}^{|\mathsf{A}|}\mathcal{S}_{\alpha'}(\mathsf{A}_j)\,.
\end{align}
In particular, the pure-scalar and single-trace integrands are given by
\begin{subequations}
	\begin{align}
	\label{eq:st1daf}
	\mathcal{I}^{\text{CHY}}_n(W_1,\ldots,W_{m+1})&=\frac{\text{PT}(W_{m+1})}{\prod_{i=1}^{m}(1-s_{W_i})}\!\sum_{\mathsf{A}\in\mathbb{T}[W_1\ldots W_m]}\!(-1)^{|\mathsf{A}|}\prod_{j=1}^{|\mathsf{A}|}\mathcal{S}_{\alpha'}(\mathsf{A}_j)\,,\\
	\label{eq:st1}
	\mathcal{I}^{\text{CHY}}_{n}(W,i_1,\ldots,i_r)&=\text{PT}(W)\!\sum_{\mathsf{A}\in\mathbb{T}[i_1\ldots i_r]}\!(-1)^{|\mathsf{A}|}\prod_{j=1}^{|\mathsf{A}|}\mathcal{S}_{\alpha'}(\mathsf{A}_j)\,.
	\end{align}
\end{subequations}
The crossing symmetry among $\{W_1,\ldots,W_m,i_1,\ldots,i_r\}$ is manifest since $\mathcal{S}_{\alpha'}$ is completely symmetric in its arguments. On the other hand, different choices of $W_{m+1}$ lead to equivalent integrand on the support of scattering equations. Some simple multitrace examples are given as follows:
\begin{subequations}
	\begin{align}
	&\text{double-trace:}& &\mathcal{I}^{\text{CHY}}_n(\sigma,\rho)=-\frac{\text{PT}(\rho)\,\mathcal{S}_{\alpha'}(\sigma)}{1-s_\sigma}=-\frac{s_{\sigma}}{1-s_{\sigma}}\,\text{PT}(\rho)\,\text{PT}(\sigma)\,, \\
	\label{eq:3tr}
	&\text{triple-trace:}& &\mathcal{I}^{\text{CHY}}_n(\sigma,\tau,\rho)=\frac{\text{PT}(\rho)\big[\mathcal{S}_{\alpha'}(\sigma)\mathcal{S}_{\alpha'}(\tau)-\mathcal{S}_{\alpha'}(\sigma,\tau)\big]}{(1-s_\sigma)(1-s_\tau)}\,,\\
	\label{eq:2trg}
	&\text{double-trace one-gluon:}& &\mathcal{I}^{\text{CHY}}_n(\tau,\rho,q)=\frac{\text{PT}(\rho)}{1-s_\tau}\big[\mathcal{S}_{\alpha'}(\tau)\mathcal{S}_{\alpha'}(q)-\mathcal{S}_{\alpha'}(\tau,q)\big]\,,
	\end{align}
\end{subequations}
where $\sigma$, $\rho$ and $\tau$ are traces and $q$ is a single gluon. The symmetrized fusions involved can be written explicitly as
\begin{gather}
\mathcal{S}_{\alpha'}(\sigma)\,\mathcal{S}_{\alpha'}(\tau)=s_{\sigma}s_{\tau}\text{PT}(\sigma)\text{PT}(\tau)\,, \qquad\mathcal{S}_{\alpha'}(\tau)\,\mathcal{S}_{\alpha'}(q)=-s_\tau\text{PT}(\tau)C_{q}\,,\nonumber \\
\mathcal{S}_{\alpha'}(\sigma,\tau)=\frac{1}{2(1-s_\rho)}\sum_{\substack{\sigma_1,\sigma_2\in\sigma \\ \tau_1,\tau_2\in\tau}}\!s_{\tau_2\sigma_1}s_{\sigma_2\tau_1}\frac{z_{\sigma_2\sigma_1}z_{\tau_2\tau_1}}{z_{\tau_2\sigma_1}z_{\sigma_2\tau_1}}\,\text{PT}(\sigma)\,\text{PT}(\tau)\,,\nonumber\\
\mathcal{S}_{\alpha'}(\tau,q)=\frac{\alpha'}{2(1-s_\rho)}\sum_{\tau_1,\tau_2\in\tau}(k_{\tau_2}\!\cdot\!f_{q}\!\cdot\!k_{\tau_1})\frac{z_{\tau_2\tau_1}}{z_{\tau_2q}z_{q\tau_1}}\text{PT}(\tau)\,.
\end{gather}
Interestingly, we can rewrite the single-trace integrand~\eqref{eq:st1} into a more familiar cycle expansion form:
\begin{align}
\label{eq:stsum}
\sum_{\mathsf{A}\in\mathbb{T}[i_1\ldots i_r]}\!(-1)^{|\mathsf{A}|}\prod_{j=1}^{|\mathsf{A}|}\mathcal{S}_{\alpha'}(\mathsf{A}_j)=(-1)^r\!\sum_{(I)(J)\ldots(K)\in S_r}\!\Psi_{(I)}\Psi_{(J)}\ldots\Psi_{(K)}\,.
\end{align}
This identity holds at the algebraic level, which can be easily checked numerically. However, the inductive proof is lengthy and we omit it here. For length-one and two cycles, the cycle factor $\Psi$ is given by
\begin{align}
& \Psi_{(i)}=C_i\,, & &\Psi_{(ij)}=-T_{ij}\,\text{PT}(ij):=-\frac{\text{tr}(f_if_j)}{2(1-s_{ij})}\,\text{PT}(ij)\,.
\end{align}
For cycles with $|I|\geqslant 3$, we have
\begin{align}
\label{eq:Psi_I}
\Psi_{(I)}=-\frac{T_I\,\text{PT}(I)}{2}:=-\frac{1}{2}\Bigg[\frac{1}{(1-s_I)}\sum_{\text{CP}}\text{tr}(F_{I_1}F_{I_2}\ldots F_{I_p})\Bigg]\text{PT}(I)\,.
\end{align}
The definition of $T_{ij}$ and $T_{I}$ can be inferred from the above two equations. Here, the summation is over all the cyclic partitions (CP) $\{I_1,I_2,\ldots,I_p\}$ of $I$ with $p\geqslant 2$. Each block $I_\ell$ of a cyclic partition must conform to the cyclic order determined by $I$.\footnote{If $|I|=n$, the number of such cyclic partitions is the Eulerian number $A(n,1)=2^n{-}n{-}1$.} For length-one blocks, $F_i^{\mu\nu}$ is just the field strength $f_i^{\mu\nu} $; for longer ones, it is recursively defined as
\begin{align}
F_{i_1i_2\cdots i_t}^{\mu\nu} =\a' k_{i_1}^\mu  T_{i_1i_2[i_3[\cdots[i_{t-1}i_t]\cdots]]} k_{i_t}^\nu\qquad (t\geqslant 2)\,,
\end{align}
where the bracket $[ij]$ stands for an antisymmetrization. For example, 
\begin{align}
T_{i_1i_2[i_3i_4]} &=T_{i_1i_2i_3i_4}-T_{i_1i_2i_4i_3}\,,\nonumber\\
T_{i_1i_2[i_3[i_4i_5]]} &=T_{i_1i_2i_3i_4i_5}-T_{i_1i_2i_3i_5i_4}-T_{i_1i_2i_4i_5i_3}+T_{i_1i_2i_5i_4i_3}\,.
\end{align} 
The cycle factor~\eqref{eq:Psi_I} is of a different form compared with the one defined in~\cite{He:2018pol}.\footnote{Comparing with the definition given in~\cite{He:2018pol}, we note that the requirement $|I_\ell|\geqslant 2$ is relaxed in Eq.~\eqref{eq:Psi_I}.} They are of course algebraically equivalent, as one can check explicitly.


Finally, we study the $\alpha'{\rightarrow} 0$ limit of the half-integrand~\eqref{eq:Imultitrace}. From the definition of $S_{\alpha'}$, one can show that, for example, $\mathcal{S}_{\alpha'}(\{\{\mathsf{a}_1,\mathsf{a}_2\},\mathsf{a}_3\})=\mathcal{S}_{\alpha'}(S_{\alpha'}(\mathsf{a}_1,\mathsf{a}_2),\mathsf{a}_3)$ gives higher $\alpha'$ order contribution than $\mathcal{S}_{\alpha'}(\{\mathsf{a}_1,\mathsf{a}_2,\mathsf{a}_3\})=\mathcal{S}_{\alpha'}(\mathsf{a}_1,\mathsf{a}_2,\mathsf{a}_3)$. As a result, the leading $\alpha'$ order is contributed solely by $\mathsf{A}=\{\mathsf{A}_1,\ldots,\mathsf{A}_{|\mathsf{A}|}\}\in\mathbb{P}[W_1,\ldots,W_m,i_1,\ldots,i_r]$, where the $\mathsf{A}_i$'s do not have nested curly brackets:
\begin{align}
\label{eq:I0}
\mathcal{I}^{\text{CHY}}_n(W_1,\ldots,W_{m+1},i_1,\ldots,i_r)_{\alpha'\rightarrow 0}=\text{PT}(W_{m+1})\!\!\sum_{\mathsf{A}\in\mathbb{P}[W_1\ldots W_m,i_1\ldots i_r]}\!\!(-1)^{|\mathsf{A}|}\prod_{j=1}^{|\mathsf{A}|}\mathcal{S}_{0}(\mathsf{A}_j)\,.
\end{align}
We have also removed all tachyon poles such that $\mathcal{S}_{\alpha'}$ reduces to $\mathcal{S}_0$. For the single-trace case, one can easily show that the summation in Eq.~\eqref{eq:I0} indeed gives a Pfaffian in the leading $\alpha'$ order. For generic cases, we can establish the equivalence between Eq.~\eqref{eq:I0} and the Yang-Mills-scalar integrand given in the squeezed form~\cite{Cachazo:2014xea} by using Eq~\eqref{eq:fusionr} and~\eqref{eq:fusionGen}.
After factorizing out the overall product $\prod_{i=1}^{m+1}\text{PT}(W_i)$, one can recognize that Eq.~\eqref{eq:I0} agrees exactly with the reduced Pfaffian under the gauge choice of deleting the two rows and columns associated to trace $W_{m+1}$.

\section{Recursive expansion of string correlator}\la{lkhfqj}
In this section, we present a recursive expansion for the open-bosonic string integrand~\eqref{eq:strInt}. To achieve this, one needs to perform the IBP reduction in a well-controlled manner.
We first show how this can be done for pure-scalar cases,
\begin{align}
\mathcal{I}_{n}^{\text{string}}=\text{PT}(W_1)\,\text{PT}(W_2)\cdots\text{PT}(W_{m+1})\,.
\end{align}
Our goal is to reduce this string integrand, via IBP relations, into a combination of logarithmic functions and the string integrands with number of traces decreased by the fusions defined in the previous section. The logarithmic function naturally takes the form of labeled trees which will be reviewed in Appendix \ref{sec:trees}.
 We can then use recursively the fewer-trace results and land on a logarithmic form integrand eventually.


As the starting point of our derivation, we always take a puncture in $W_{m+1}$ to infinity. Under this gauge, we can break another subcycle $W$ at a chosen puncture $z_a\in W$ by the following IBP relation~\cite{Schlotterer:2016cxa}:
\begin{align}\label{stibp}
{\rm PT}(W)(\cdots)\overset{\rm IBP}{\cong}  \frac{1}{1-s_W} \sum_{j\notin W}\, \underbrace{{\rm PT}(W)\sum_{b\in W} \frac{s_{b j}z_{ba}}{z_{b j}}}_{C^{W}_{a,j}}\,(\cdots)\,,
\end{align}
where $(\cdots)$ does not involve any punctures in $W$ except for $z_a$. It is convenient to represent $C^W_{a,j}$ by the following diagram
\begin{align}
C^W_{a,j}=\sum_{b\in W}\begin{tikzpicture}[baseline={([yshift=-.5ex]current bounding box.center)},every node/.style={font=\footnotesize,},vertex/.style={inner sep=0,minimum size=3pt,circle,fill},wavy/.style={decorate,decoration={coil,aspect=0, segment length=2.2mm, amplitude=0.5mm}},dir/.style={decoration={markings, mark=at position \halfway with {\arrow{Latex}}},postaction={decorate}}]
\node at (-0,0) [label={above:{${a}$}},vertex] {};
\node at (1.5,0) [label={above:{${b}$}},vertex] {};
\draw[thick,wavy] (0,0) -- (1.5,0) node[pos=0.5,below=0pt]{$\vphantom{A\shuffle B^T}$};
\draw[thick,dir] (1.5,0) -- (2.5,0) node[label={above:{$j$}},vertex] {};
\end{tikzpicture}\,s_{bj}\,,
\end{align}
where each term in the summation can be viewed as a chain (dressed with an additional factor $s_{b j}$). 
This diagrammatic representation is consistent with the one shown in figure~\ref{fig:fusion2w}. If the $(\cdots)$ contains no subcycles, for example, the double-trace case,
\begin{align}\label{doubleTraceLog}
\text{PT}(W_1)\text{PT}(W_2)\overset{\text{IBP}}{\cong}\frac{1}{1-s_{W_1}}\sum_{j_2\in W_2}C^{W_1}_{a_1,j_2}\text{PT}(W_2):=\frac{\text{PT}(W_2)}{1-s_{W_1}}\mathcal{T}_{W_2}(W_1)\,,
\end{align}
the result is already logarithmic and our IBP reduction finishes.

However, for triple-trace and beyond, the above no-subcycle condition no longer holds after breaking one subcycle, say $W_1$. Thus further IBP reduction is necessary. We first show by some examples on how to proceed in this situation. We then provide a systematic solution that leads to a recursive expansion for generic pure-scalar integrands. 
Finally, we write down the expansion for the case with gluons, which is an analog of the pure-scalar case, and leave the derivation to Appendix~\ref{sec:appb}. 

\subsection{Pure-scalar examples}
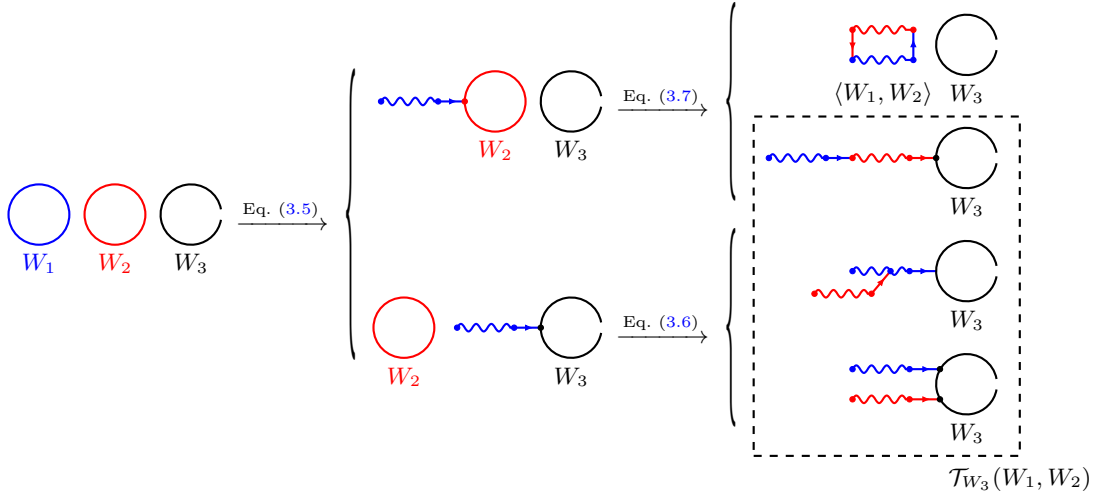
\begin{figure}[t]
\centering
\begin{tikzpicture}[every node/.style={font=\footnotesize,},vertex/.style={inner sep=0,minimum size=3pt,circle,fill},wavy/.style={decorate,decoration={coil,aspect=0, segment length=2mm, amplitude=0.5mm}},dir/.style={decoration={markings, mark=at position \halfwayb with {\arrow[scale=0.5]{Latex}}},postaction={decorate}},scale=1]

\begin{scope}[xshift=0cm,yshift=0cm]
\draw[thick,blue] (0,0) circle (0.4);
\draw[thick,red] (1,0) circle (0.4);
\draw[thick] {(2,0)+({.4*cos(10)},{.4*sin(10)})} arc (10:350:.4);
\node at (0,-0.65) [blue] {$W_1$};
\node at (1,-0.65) [red] {$W_2$};
\node at  (2,-.65)   {$W_3$};
\node at (3.2,0) {$\xrightarrow{\text{Eq.~}\eqref{threeaf}}$};
\node at (4.1,0) {$\stretchto[2000]{\{}{110pt}$};
\end{scope}

\begin{scope}[xshift=5cm,yshift=1.5cm]
\draw[thick,wavy,blue] (-.5,0) -- (.25,0);
\filldraw[blue] (-0.5,0) circle (1pt) (0.25,0) circle (1pt);
\draw[thick,blue,dir] (.25,0)--(0.6,0);
\draw[thick,red] (1,0) circle (0.4);
\filldraw[red] (0.6,0) circle (1pt);
\draw[thick] {(2,0)+({.4*cos(10)},{.4*sin(10)})} arc (10:350:.4);
\node at (1,-0.65) [red] {$W_2$};
\node at  (2,-.65)   {$W_3$};
\node at (3.2,0) {$\xrightarrow{\text{Eq.~}\eqref{three2}}$};
\node at (4.1,0) {$\stretchto[2000]{\{}{75pt}$};
\end{scope}

\begin{scope}[xshift=5cm,yshift=-1.5cm]
\draw[thick,wavy,blue] (.5,0) -- (1.25,0);
\draw[blue,thick,dir] (1.25,0)--(1.6,0);
\filldraw[blue] (0.5,0) circle (1pt) (1.25,0) circle (1pt);
\draw[thick,red] (-.2,0) circle (0.4);
\draw[thick] {(2,0)+({.4*cos(10)},{.4*sin(10)})} arc (10:350:.4);
\filldraw (1.6,0) circle (1pt);
\node at (-0.2,-0.65) [red] {$W_2$};
\node at  (2,-.65)   {$W_3$};
\node at (3.2,0) {$\xrightarrow{\text{Eq.~}\eqref{three3}}$};
\node at (4.1,0) {$\stretchto[2000]{\{}{75pt}$};
\end{scope}

\begin{scope}[xshift=10.2cm,yshift=2.25cm]
\draw[thick,wavy,blue] (.5,-.2) -- (1.3,-.2);
\draw[thick,blue,dir] (1.3,-.2)--(1.3,.2);
\draw[thick,wavy,red] (.5,.2) -- (1.3,.2);
\draw[thick,red,dir] (0.5,0.2) -- (0.5,-0.2);
\filldraw[blue] (0.5,-0.2) circle (1pt) (1.3,-0.2) circle (1pt);
\filldraw[red] (0.5,0.2) circle (1pt) (1.3,0.2) circle (1pt);
\node at  (.9,-.65)   {$\langle W_1,W_2 \rangle$};

\draw[thick] {(2,0)+({.4*cos(10)},{.4*sin(10)})} arc (10:350:.4);
\node at  (2,-.65)   {$W_3$};
\end{scope}

\begin{scope}[xshift=10.2cm,yshift=0.75cm]

\draw[thick,wavy,blue] (-0.6,0) -- (.15,0);
\draw[thick,blue,dir] (.15,0)--(.5,0);
\filldraw[blue] (-0.6,0) circle (1pt) (.15,0) circle (1pt);
\draw[thick,wavy,red] (.5,0) -- (1.25,0);
\draw[thick,red,dir] (1.25,0)--(1.6,0);
\filldraw[red] (1.25,0) circle (1pt) (0.5,0) circle (1pt);
\draw[thick] {(2,0)+({.4*cos(10)},{.4*sin(10)})} arc (10:350:.4);
\filldraw (1.6,0) circle (1pt);
\node at  (2,-.65)   {$W_3$};
\end{scope}

\begin{scope}[xshift=10.2cm,yshift=-0.75cm]
\draw[thick,wavy,blue] (.5,0) -- (1.25,0);
\draw[thick,blue,dir] (1.25,0)--(1.6,0);
\draw[thick,wavy,red] (0,-0.3) -- (0.75,-0.3);
\draw[thick,red,dir] (0.75,-0.3)--(1,0);
\filldraw[red] (0,-0.3) circle (1pt) (0.75,-0.3) circle (1pt);
\filldraw[blue] (.5,0) circle (1pt) (1.25,0) circle (1pt) (1,0) circle (1pt);
\draw[thick] {(2,0)+({.4*cos(10)},{.4*sin(10)})} arc (10:350:.4);
\node at  (2,-.65)   {$W_3$};
\end{scope}

\begin{scope}[xshift=10.2cm,yshift=-2.25cm]
\draw[thick,wavy,blue] (.5,0.2) -- (1.25,0.2);
\draw[thick,blue,dir] (1.25,0.2)--(1.65,0.2);
\filldraw[blue] (.5,0.2) circle (1pt) (1.25,0.2) circle (1pt);
\draw[thick,wavy,red] (.5,-.2) -- (1.25,-.2);
\draw[thick,red,dir] (1.25,-.2)--(1.65,-.2);
\filldraw[red] (.5,-.2) circle (1pt) (1.25,-.2) circle (1pt);
\filldraw (1.65,0.2) circle (1pt) (1.65,-0.2) circle (1pt);
\draw[thick] {(2,0)+({.4*cos(10)},{.4*sin(10)})} arc (10:350:.4);
\node at  (2,-.65)   {$W_3$};
\end{scope}

\begin{scope}[xshift=9.4cm,yshift=1.3cm]
\draw[dashed,thick] (0,0) rectangle (3.5,-4.5) node[below]{$\mathcal{T}_{W_3}(W_1,W_2)$};
\end{scope}

\end{tikzpicture}
\caption{
	The IBP reduction of the triple-trace string integrand $\text{PT}(W_1)\text{PT}(W_2)\text{PT}(W_3)$. The subcycle $\text{PT}(W_3)$ is broken since we set one of its punctures to infinity.}
\label{threefafd}
\end{figure}

We start with the triple-trace integrand $\text{PT}(W_1)\,\text{PT}(W_2)\,\text{PT}(W_3)$, the simplest nontrivial example. As the common first step, we break $W_1$ using the relation~\eqref{stibp}, which leads to
\begin{align}\label{threeaf}
{\rm PT}(W_1){\rm PT}(W_2){\rm PT}(W_3) \overset{\rm IBP}{\cong}   
\frac{1}{1-s_{W_1}} \Bigg[    \sum\limits_{\substack{j_3\in W_3}}C^{W_1}_{a_1,j_3} +  \sum\limits_{\substack{j_2\in W_2}}C^{W_1}_{a_1,{j_2}}   \Bigg]  {\rm PT}(W_2){\rm PT}(W_3)\,.
\end{align}
The structure of Eq.~\eqref{threeaf} is represented by the middle column of figure~\ref{threefafd}. For the first term on the right hand side, the chain $C^{W_1}$ is attached to $W_3$ and forms a labeled tree. We then continue to break $W_2$ at a puncture $z_{a_2}\in W_2$ using Eq.~\eqref{stibp}:
\begin{align}\label{three3}
{{\rm PT}(W_2)\,{\rm PT}(W_3)} \sum\limits_{\substack{j_3\in W_3}}C^{W_1}_{a_1,j_3}
\overset{\rm IBP}{\cong}   \frac{ {\rm PT}(W_3)}{1-s_{W_2}}   \sum\limits_{\substack{j_3\in W_3}}C^{W_1}_{a_1,j_3}   \sum\limits_{\substack{j\notin W_2}}C^{W_2}_{a_2,j}\,,
\end{align}
which becomes a combination of labeled trees. We note that in Eq.~\eqref{three3} the choice of $a_2$ is arbitrary. In contrary, in the second term of Eq.~\eqref{threeaf}, the chain $C^{W_1}$ is attached to $W_2$ and forms a \emph{normal tadpole}. We then break the subcycle $W_2$ at the attach point, namely, we must choose $a_2=j_2$ for each $C^W_{a_1,j_2}$ when using the relation~\eqref{stibp}. This leads to
\begin{align}\label{three2}
&{\rm PT}(W_3)\!\sum_{j_2\in W_2}\!C^{W_1}_{a_1,j_2}  {\rm PT}(W_2)\overset{\rm IBP}{\cong}   \frac{ {\rm PT}(W_3)}{1-s_{W_2}}  \Bigg[ \sum_{\substack{j_2\in W_2\\ j_3 \in W_3}}\!C^{W_1}_{a_1,j_2} C^{W_2}_{j_2,j_3} +\sum\limits_{\substack{j_2\in W_2\\ j_1 \in W_1}}\!C^{W_1}_{a_1,j_2} C^{W_2}_{j_2,j_1}  \Bigg]\,,
\end{align}
where the first term is a combination of labeled trees. The second term consists of \emph{induced tadpoles}, in which the subcycle involves punctures from different $W_i$'s due to IBP. Remarkably it reproduces the fusion between $W_1$ and $W_2$ after some algebraic manipulation:
\begin{align}\label{fusionCC}
\sum_{\substack{j_2\in W_2\\ j_1 \in W_1}}   
C^{W_1}_{a_1,j_2} C^{W_2}_{j_2,j_1}=\sum_{\substack{j_2\in W_2\\ j_1 \in W_1}}\begin{tikzpicture}[baseline={([yshift=-0ex]current bounding box.center)},every node/.style={font=\footnotesize,},vertex/.style={inner sep=0,minimum size=3pt,circle,fill},wavy/.style={decorate,decoration={coil,aspect=0, segment length=2.2mm, amplitude=0.5mm}},dir/.style={decoration={markings, mark=at position \halfway with {\arrow{Latex}}},postaction={decorate}}]
\node at (-0,0) [label={below:{${a_1}$}},vertex] {};
\node at (1,0) [label={below:{${j_1}$}},vertex] {};
\node at (1.5,0) [vertex] {};
\node at (4,0) [vertex] {};
\draw[thick,wavy] (0,0) -- (1,0) -- (1.5,0) (2.5,0) -- (4,0);
\draw[thick,dir] (1.5,0) -- (2.5,0) node[label={below:{$j_2$}},vertex] {};
\draw[thick,dir] (4,0) .. controls (3.5,0.6) and (1.5,0.6) .. (1,0);
\end{tikzpicture}=\langle W_1,W_2\rangle\,.
\end{align}
This identity can be proved by writing out the definition of the $C$'s in full and then average over different ways of assigning dummy indices. Combining Eq.~\eqref{three3} and~\eqref{three2}, we get
\begin{align}\label{3trace2}
{\rm PT}(W_1)\,{\rm PT}(W_2)\,{\rm PT}(W_3)
&\overset{\rm IBP}{\cong}
\frac{ {\rm PT}(W_3)}{(1-s_{W_1})(1-s_{W_2})}\Big[\alpha'^2\mathcal{T}_{W_3}(W_1,W_2)+\langle W_1,W_2\rangle\Big]\,,  
\end{align}
where $\mathcal{T}_{W_3}(W_1,W_2)$, when dressed with $\text{PT}(W_3)$, is a combination of logarithmic functions:
\begin{align}
\la{fqefhpqo}
\alpha'^2\mathcal{T}_{W_3}(W_1,W_2)=    \sum_{j_3\in W_3}C^{W_1}_{a_1,j_3}     \sum_{j\notin W_2}C^{W_2}_{a_2,j}
+ \sum_{j_2\in W_2}\sum_{j_3 \in W_3}   C^{W_1}_{a_1,j_2} C^{W_2}_{j_2,j_3}\,.
\end{align}
We can interpret $\mathcal{T}_{W_3}(W_1,W_2)$ as a set of labeled trees rooted on $W_3$ and evaluated under the \emph{reference order} that $W_1$ proceeds $W_2$, denoted as $W_{1}\prec W_{2}$. These labeled trees are illustrated in the last column of figure~\ref{threefafd}. The generic construction of these labeled trees will be given in Appendix~\ref{sec:trees}. A remarkable feature of Eq.~\eqref{3trace2} is that the triple-trace integrand is given as a linear combination of logarithmic functions and double-trace integrands. 

The recursive nature of this derivation is more obvious when we carry on to four traces. After setting a puncture in $W_4$ to infinity, we break the subcycle $W_1$ by Eq.~\eqref{stibp}. Now the chain $C^{W_1}$ can either connect to another subcycle ($W_2$ or $W_3$), forming a normal tadpole, or to the root $W_4$. Next, we break all the normal tadpoles by Eq.~\eqref{stibp} at the attach point of the tail, including those generated in the process. At the end, the chain $C^{W_1}$ is either connected to the root (maybe through another chain), or appears in a induced tadpole. We then repeat these steps for subcycle $W_2$ (if exists), followed by $W_3$ (if exists). This prescription introduces a reference order $W_1\prec W_2\prec W_3$ for the subcycles as the priority rank of being broken by the IBP relation~\eqref{stibp}. The result of the above IBP reduction is
\begin{align}\label{4tracea}
\prod_{i=1}^{4}\text{PT}(W_i)\overset{\text{IBP}}{\cong}\frac{\text{PT}(W_4)}{\prod_{i=1}^3(1-s_{W_i})}&\Big[\alpha'^3\mathcal{T}_{W_4}(W_1,W_2,W_3)+\langle W_1,W_2,W_3\rangle+\langle W_1,W_3,W_2\rangle\nonumber\\
&+\big((1-s_{W_1})\,\text{PT}(W_1)\langle W_2,W_3\rangle+\text{cyclic}\big)\Big]\,,
\end{align}
where $\mathcal{T}_{W_4}(W_1,W_2,W_3)$ is a set of labeled trees rooted on $W_4$ and evaluated under the reference order $W_1\prec W_2\prec W_3$. Again we refer the readers to Appendix~\ref{sec:trees} for the construction of these labeled trees. Different reference orders will give different $\mathcal{T}_{W_4}$'s that are equivalent up to an IBP relation. The fusions in the first line of Eq.~\eqref{4tracea} come from induced tadpoles after using the algebraic identity
\begin{align}
\sum_{j_1\in W_1}\sum_{j_2\in W_2}\sum_{j_3\in W_3}\!\!\left(C^{W_1}_{a_1,j_2}C^{W_2}_{j_2,j_3}C^{W_3}_{j_3,j_1}+(2\leftrightarrow 3)\right)=\langle W_1,W_2,W_3\rangle+\langle W_1,W_3,W_2\rangle\,,
\end{align}
while the fusions in the second line of Eq.~\eqref{4tracea} are obtained by first using Eq.~\eqref{fusionCC} and then the inversion of the relation~\eqref{stibp}, for example:
\begin{align}
\sum_{j\notin W_1}C^{W_1}_{a_1,j}\langle W_2,W_3\rangle\,\text{PT}(W_4)\overset{\text{IBP}}{\cong}(1-s_{W_1})\,\text{PT}(W_1)\langle W_2,W_3\rangle\,\text{PT}(W_4)\,.
\end{align}
Although the above manipulation seems to be a move in the opposite direction, the benefit is that now we can use directly the double- and triple-trace results in Eq.~\eqref{4tracea}. 


\subsection{General pure-scalar cases }\label{ciqelhf}
Now we provide the IBP reduction algorithm for generic $(m+1)$-trace string integrands, where we gauge fix one of the punctures in $W_{m+1}$ to infinity. We first pick an arbitrary reference order, say $\pmb{R}=W_1\prec W_2\prec\ldots\prec W_m$, as the priority rank of being broken by the IBP relation~\eqref{stibp}. For each term in the integrand, we carry out the following algorithm:
\begin{enumerate}[label=(t\arabic*)]
	\item Break the first trace in the reference order (here $W_1$) by Eq.~\eqref{stibp}, which turns the trace $W_1$ into a chain $C^{W_1}$ that is attached to another subcycle or the root $W_{m+1}$. \label{step1}
	\item If the chain $C^{W_1}$ appears in the tail of a normal tadpole, break that subcycle using Eq.~\eqref{stibp} at the attach point of the tail. Repeat this step until $C^{W_1}$ is connected to the root $W_{m+1}$ (maybe through another chain) or appears in an induced tadpole.\label{step2}
	\item Repeat step~\ref{step1} and \ref{step2} for the next subcycle $W_i$ in the reference order that remains in the original $\text{PT}(W_i)$ configuration.\label{step3}
\end{enumerate}
The algorithm leads to a remarkable recursive expansion for the string integrand:
\begin{align}\label{ntracea}
\prod_{i=1}^{m+1}\!\text{PT}(W_i)\overset{\text{IBP}}{\cong}\frac{\text{PT}(W_{m+1})}{\prod_{i=1}^m(1-s_{W_i})}\Bigg[&\alpha'^{m}\mathcal{T}_{W_{m+1}}(\pmb{R})-\!\!\sum_{\substack{\mathsf{A}\in\mathbb{P}[W_1\ldots W_m] \\ |\mathsf{A}|<m}}\!\!(-1)^{|\mathsf{A}|}\mathcal{J}[\mathsf{A}]\Bigg],
\end{align}
where $\mathcal{T}_{W_{m+1}}(\pmb{R})$ consists of labeled trees only, and thus logarithmic. It is actually the logarithmic form CHY integrand for the pure-scalar sector of Yang-Mills scalar amplitudes. The explicit form depends on the reference order $\pmb{R}$ (for example, $W_1\prec W_2\prec\ldots\prec W_m$), and we defer the details to Appendix~\ref{sec:trees}. The second term of Eq.~\eqref{ntracea} is independent of the reference order, in which the summation is over all the partitions $\mathsf{A}$ of $\{W_1,W_2,\ldots,W_m\}$ whose number of blocks is less than $m$. Suppose $\mathsf{A}$ has $s$ singleton blocks and the rest non-singletons, namely, $\mathsf{A}=\{\mathsf{a}_1,\ldots,\mathsf{a}_s,\mathsf{A}_{s+1},\ldots,\mathsf{A}_{|\mathsf{A}|}\}$, we can write $\mathcal{J}$ as
\begin{align}\label{eq:Jtrace}
\mathcal{J}[\mathsf{A}]=(-1)^s\Bigg[\prod_{j=1}^{s}(1-s_{\mathsf{a}_j})\,\text{PT}(\mathsf{a}_j)\Bigg]\Bigg[\prod_{j=s+1}^{|\mathsf{A}|}\mathcal{S}_0(\mathsf{A}_j)\Bigg]\,.
\end{align}
They come from the induced tadpoles generated in step \ref{step2} after we use some algebraic identities like
\begin{align}\la{cquef}
\sum_{j_2\in W_2}\!C^{W_1}_{a_1,j_2}\sum_{j_3\in W_3}\!C^{W_2}_{j_2,j_3}\cdots &\sum_{j_r\in W_r}\!C^{W_r}_{j_{r-1},j_r}+\text{perm}\,(2,3,\ldots,r)=\mathcal{S}_0(W_1,W_2,\ldots,W_r)\,.
\end{align}
In addition, we need to use the inverse of Eq.~\eqref{stibp} to obtain the $(1-s_{\mathsf{a}_j})\,\text{PT}(\mathsf{a}_j)$ factor. 

We note that each singleton block $\mathsf{a}_i$ contributes factor proportional to the original PT factor $\text{PT}(\mathsf{a}_i)$. The condition $|\mathsf{A}|<m$ in Eq.~\eqref{ntracea} guarantees that there exists at least one non-singleton block $\mathsf{A}_j=\{W_{j_1},W_{j_2},\ldots\}$. Each $\mathcal{S}_0(\mathsf{A}_j)$ then merges the traces in the non-singleton block $\mathsf{A}_j$ into a single trace according to Eq.~\eqref{eq:S0}. In factor, $\mathcal{S}_{0}(\mathsf{A}_j)$ is a linear combination of $\text{PT}(\rho)$ with $\rho$ belong to a subset of $\text{perm}(\mathsf{A}_j)=\text{perm}(W_{J_1}\cup W_{j_2}\cup\ldots)$. The symmetrization defined in $\mathcal{S}_0$ takes care of the bosonic exchange symmetry between the original traces.

Therefore, $\mathcal{J}[\mathsf{A}]$ is a product of PT factors taking value in each block of $\mathsf{A}$ respectively, the physical meaning of  $\mathcal{J}[\mathsf{A}]$ is then clear: \emph{it is a linear combination of the string integrands with less number of traces,} \emph{cf.} the left hand side of Eq.~\eqref{ntracea}. It is non-logarithmic because of the existence of subcycles. Nevertheless, we can recursively use Eq.~\eqref{ntracea} to eventually obtain a logarithmic integrand. On the other hand, further IBP reduction on $\mathcal{J}[\mathsf{A}]$ only leads to contributions to higher order of $\alpha'$. Thus in the limit $\alpha'\rightarrow 0$, we have
\begin{align}
\prod_{i=1}^{m+1}\!\text{PT}(W_i)\overset{\text{IBP}}{\cong}\alpha'^{m}\text{PT}(W_{m+1})\mathcal{T}_{W_{m+1}}(\pmb{R})+\mathcal{O}(\alpha'^{m+1})\,,
\end{align}
where the first term is the result if we would have started with a type-I superstring correlator and obtained the multitrace structure through a compactification.

Our algorithm can reduce any multitrace correlator to a logarithmic function. In the arXiv submission of this paper, we implement the recursive expansion~\eqref{ntracea} in the ancillary Mathematica notebook {\tt IBP.nb}.
Given a reference order, the result can either be exported as a linear combination of labeled trees or further expanded in terms of Parke-Taylor factors in the DDM basis~\cite{DelDuca:1999rs}. Run on a laptop, our algorithm can process eleven points with five traces in a few minutes.

\subsection{Inserting one gluon}
The single-gluon string integrand $C_i\prod_{j=1}^{m+1}\text{PT}(W_j)$ is only slightly more general than the pure-scalar one: each term in the integrand contains exactly one tadpole whose tail is a single gluon. We start with breaking the tadpole at the gluon attach point by Eq.~\eqref{stibp}, and then follow the same prescription as the pure-scalar case. In other words, the subcycle connected with the gluon is always prioritized. This essentially means that we choose the reference order $\pmb{R}=i\prec W_1\prec\ldots\prec W_m$. The result can again be written as
\begin{align}\label{ntracafwweea}
C_i\prod_{i=1}^{m+1}\!\text{PT}(W_i)\overset{\text{IBP}}{\cong}\frac{\text{PT}(W_{m+1})}{\prod_{i=1}^m(1-s_{W_i})}\Bigg[&\alpha'^{m}\mathcal{T}_{W_{m+1}}(i,W_1,\ldots,W_{m})-\!\!\!\sum_{\substack{\mathsf{A}\in\mathbb{P}[i,W_1\ldots W_m] \\ |\mathsf{A}|<m+1}}\!\!\!(-1)^{|\mathsf{A}|}\mathcal{J}[\mathsf{A}]\Bigg],
\end{align} 
where the definition of $\mathcal{J}[\mathsf{A}]$ is extended to include one single gluon. In particular, if there is a gluon singleton block, namely, $\mathsf{A}=\{i,\mathsf{a}_2,\ldots,\mathsf{a}_s,\mathsf{A}_{s+1},\ldots,\mathsf{A}_{|\mathsf{A}|}\}$, 
\begin{align}\label{J1g}
\mathcal{J}[\mathsf{A}]=(-1)^sC_i\Bigg[\prod_{j=2}^{s}(1-s_{\mathsf{a}_j})\,\text{PT}(\mathsf{a}_j)\Bigg]\Bigg[\prod_{j=s+1}^{|\mathsf{A}|}\mathcal{S}_0(\mathsf{A}_j)\Bigg]\,.
\end{align}
Otherwise, the definition is the same as Eq.~\eqref{eq:Jtrace} but with $\mathcal{S}_0$ involving fusion between the gluon and traces, which is obtained through algebraic identities like
\begin{align}
\sum_{j_1\in W_1}\!C_{ij_1}\!\!\sum_{j_2\in W_2}\!C^{W_1}_{j_1,j_2}\cdots\!\sum_{j_{r}\in W_{r}}\!C^{W_{r-1}}_{j_{r-1},j_{r}} C^{W_r}_{j_{r},i}+\text{perm}(1,2,\ldots,r)= \mathcal{S}_{0}(i,W_1,\cdots,W_r)\,.
\end{align}
As a very simple example of Eq.~\eqref{ntracafwweea}, we show the expansion of the double-trace single-gluon integrand under the reference order $i\prec W_1$,
\ba\la{chqpof}
C_i \PT(W_1)\PT(W_2) \overset{\text{IBP}}{\cong} &\Big[{\cal T}_{W_2}(i,W_1)+ \<i,W_1\>\Big] \PT(W_2)\,,
\ea
where $\mathcal{T}_{W_2}(1,W_1)$ is a combination of logarithmic functions:
\begin{align}
\mathcal{T}_{W_2}(i,W_1)=    \sum_{j_2\in W_2}C_{ij_2}    \sum_{j\notin W_1}C_{a_1j}^{W_1}
+ \sum_{j_1 \in W_1} \sum_{j_2\in W_2}  C_{ij_1}  C_{j_1,j_2}^{W_1}   \,.
\end{align}
We note that to obtain Eq.~\eqref{ntracafwweea} with a more generic reference ordering in which the gluon $i$ appears between traces, we need to use some new IBP relations that allow us to break a subcycle at a point different from the gluon attach point. We will discuss these IBP relations in Appendix~\ref{sec:multibranch}.

\subsection{General cases with gluons}
For the most generic string integrand~\eqref{eq:strInt}, we encounter graphs with multiple tails attached to a subcycle that consists of a single trace $W_i$ or a set of gluons. We thus need to perform IBP reductions on these \emph{multibranch graphs} \cite{He:2018pol}, 
which will be discussed in details in Appendix~\ref{sec:multibranch}. However, using the intuition developed in the single-gluon formula~\eqref{ntracafwweea}, one can be convinced that the generic recursive expansion is
\begin{align}\label{stringExpansion}
R(i_1,\ldots,i_r)&\prod_{j=1}^{m+1}\text{PT}(W_j)\nonumber\\
&\overset{\text{IBP}}{\cong}
\frac{\text{PT}(W_{m+1})}{\prod_{j=1}^m(1-s_{W_j})}\Bigg[\alpha'^{m}\mathcal{T}_{W_{m+1}}(\pmb{R})-\!\!\sum_{\substack{\mathsf{A}\in\mathbb{P}[i_1,\ldots,i_r,W_1\ldots W_m]\\ |\mathsf{A}|<r+m }}\!\!\!(-1)^{|\mathsf{A}|}\mathcal{J}[\mathsf{A}]\Bigg]\,,
\end{align}
where the first term is the logarithmic form CHY integrand for generic Yang-Mills-scalar amplitudes written under the reference order $\pmb{R}$. A very convenient choice is to put all gluons before the traces:
\begin{equation}\label{reference}
\pmb{R}=i_1\prec\ldots\prec i_r\prec W_1\prec\ldots\prec W_m\,.
\end{equation}
In the string context, it corresponds to the contribution from a compactified superstring.

The non-logarithmic function $\mathcal{J}$ is generalized from Eq.~\eqref{J1g} to include more gluons. For a partition $\mathsf{A}$ that contains $s$ singleton blocks, in which $t$ of them are gluons and the rest traces, namely, $\mathsf{A}=\{\mathsf{a}_1,\ldots\mathsf{a}_t,\mathsf{b}_{t+1},\ldots,\mathsf{b}_{s},\mathsf{A}_{s+1},\ldots\mathsf{A}_{|\mathsf{A}|}\}$, we have
\begin{align}
\mathcal{J}[\mathsf{A}]=(-1)^{s}R(\mathsf{a}_1,\ldots,\mathsf{a}_t)\Bigg[\prod_{j=t+1}^s(1-s_{\mathsf{b}_j})\,\text{PT}(\mathsf{b}_j)\Bigg]\Bigg[\prod_{j=s+1}^{|\mathsf{A}|}\mathcal{S}_0(\mathsf{A}_j)\Bigg]\,.
\end{align}
Similar to the discussion in section~\ref{ciqelhf}, the $|\mathsf{A}|<r+m$ condition in Eq.~\eqref{stringExpansion} guarantees that there must be at least one non-singleton block $\mathsf{A}_j$ in $\mathsf{A}$ and thus at least one nontrivial fusion $\mathcal{S}_{0}(\mathsf{A}_j)$. It is a linear combination of PT factors taking value in a subset of $\text{perm}(\mathsf{A}_j)$, where the coefficients contain the polarization vectors if $\mathsf{A}_j$ contain gluons. 

Now combining the contributions from every block in $\mathsf{A}$, we can see that 
$\mathcal{J}[\mathsf{A}]$ is a string integrand with total number of traces and gluons decreased due to the of nontrivial fusions. Consider the triple-trace three-gluon integrand $R(i_1,i_2,i_3)\text{PT}(W_1)\text{PT}(W_2)\text{PT}(W_3)$, with total number of gluons and traces being six. After pulling out the overall factor $\text{PT}(W_3)$, we need to consider the partitions of the set $\{i_1,i_2,i_3,W_1,W_2\}$. Two such examples are
\begin{align}
  & \mathsf{A}=\{i_2,i_3, W_1 ,\{i_1,W_2\}  \} & & \Rightarrow & & \mathcal{J}[\mathsf{A}]=-R(i_2,i_3) (1-s_{W_1})\PT(W_1)\mathcal{S}_{0}(i_1,W_2)\,, \nonumber\\
  & \mathsf{A}=\{i_1, i_2,i_3,\{ W_1,W_2\}  \} & & \Rightarrow & & \mathcal{J}[\mathsf{A}]=-R(i_1,i_2,i_3)\mathcal{S}_{0}(W_1,W_2)\,,
\end{align}
where the first line gives a linear combination of triple-trace two-gluon integrands and the second line double-trace three-gluon integrands. For both cases, the total number of gluons and traces is five.
When there are no gluons, this $\mathcal{J}[\mathsf{A}]$ reduces trivially to the one in Eq.~\eqref{eq:Jtrace}.

We give several examples for this recursive expansion under the reference order~\eqref{reference}. First, The recursive expansion for single-trace integrands with two and three gluons will be worked out in detail in Appendix~\ref{sec:examples}. As a more involving case, the single-trace four-gluon integrand can be expanded as
\begin{align}
	R(1,2,3,4)\text{PT}(W_1)\overset{\text{IBP}}{\cong}\text{PT}(W_1)&\Big[{\cal T}_{W_1}(1,2,3,4)+ {\cal S}_0(1,2,3,4)+\big( {\cal S}_0(1,2,3) C_4 +\text{cyclic}\big) \nonumber\\
	&-\mathcal{S}_0(1,2)\mathcal{S}_0(3,4)-\mathcal{S}_0(1,3)\mathcal{S}_0(2,4)-\mathcal{S}_0(1,4)\mathcal{S}_0(2,3)\nonumber\\
	&+R(1,2)\mathcal{S}_0(3,4) + R(3,4)\mathcal{S}_0(1,2) + R(1,3)\mathcal{S}_0(2,4) \nonumber\\
	&+ R(2,4)\mathcal{S}_0(1,3)+R(1,4)\mathcal{S}_0(2,3) + R(2,3)\mathcal{S}_0(1,4)\Big].
\end{align}
The next example is the expansion of the double-trace two-gluon integrand,
\begin{align}
	R(1,2)\text{PT}(W_1)\text{PT}(W_2)\overset{\text{IBP}}{\cong}\frac{\text{PT}(W_2)}{1-s_{W_1}}&\Big[\alpha'\mathcal{T}_{W_2}(1,2,W_1)+\mathcal{S}_{0}(1,2,W_1)+C_1\mathcal{S}_0(2,W_1)\nonumber\\
	&+C_2\mathcal{S}_0(1,W_1)+(1-s_{W_1})\text{PT}(W_1)\mathcal{S}_0(1,2)\Big]\,.
\end{align}
In Appendix~\ref{sec:appb}, we will provide more details on the derivation of the generic formula~\eqref{stringExpansion}.

\section{Derivation of the CHY integrand}\la{fuqiwepfh}
The recursive expansion~\eqref{stringExpansion} of string integrands might have a very wide application. In this section, we show how to use it to derive inductively the CHY integrand of the pure-scalar sector of $(DF)^2+\text{YM}+\phi^3$, Eq.~\eqref{eq:st1daf}. The derivation of the most generic integrand~\eqref{eq:Imultitrace} is very similar and we will comment on it at the end. 

The induction starts at double trace. The logarithmic function~\eqref{doubleTraceLog} can be further simplified by SE as
\begin{align}\label{SEred}
\text{PT}(W_1)\,\text{PT}(W_2)\overset{\text{IBP}}{\cong}\frac{\text{PT}(W_2)}{1-s_{W_1}}\sum_{j_2\in W_2}C^{W_1}_{a_1,j_2}\overset{\text{SE}}{\cong}-\frac{s_{W_1}}{1-s_{W_1}}\,\text{PT}(W_1)\text{PT}(W_2)\,.
\end{align}
We can combine the two-step process and write
\begin{align}\label{2trRed}
\text{PT}(W_1)\,\text{PT}(W_2)\overset{\text{IBP}+\text{SE}}{\cong}-\frac{s_{W_1}}{1-s_{W_1}}\,\text{PT}(W_1)\,\text{PT}(W_2)\,.
\end{align}
This CHY integrand was first identified in ~\cite{Azevedo:2018dgo}. We use ``IBP$+$SE'' to stand for the process of IBP reduction to logarithmic functions followed by a SE simplification.

To derive the triple-trace CHY integrand, we can directly apply the double-trace result~\eqref{2trRed} to the second term of Eq.~\eqref{3trace2}:
\begin{align}\label{recursivep}
\langle W_1,W_2\rangle\,{\rm PT}(W_3) \overset{\text{IBP}+\text{SE}}{\cong}&\,\frac{ -s_{W_1W_2}\langle W_1,W_2\rangle{\rm PT}(W_3)}{1-s_{W_1,W_2}}\nonumber\\
\overset{\hphantom{\text{IBP}+\text{SE}}}{=}&\,\Big[\langle W_1,W_2\rangle-\mathcal{S}_{\alpha'}(W_1,W_2)\Big]\text{PT}(W_3)\,.
\end{align}
Finally, the $\langle W_1,W_2\rangle$ in the above equation, when combined with $\mathcal{T}_{W_3}(W_1,W_2)$, produces the last piece of the triple-trace CHY integrand:
\begin{align}\label{EYMtree3}
\Big[\alpha'^2\mathcal{T}_{W_3}(W_1,W_2)+\langle W_1,W_2\rangle\Big]\text{PT}(W_3)\overset{\text{SE}}{\cong}s_{W_1}s_{W_2}\text{PT}(W_1)\,\text{PT}(W_2)\,\text{PT}(W_3)\,.
\end{align}
This completes the derivation of the triple-trace CHY integrand from the string integrand
\begin{align}\label{threettt}
\prod_{i=1}^3 {\rm PT}(W_i)
\overset{\text{IBP}+\text{SE}}{\cong}
\frac{{\rm PT}(W_3)}{(1-s_{W_1})(1-s_{W_2})}\Big[\mathcal{S}_{\alpha'}(W_1)\mathcal{S}_{\alpha'}(W_2)  - \mathcal{S}_{\alpha'}(W_1,W_2)  \Big]\,,
\end{align}
which agrees with Eq.~\eqref{eq:st1daf}.


The calculation at four traces is also similar. The last two terms in the first line of Eq.~\eqref{4tracea} are double-trace string integrands, such that Eq.~\eqref{2trRed} leads to
\begin{align}\label{4tracel1}
\text{PT}(W_4)&\Big(\langle W_1,W_2,W_3\rangle+\langle W_1,W_3,W_2\rangle\Big)\nonumber\\
&\overset{\text{IBP}+\text{SE}}{\cong}\text{PT}(W_4)\Big(\langle W_1,W_2,W_3\rangle+\langle W_1,W_3,W_2\rangle-\mathcal{S}_{\alpha'}(W_1,W_2,W_3)\Big)\,.
\end{align}
On the other hand, the terms in the second line of Eq.~\eqref{4tracea} are all triple-trace integrands, from which we can generate nested fusions using Eq.~\eqref{threettt}. For example,
\begin{align}\label{4tracel2}
(1-s_{W_1})\,\text{PT}(W_1)\langle W_2,W_3\rangle&\,\text{PT}(W_4)\nonumber\\
\overset{\text{IBP}+\text{SE}}{\cong}\text{PT}(W_4)\Big[&-s_{W_1}\text{PT}(W_1)\langle W_2,W_3\rangle+\mathcal{S}_{\alpha'}(W_1)\mathcal{S}_{\alpha'}(W_2,W_3)\nonumber\\
&-\mathcal{S}_{\alpha'}\big(W_1,\mathcal{S}_{\alpha'}(W_2,W_3)\big)\Big]\,,
\end{align}
and the rest are obtained by cyclic permutations. If we plug the above two equations back to Eq.~\eqref{4tracea}, add and subtract $\prod_{i=1}^3\big[s_{W_i}\text{PT}(W_i)\big]$, the labeled trees are exactly canceled due to the relation
\begin{align}\label{EYMtree4}
\alpha'^3\mathcal{T}_{W_4}(W_1,W_2,W_3)\overset{\text{SE}}{\cong}&-\prod_{i=1}^{3}\big[s_{W_i}\text{PT}(W_i)\big]-\langle W_1,W_2,W_3\rangle-\langle W_1,W_3,W_2\rangle\nonumber\\
&+\Big(s_{W_1}\text{PT}(W_1)\langle W_2,W_3\rangle+\text{cyclic}\Big)\,,
\end{align} 
where we use the definition~\eqref{eq:I0} for the Yang-Mills-scalar integrand on the right hand side. Collecting all the relevant terms, we get the four-trace CHY integrand
\begin{align}
\prod_{i=1}^4\text{PT}(W_i)\overset{\text{IBP}+\text{SE}}{\cong}&\,\frac{\text{PT}(W_4)}{\prod_{i=1}^3(1-s_{W_i})}\Bigg[-\prod_{i=1}^3\mathcal{S}_{\alpha'}(W_i)-\mathcal{S}_{\alpha'}(W_1,W_2,W_3)\nonumber\\
&+\Big(\mathcal{S}_{\alpha'}(W_1)\mathcal{S}_{\alpha'}(W_2,W_3)-\mathcal{S}_{\alpha'}\big(W_1,\mathcal{S}_{\alpha'}(W_2,W_3)\big)+\text{cyclic}\Big)\Bigg]\,,
\end{align}
which again agrees with Eq.~\eqref{eq:st1daf}.

In fact, starting from Eq.~\eqref{ntracea}, we can derive the $(m+1)$-trace CHY integrand~\eqref{eq:st1daf} inductively. Since the second term of Eq.~\eqref{ntracea} is a linear combination of string integrands with fewer traces, we can simplify it using Eq.~\eqref{eq:st1daf} as our inductive assumption. After some algebras, one can show that the result is
\begin{align}\label{eq:induction1}
&-\!\!\!\!\sum_{\substack{\mathsf{A}\in\mathbb{P}[W_1\ldots W_m] \\ |\mathsf{A}|<m}}\!\!\!(-1)^{|\mathsf{A}|}\mathcal{J}[\mathsf{A}]\overset{\text{IBP}+\text{SE}}{\cong}\!\!\!\!\!\!\sum_{\mathsf{A}\in\mathbb{T}[W_1\ldots W_m]}\!\!(-1)^{|\mathsf{A}|}\!\prod_{j=1}^{|\mathsf{A}|}\mathcal{S}_{\alpha'}(\mathsf{A}_j)-\!\!\!\sum_{\mathsf{A}\in\mathbb{P}[W_1\ldots W_m]}\!\!(-1)^{|\mathsf{A}|}\!\prod_{j=1}^{|\mathsf{A}|}\mathcal{S}_{0}(\mathsf{A}_j)\,.
\end{align}
While the first term is precisely the desired $(m+1)$-trace integrand~\eqref{eq:st1daf}, the second term exactly cancels the labeled trees:
\begin{align}\label{eq:induction2}
\alpha'^{m}\mathcal{T}_{W_{m+1}}(\pmb{R})\overset{\text{SE}}{\cong}\sum_{\mathsf{A}\in\mathbb{P}[W_1\ldots W_m]}\!\!(-1)^{|\mathsf{A}|}\!\prod_{j=1}^{|\mathsf{A}|}\mathcal{S}_{0}(\mathsf{A}_j)\,,
\end{align}
since both of them are valid CHY integrands for the same amplitude and thus must equal on the support of the scattering equations. Finally, we note that the most generic CHY integrand~\eqref{eq:Imultitrace} for multitrace $(DF)^2+\text{YM}+\phi^3$ can be inductively derived following a procedure similar to Eq.~\eqref{eq:induction1} and~\eqref{eq:induction2}.


\section{Special massless factorizations}
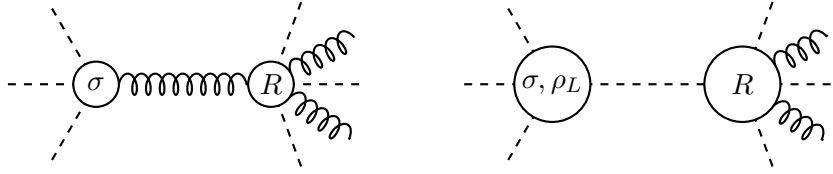
\begin{figure}[t]
\centering
\begin{tikzpicture}
\draw[thick,dashed] (0,0) -- (-1.2,0) (0,0) -- (120:1.2) (0,0) -- (-120:1.2);
\draw[thick,dashed] (2.3,0) -- ++ (1.2,0) (2.3,0) -- ++ (70:1.2) (2.3,0) -- ++ (-70:1.2);
\draw[thick,decorate,decoration={coil, amplitude=4pt,segment length=5.15pt}] (0.3,0) -- (2,0);
\draw[thick,decorate,decoration={coil, amplitude=4pt,segment length=5.15pt}] (2.3,0) ++ (30:0.3) -- ++(30:0.97);
\draw[thick,decorate,decoration={coil, amplitude=4pt,segment length=5.15pt}] (2.3,0) ++ (-35:0.3) -- ++(-35:0.97);
\filldraw [fill=white,draw=black,thick] (0,0) circle (0.3cm);
\filldraw [fill=white,draw=black,thick] (2.3,0) circle (0.3cm);
\node at (0,0) {$\sigma$};
\node at (2.3,0) {$R$};
\begin{scope}[xshift=6cm]
\draw[thick,dashed] (0,0) -- (-1.2,0) (0,0) -- (120:1.2) (0,0) -- (-120:1.2);
\draw[thick,dashed] (2.5,0) -- ++ (1.2,0) (2.5,0) -- ++ (70:1.2) (2.5,0) -- ++ (-70:1.2) (0,0) -- (2.5,0);
\draw[thick,decorate,decoration={coil, amplitude=4pt,segment length=5.15pt}] (2.5,0) ++ (30:0.5) -- ++(30:0.8);
\draw[thick,decorate,decoration={coil, amplitude=4pt,segment length=5.15pt}] (2.5,0) ++ (-35:0.5) -- ++(-35:0.8);
\filldraw [fill=white,draw=black,thick] (0,0) circle (0.5cm);
\filldraw [fill=white,draw=black,thick] (2.5,0) circle (0.5cm);
\node at (0,0) {$\sigma,\rho_L$};
\node at (2.5,0) {$R$};
\end{scope}
\end{tikzpicture}
\caption{Two special massless factorization channels of our integrand. The dashed lines represent scalars and the curly lines represent gluons. Note that there are no original external gluons on the left.}
\label{fig:factorize}
\end{figure}

\la{fq3uofi}
As an important consistency check, our integrand~\eqref{eq:Imultitrace} should demonstrate the correct factorization behavior. In particular, we consider two special massless factorization channels as shown in figure~\ref{fig:factorize}: \begin{enumerate*}[label=(\roman*)]
\item we cut out exactly a single trace $\sigma$. The on-shell internal propagator is thus a gluon;
\item we also cut out part of a second trace $\rho_L\subset\rho$. The on-shell internal propagator is thus a scalar.
\end{enumerate*}

We start with introducing essential tools for studying factorization in the CHY framework.  We consider a generic physical factorization limit $q_L^2\rightarrow 0$, where
\begin{align}\label{eq:channel1}
&\sum_{i=1}^{n_L}k_i=-q_L\,,\qquad\sum_{i=n_L+1}^{n}k_i=-q_R=q_L\,.
\end{align}
We follow the prescription of~\cite{Cachazo:2013iea} and change the variables to
\begin{align}\label{eq:newvar}
& z_a=\frac{\zeta}{u_a}\qquad a\in L=\{1,2\ldots n_L\}\,,\nonumber\\
&z_a=\frac{v_a}{\zeta}\qquad a\in R=\{n_L+1,n_L+2\ldots n\}\,,
\end{align}
where we have fixed one of the $v$'s, say, $v_{n-1}=v^*_{n-1}$. In terms of the new variables, the scattering equations for $L$ and $R$ are independent of each other at the zeroth order of $\zeta^2$:
\begin{subequations}
\begin{align}
& a\in L:& &0=\sum_{b\in L\cup\{q_L\}\backslash\{a\}}\frac{k_a\!\cdot\!k_b}{u_{ab}}+\mathcal{O}(\zeta^2)\,,\\
& a\in R:& &0=\sum_{b\in R\cup\{q_R\}\backslash\{a\}}\frac{k_a\!\cdot\!k_b}{v_{ab}}+\mathcal{O}(\zeta^2)\,,
\end{align}
\end{subequations}
where we have used the gauge choice $u_{q_L}=v_{q_R}=0$. On the other hand, $\zeta^2$ satisfy the following equation,
\begin{align}
0=-\frac{q_L^2}{2}+\zeta^2\sum_{a\in R}\sum_{b\in L}\frac{k_a\!\cdot\!k_b}{v_au_b-\zeta^2}\equiv-\frac{q_L^2}{2}+\frac{\zeta^2}{2}F(u,v,k_i)+\mathcal{O}(\zeta^4)\,,
\end{align}
where $F:=\sum_{a\in R}\sum_{b\in L}\frac{2k_a \cdot k_b}{v_au_b}$ is independent of $\zeta$. In the limit $q_L^2\rightarrow 0$, there always exists a singular solution
\begin{align}\label{eq:zeta2}
\zeta^2=\frac{q_L^2}{F(u,v,k_i)}+\mathcal{O}(q_L^4)\,.
\end{align}
We can ignore other solutions of $\zeta^2$ since they are only relevant to subleading orders in the factorization limit.

The benefit of these new worldsheet variables is that, at the leading order of $\zeta$, the CHY integration measure factorizes nicely as~\cite{Cachazo:2013iea}
\begin{align}\label{eq:measureCHY}
d\mu_{\text{CHY}}\sim d\mu_L d\mu_R\frac{d\zeta^2}{\zeta^2}\delta(\zeta^2 F-q_L^2)\frac{\zeta^{2n_L-2n_R-4}}{\mathcal{U}^4}\,,
\end{align}
where $\mathcal{U}:=\prod_{a=1}^{n_L}u_a$, and
\begin{align}
& d\mu_L=(u_1u_2u_{12})^2\Bigg[\prod_{a=3}^{n_L}du_a\delta(E_a^L)\Bigg]\,, & &d\mu_R=(v_{n-1}v_{n}u_{n-1,n})^2\Bigg[\prod_{a=n_L+1}^{n-2}dv_a\delta(E_a^R)\Bigg]\,,\nonumber\\
& E_a^L=\sum_{b\in L\cup\{q_L\}\backslash\{a\}}\frac{k_a\!\cdot\!k_b}{u_{ab}}\,,& & E_a^R=\sum_{b\in R\cup\{q_R\}\backslash\{a\}}\frac{k_a\!\cdot\!k_b}{v_{ab}}\,.
\end{align}
Namely, $d\mu_L$ and $d\mu_R$ are nothing but the integration measure for $L\cup\{q_L\}$ and $R\cup\{q_R\}$ respectively, where $q_L^2=q_R^2=0$. As a universal building block for gauge amplitudes, the Parke-Taylor factor becomes
\begin{align}
\text{PT}(1,2,\ldots,n)\sim (-1)^{n_L}\zeta^{n_R-n_L+2}\,\mathcal{U}^2\,\text{PT}(1,2,\ldots,n_L,q_L)\,\text{PT}(q_R,n_L+1,\ldots,n)\,,
\end{align}
where the two PT's on the right hand side are given by $u$ and $v$ variables respectively.\footnote{In the following, it is understood that in a factorization analysis, particles in $L$ are always associated with the $u$'s while particles in $R$ with the $v$'s.} 
We expect a valid CHY half-integrand $\mathcal{I}_n^{\text{CHY}}$ to behave as
\begin{align}\label{eq:CHYfactorize}
\mathcal{I}_n^{\text{CHY}}\sim(-1)^{n_L}\zeta^{n_R-n_L+2}\,\mathcal{U}^2\,\sum_{\text{states}}\mathcal{I}_L^{\text{CHY}}(1,2,\ldots,n_L,q_L)\,\mathcal{I}_R^{\text{CHY}}(q_R,n_L+1,\ldots,n)\,,
\end{align}
where the summation is over the on-shell states on the factorization channel $q_L^2=0$. If this is true, then the $\zeta^{2n_R-2n_L+4}/\mathcal{U}^4$ factor in the measure~\eqref{eq:measureCHY} will be cancel, such that the $\zeta^2$ integration will provide the desired massless pole:
\begin{align}
\int\frac{d\zeta^2}{\zeta^2}\delta(\zeta^2F-q_L^2)=\frac{1}{q^2_L}\,,
\end{align}
We thus obtain the correct factorization behavior
\begin{align}
A(1,2,\ldots,n)\sim\sum_{\text{states}} A_L(1,2,\ldots,n_L,q_L)\,\frac{1}{q^2_L}\,A_R(q_R,n_{L+1},\ldots,n)\,,
\end{align}
where $A_L$ and $A_R$ are the amplitude given by the half-integrand $\mathcal{I}_L$ and $\mathcal{I}_R$.

\subsection{An example: factorization of the triple-trace scalar integrand}
We apply the above construction to study the factorization behavior of the triple-trace scalar integrand~\eqref{eq:3tr}.
We first consider the scalar factorization channel. In particular, we cut through the trace $\rho$ such that $L=\sigma\cup\rho_L$ and $R=\tau\cup\rho_R$, where $\rho_L\cup\rho_R=\rho$ and $\rho_L\cap\rho_R=\emptyset$. The leading order is contributed only by the first term of~\eqref{eq:3tr}. Using Eq.~\eqref{eq:newvar}, we find that
\begin{align}
\mathcal{I}_n^{\text{CHY}}(\sigma,\tau,\rho)&\sim\zeta^{n_R-n_L+2}\,\mathcal{U}^2\Bigg[\frac{\text{PT}(\rho_1)s_\sigma\text{PT}(\sigma)}{1-s_\sigma}\Bigg]\Bigg[\frac{\text{PT}(\rho_2)s_\tau\text{PT}(\tau)}{1-s_\tau}\Bigg]\nonumber\\
&=\zeta^{n_R-n_L+2}\,\mathcal{U}^2\,\mathcal{I}^{\text{CHY}}_{L}(\sigma,\rho_1)\,\mathcal{I}^{\text{CHY}}_R(\tau,\rho_2)\,,
\end{align}
which is the correct factorization behavior according to Eq.~\eqref{eq:CHYfactorize}. We note that in our factorization analyses, we always omit a possible overall sign, but keep track carefully the relative signs in our integrands.

Next, we consider the gluon factorization channel $L=\sigma$ and $R=\tau\cup\rho$. The first term of Eq.~\eqref{eq:3tr} behaves as
\begin{align}\label{eq:3tra}
\frac{\text{PT}(\rho)\mathcal{S}_{\alpha'}(\sigma)\mathcal{S}_{\alpha'}(\tau)}{(1-s_\sigma)(1-s_\tau)}&\sim \zeta^{n_R-n_L+2}\,\mathcal{U}^2\,\text{PT}(\sigma)\sum_{a\in R\,,\, b\in L}\frac{s_{ab}}{v_au_b}\frac{s_\tau\text{PT}(\tau)\text{PT}(\rho)}{1-s_\tau}\\
&\sim\alpha'\zeta^{n_R-n_L+2}\,\mathcal{U}^2\sum_{\text{states}}\Bigg[\mathcal{S}_{\alpha'}(q_L)\text{PT}(\sigma)\Bigg]\Bigg[\frac{\text{PT}(\rho)\mathcal{S}_{\alpha'}(\tau)\mathcal{S}_{\alpha'}(q_R)}{1-s_\tau}\Bigg]\,.\nonumber
\end{align}
Besides changing the variables to~\eqref{eq:newvar}, we also replace $s_{\sigma}$ by Eq.~\eqref{eq:zeta2}, which leads to the first line.
To achieve the second line, we first insert the on-shell completeness relation\footnote{The arrow means that we have excluded terms that vanish on-shell.} 
\begin{align}\label{eq:state}
\sum_{\text{states}}\epsilon_{\mu}(q_L)\epsilon_{\nu}(q_R)\rightarrow\eta_{\mu\nu}
\end{align}
into $s_{ab}=\alpha'k_a^\mu\eta_{\mu\nu}k_b^\nu$.
Similarly, the second term of Eq.~\eqref{eq:3tr} behaves as
\begin{align}\label{eq:3trb}
\frac{\text{PT}(\rho)\mathcal{S}_{\alpha'}(\sigma,\tau)}{(1-s_\sigma)(1-s_\tau)}&\sim\frac{\zeta^{n_R-n_L+2}\,\mathcal{U}^2\,\text{PT}(\sigma)\text{PT}(\tau)\text{PT}(\rho)}{2(1-s_\tau)(1-s_\rho)}\!\!\sum_{\substack{\sigma_1,\sigma_2\in\sigma \\ \tau_1,\tau_2\in\tau}}\!\Big(\frac{s_{\tau_2\sigma_1}}{u_{\sigma_1}}s_{\sigma_2\tau_1}-\frac{s_{\sigma_2\tau_1}}{u_{\sigma_2}}s_{\tau_2\sigma_1}\Big)\frac{v_{\tau_1\tau_2}}{v_{\tau_1}v_{\tau_2}}\nonumber\\
&\sim\alpha'^2\zeta^{n_R-n_L+2}\,\mathcal{U}^2\!\sum_{\text{states}}\Bigg[\frac{C_{q_L}\text{PT}(\sigma)\text{PT}(\tau)\text{PT}(\rho)}{2(1-s_\tau)(1-s_\rho)}\!\sum_{\tau_1,\tau_2\in\tau}\!\frac{(k_{\tau_2}\!\cdot\!f_{q_R}\!\cdot\!k_{\tau_1})v_{\tau_1\tau_2}}{v_{\tau_2}v_{\tau_1}}\Bigg]\nonumber\\
&\sim\alpha'\zeta^{n_R-n_L+2}\,\mathcal{U}^2\,\sum_{\text{states}}\Bigg[\mathcal{S}_{\alpha'}(q_L)\text{PT}(\sigma)\Bigg]\Bigg[\frac{\text{PT}(\rho)\,\mathcal{S}_{\alpha'}(\tau,q_R)}{1-s_\tau}\Bigg]\,.
\end{align}
We have used the completeness relation~\eqref{eq:state}, the momentum conservation~\eqref{eq:channel1} and the gauge choice $v_{q_R}=0$ to get the final result.

Combining Eq.~\eqref{eq:3tra} and~\eqref{eq:3trb}, we find that the triple-trace integrand~\eqref{eq:3tr} indeed factorizes into a single-trace one-gluon and a double-trace one-gluon integrand, namely,
\begin{align}
\mathcal{I}_n^{\text{CHY}}(\sigma,\tau,\rho)\sim\alpha'\zeta^{n_R-n_L+2}\,\mathcal{U}^2\sum_{\text{states}}\mathcal{I}^{\text{CHY}}_L(\sigma,q_L)\,\mathcal{I}^{\text{CHY}}_R(\tau,\rho,q_R)\,,
\end{align}
where $\mathcal{I}^{\text{CHY}}_L(\sigma,q_L)=C_{q_L}\text{PT}(\sigma)=-\mathcal{S}_{\alpha'}(q_L)\,\text{PT}(\sigma)$ and $\mathcal{I}^{\text{CHY}}_R$ is given by Eq.~\eqref{eq:2trg}. Our result thus agrees with the general requirement~\eqref{eq:CHYfactorize}.

\subsection{Generic integrands}
We briefly talk about how the above two special factorization channels work for generic multitrace integrand~\eqref{eq:Imultitrace}. We first consider the scalar channel $L=W_1\cup W_{m+1}^L$, where $W_{m+1}^L$ is part of the trace $W_{m+1}$ treated specially in Eq.~\eqref{eq:Imultitrace}. One can show that the more mixed $L$ and $R$ are in a PT factor, the higher order of $\zeta$ it will give rise to:
\begin{align*}
&\text{PT}(L)\,\text{PT}(R)\sim\zeta^{n_R-n_L}\,,& &\text{PT}(LR)\sim\zeta^{n_R-n_L+2}\,,& &\text{PT}(LRLR)\sim\zeta^{n_R-n_L+4}\,,\text{ etc.}
\end{align*}
Since the overall factor $\text{PT}(W_{m+1})$ already mixes $L$ and $R$ once, the leading order $\zeta^{n_R-n_L+2}$ must be contributed by the terms 
containing also the stand-alone factor $\text{PT}(W_1)$. Therefore, only those total partitions that have a singleton block $W_1$ is relevant at the leading order. This immediately leads to the correct factorization behavior
\begin{align}
\mathcal{I}_n^{\text{CHY}}(W_1,\ldots,W_{m+1},i_1,\ldots,i_r)\sim &\,\zeta^{n_R-n_L+2}\mathcal{U}^2\,\mathcal{I}^{\text{CHY}}_{L}(W_1,W_{m+1}^L)\nonumber\\
&\times\mathcal{I}^{\text{CHY}}_{R}(W_2,\ldots,W_m,W_{m+1}^R,i_1,\ldots,i_r)\,,
\end{align}
where $W_{m+1}^L\cup W_{m+1}^R=W_{m+1}$ and $W_{m+1}^L\cap W_{m+1}^R=\emptyset$.

The analysis of the gluon channel $L=W_1$ is only slightly more difficult. The relevant terms are either of the form $\text{PT}(LR)$ or $s_L\text{PT}(L)$. These terms are generated by $\mathcal{S}_{\alpha'}(W_1,\ldots)$, where the ``$\ldots$'' may contain traces, gluons, or their symmetrized fusion $\mathcal{S}_{\alpha'}$. A calculation similar to Eq.~\eqref{eq:3tra} and~\eqref{eq:3trb} shows that
\begin{align}\label{eq:W1factorize}
\mathcal{S}_{\alpha'}(W_1,\ldots)\sim\zeta^{\#-n_L+2}\,\mathcal{U}^2\sum_{\text{states}}\mathcal{I}^{\text{CHY}}_L(W_1,q_L)\,\mathcal{S}_{\alpha'}(q_R,\ldots)\,,
\end{align}
where $\#$ is the number of particles contained in ``$\ldots$''. In other words, we can simply replace $W_1$ by $q_R$. We note that the prescription still holds when $\mathcal{S}_{\alpha'}(W_1,\ldots)$ is nested in some other symmetrized fusions: 
\begin{align}\label{eq:W1factorize2}
\mathcal{S}_{\alpha'}\big(\ldots,\mathcal{S}_{\alpha'}(W_1,\ldots)\big)\sim\zeta^{\#-n_L+2}\,\mathcal{U}^2\sum_{\text{states}}\mathcal{I}^{\text{CHY}}_L(W_1,q_L)\,\mathcal{S}_{\alpha'}\big(\ldots,\mathcal{S}_{\alpha'}(q_R,\ldots)\big)\,.
\end{align}
Although the outer level $\mathcal{S}_{\alpha'}$ may further mix $W_1$ with the others, the contribution is subleading and thus can be ignored. On the other hand, the terms still of the form $\text{PT}(LR)$ are exactly captured by~\eqref{eq:W1factorize2}. The final result of this channel is 
\begin{align}
\mathcal{I}^{\text{CHY}}_n(W_1,\ldots,W_{m+1},i_1,\ldots,i_r)\sim &\,\zeta^{n_R-n_L+2}\,\mathcal{U}^2\!\sum_{\text{states}}\!\mathcal{I}^{\text{CHY}}_L(W_1,q_L)\nonumber\\
&\times\mathcal{I}^{\text{CHY}}_R(W_2,\ldots,W_{m+1},i_1,\ldots,i_r,q_R)\,,
\end{align}
agreeing with Eq.~\eqref{eq:CHYfactorize}. Schematically, we can obtain $\mathcal{I}^{\text{CHY}}_R$ by replacing the trace $W_1$ by $q_R$ in the original $\mathcal{I}^{\text{CHY}}_n$.

By iterating the two cuts discussed above, we eventually land on the single-trace integrand~\eqref{eq:st1}. This factorization analysis provides a simple but nontrivial consistency check to our integrands.

\section{Conclusion and discussion}

In this work we have continued our study of the two-step method proposed in~\cite{He:2018pol}: namely (1) IBP reduction of correlators of string amplitudes to a logarithmic function, and (2) rewriting the logarithmic function into a closed-form CHY half-integrand for field-theory amplitudes using scattering equations. We present two main results regarding heterotic and compactified bosonic strings for arbitrary multiplicities and number of traces. The final outcome of our calculation is remarkably simple CHY formulas for general $(DF)^2+{\rm YM}+\phi^3$ amplitudes, which extend our previous formula for the single-trace case greatly. As a paraphrase, our result gives the half-integrand needed in a rewriting of open-string amplitude as a CHY formula. In order to derive the formula, we find that the key new result is a recursive expansion for multi-trace string correlators. It provides an efficient algorithm for reducing multi-trace correlator to logarithmic functions, which is useful for other purposes as well. Note that we have left out one special case, which is the formula for pure-graviton case: while it can be obtained from factorization already from our single-trace formula, it would be highly desirable to obtain a closed-form result for it as well. 

Our results opens various interesting avenues for further investigations. First of all, they may bring new insight into the $(DF)^2+{\rm YM}+\phi^3$ theory especially in the multi-trace sector, as well as conformal (super-)gravity which can be obtained from a double copy with (super)-YM~\cite{Johansson:2017srf,Johansson:2018ues}. Moreover, it would be interesting to compute matrix elements with higher-dimensional operators from bosonic/heterotic string corrections, and our formulas can serve as a starting point for extracting such corrections, in a way similar to~\cite{He:2016iqi,Garozzo:2018uzj}. Of course having such a general formula for a large class of amplitudes, including those in Einstein-Yang-Mills theory, provides more applications. For example, one could use it for extracting BCJ numerators and discovering new amplitude relations, which have been recently studied further in~\cite{Du:2017gnh,Du:2017kpo, Du:2018khm, Hou:2018bwm}. 

The recursive expansion certainly has more applications. Most directly it gives the BCJ numerators for the $(DF)^2+{\rm YM}+\phi^3$ theory. From a more mathematical point of view, it allows us to reduce non-logarithmic functions with multiple cycles to logarithmic ones, both for IBP reduction of string correlator and, in the $\alpha'\to \infty$ limit, for manipulating CHY integrand using SE. As shown in the paper and in the ancillary file, the recursive expansion allows us to do such calculations in a very efficient way. It is also interesting to relate our general procedure to various ideas in the literature, such as intersection theory~\cite{Mizera:2017rqa,Mizera:2019gea}, studies of disk/sphere integrals from a mathematical point of view~\cite{Schlotterer:2018zce, Brown:2018omk, Vanhove:2018elu}, and positive geometries related to string worldsheet~\cite{Arkani-Hamed:2017mur,He:2018pue, 
	Arkani-Hamed2}.   

As we have pointed out in~\cite{He:2018pol}, our method applies to any string correlator for massless external states with the correct SL$(2)$ weight, and it would be interesting to study more examples beyond type I, bosonic and heterotic cases, such as the dual model proposed in ref.~\cite{Baadsgaard:2016fel}. More importantly, it would be highly desirable to apply our method to string correlators with massive states, as well as to cases at genus one~\cite{Tsuchiya:1988va,Dolan:2007eh,Mafra:2017ioj,Mafra:2018nla,Mafra:2018pll,Mafra:2018qqe,Gerken:2018jrq}. On the other hand, the CHY half-integrands here contain explicit $\alpha'$ dependence, and it would be interesting to understand if there are worldsheet models, such as ambitwistor string~\cite{Mason:2013sva,Casali:2015vta,Geyer:2018xwu} theory, underpin it (see~\cite{Azevedo:2017yjy} for some progress which gives at least correct three-point amplitudes). Investigations along these lines may shed new light into the universality and origin of CHY/ambitwistor string constructions.

\begin{acknowledgments}
It is our pleasure to thank Thales Azevedo, Yi-Jian Du, Bo Feng, Henrik Johansson, Oliver Schlotterer and Ellis Yuan for inspiring discussions. 
S.H.'s research is supported in part by the Thousand Young Talents program, the Key Research Program of Frontier Sciences of CAS under Grant No. QYZDB-SSW-SYS014 and Peng Huanwu center under Grant No. 11747601.
F.T. is supported by the Knut and Alice Wallenberg Foundation under grant KAW 2013.0235, and the
Ragnar S\"{o}derberg Foundation (Swedish Foundations’ Starting Grant).
\end{acknowledgments}

\appendix

\section{Labeled trees and logarithmic Yang-Mills-scalar integrands}\label{sec:trees}
As shown in ref.~\cite{Gao:2017dek}, labeled trees form a basis for logarithmic functions on the worldsheet. In this section, we give the rules to write down the logarithmic Yang-Mills-scalar CHY integrand $\mathcal{T}_{W_{m+1}}(\pmb{R})$ in terms of a labeled-tree expansion:
\begin{align}\label{treeExpansion}
\mathcal{T}_{W_{m+1}}(\pmb{R})=\sum_{T\in\mathbf{T}(W_{m+1})}N_{\pmb{R}}(T)\,\mathcal{C}(T)\,.
\end{align}
To carry out this expansion, we need to first construct the relevant labeled trees, and then define the map $N_{\pmb{R}}$ and $\mathcal{C}$ for each labeled tree.
The summation in Eq.~\eqref{treeExpansion} is over $\mathbf{T}(W_{m+1})$, the labeled trees with roots in $W_{m+1}$, the nodes of which are labels of all the external particles. The function $\mathcal{C}$ maps a tree $T$ into a rational function of worldsheet variables:\footnote{In~\cite{Gao:2017dek}, it is called the Cayley function. For a trivial tree with a single node, we define $\mathcal{C}[\,\tikz[baseline={([yshift=-0.75ex]current bounding box.center)}]{\filldraw (0,0) circle (1pt);}\,]=1$.} each edge is mapped to a $z_{ij}$ factor in the denominator, where $i$ and $j$ are the labels of the nodes connected by the edge. Each tree also carries a dual kinematic factor $N_{\pmb{R}}$, the evaluation of which depends on the choice of reference order $\pmb{R}$. Very interestingly, these dual kinematic factors form a basis for the DDM basis BCJ numerators~\cite{Teng:2017tbo,Du:2017gnh}, while the reference order $\pmb{R}$ characterizes certain generalized gauge degrees of freedom. 

We start with constructing the relevant labeled trees $\mathbf{T}(W_{m+1})$. We first treat the gluons and traces on the same footing, and draw all the rooted trees on $W_{m+1}$ with nodes $\{i_1,\ldots, i_r , W_1 , \ldots , W_m \}$. In all there are $(r+m+1)^{r+m-1}$ such trees. 
For example, the spanning trees for the double-trace single-gluon case are
\begin{equation}\label{treeExample}
	\begin{tikzpicture}[baseline={([yshift=-2ex]current bounding box.center)},every node/.style={font=\footnotesize},scale=1.1]
	\draw [thick] (0,0) -- (0,0.75) -- (0.75,0.75);
	\filldraw [thick] (0,0) circle (1pt) node[left=0pt]{$W_2$} (0,0.75) circle (1pt) node[above=0pt]{$W_1$};
	\filldraw [thick] (0.75,0.75) circle (1pt) node[above=0pt]{$i$};
	\begin{scope}[xshift=2.5cm]
	\draw [thick] (0,0) -- (0,0.75);
	\draw [thick] (0,0.75) -- (0.75,0.75);
	\filldraw [thick] (0,0) circle (1pt) node[left=0pt]{$W_2$} (0,0.75) circle (1pt) node[above=0pt]{$i$};
	\filldraw [thick] (0.75,0.75) circle (1pt) node[above=0pt]{$W_1$};
	\end{scope}
	\begin{scope}[xshift=5cm]
	\draw [thick] (0,0) -- (0.75,0.75);
	\draw [thick] (0,0) -- (0,0.75);
	\filldraw [thick] (0,0) circle (1pt) node[left=0pt]{$W_2$} (0,0.75) circle (1pt) node[above=0pt]{$W_1$};
	\filldraw [thick] (0.75,0.75) circle (1pt) node[above=0pt]{$i$};
	\end{scope}
	\end{tikzpicture}\,.
\end{equation}
Next, given a reference order $\pmb{R}$, we decompose each tree into a collection of paths and blow up the traces according to the following procedures:
\begin{enumerate}[label=(\arabic*)]
\item draw a path from the first element of $\pmb{R}$ to the root. Then draw another path towards the root from the first element of $\pmb{R}$ that has not been traversed. This path will end on a previous path. Repeat the process until all nodes are traversed. This decomposes each tree into a set of paths, denoted as $\mathcal{P}[T]$.\label{path}
\item replace the root by a chain evaluated to $\text{PT}(W_{m+1})$ after restoring the gauge:
\begin{align}
\begin{tikzpicture}[baseline={([yshift=-1.1ex]current bounding box.center)},every node/.style={font=\footnotesize,},vertex/.style={inner sep=0,minimum size=3pt,circle,fill},wavy/.style={decorate,decoration={coil,aspect=0, segment length=2.2mm, amplitude=0.5mm}},dir/.style={decoration={markings, mark=at position \Halfway with {\arrow{Latex}}},postaction={decorate}}]
\filldraw (-1,0) circle (1pt) node[above=0pt]{$W_{m+1}$};
\node at (-0.5,0) {$\rightarrow$};
\draw [thick] (0.35,0) ++(-110:0.35) arc (-110:-430:0.35);
\end{tikzpicture}\,,\qquad\mathcal{C}\!\left(\begin{tikzpicture}[baseline={([yshift=-1.1ex]current bounding box.center)},every node/.style={font=\footnotesize,},vertex/.style={inner sep=0,minimum size=3pt,circle,fill},wavy/.style={decorate,decoration={coil,aspect=0, segment length=2.2mm, amplitude=0.5mm}}]
\draw [thick] (0.35,0) ++(-110:0.35) arc (-110:-430:0.35);
\end{tikzpicture}\right)\xrightarrow{\text{restore gauge}}\text{PT}(W_{m+1})\,.
\end{align}
\item if a trace $W_i$ appears in the mid of a path, blow it up according to
\begin{align}\label{traceBlowUp}
\begin{tikzpicture}[baseline={([yshift=-1.ex]current bounding box.center)},every node/.style={font=\footnotesize,},vertex/.style={inner sep=0,minimum size=3pt,circle,fill},wavy/.style={decorate,decoration={coil,aspect=0, segment length=2.2mm, amplitude=0.5mm}}]
\filldraw (-1,0) circle (1pt) node[above=0pt]{$W_i$};
\node at (-0.5,0) {$\rightarrow$};
\node at (-0,0) [label={above:{${a_i}$}},vertex] {};
\node at (1.5,0) [label={above:{${b_i}$}},vertex] {};
\draw[thick,wavy] (0,0) -- (1.5,0);
\end{tikzpicture},
\end{align}
where by our convention $b_i$ is the end closer to the root. We will sum over all pairs of $a_i$ and $b_i$ in $W_i$.
\item if a trace $W_i$ appears at the start of a path, still blow it up as~\eqref{traceBlowUp}. However, only $b_i$ will be summed in $W_i$, while $a_i\in W_i$ is arbitrary but fixed. Across our construction, we keep the same choice of $a_i$ if this situation happens.\footnote{This choice eliminates some redundancy in the construction. In~\cite{Du:2017gnh}, $a_i$ is called a fiducial particle.}
\item if a path ends on a trace $W_i$, then the end point can take any value in $W_i$. 
\end{enumerate}
Accordingly, the three spanning trees in Eq.~\eqref{treeExample} generate the following labeled trees relevant to the logarithmic CHY integrand:
\begin{align}\label{labeledTreesExample}
	\mathbf{T}(W_2):\qquad\begin{tikzpicture}[baseline={([yshift=-1.1ex]current bounding box.center)},every node/.style={font=\scriptsize},wavy/.style={decorate,decoration={coil,aspect=0, segment length=2mm, amplitude=0.5mm}},dir/.style={decoration={markings, mark=at position \halfway with {\arrow{Latex[scale=0.9]}}},postaction={decorate}}]
	\draw [thick,blue,wavy] (0,0.75) -- (-0.75,0.75);
	\draw [thick,blue,dir] (0.75,0.75) -- (0,0.75) ;
	\draw [thick,blue,dir] (-0.75,0.75) -- (-0.75,0);
	\filldraw [thick] (-0.75,0) circle (1pt) node[below=0pt]{$j_2$} (0,0.75) circle (1pt) node[above=0pt]{$a_1$} (-0.75,0.75) circle (1pt) node[above=0pt]{$b_1$};
	\filldraw [thick] (0.75,0.75) circle (1pt) node[above=0pt]{$i$};
	\draw [thick] (-0.75,-0.35) ++(-110:0.35) arc (-110:-430:0.35);
	\node at (0,-1.25) [align=center] {$a_1,b_1\in W_1$ \\ $j_2\in W_2$};
	\begin{scope}[xshift=2.5cm]
	\draw [thick,blue,dir] (0,0.75) -- (0,0);
	\draw [thick,red,dir] (0.75,0.75) -- (0,0.75);
	\draw [thick,red,wavy] (0.75,0.75) -- (1.5,0.75);
	\filldraw [thick] (0,0) circle (1pt) node[below=0pt]{$j_2$} (0,0.75) circle (1pt) node[above=0pt]{$i$};
	\filldraw [thick] (0.75,0.75) circle (1pt) node[above=0pt]{$b_1$} (1.5,0.75) circle (1pt) node[above=0pt]{$a_1$};
	\draw [thick] (-0,-0.35) ++(-110:0.35) arc (-110:-430:0.35);
	\node at (0.75,-1.25) [align=center] {$b_1\in W_1$ \\ $j_2\in W_2$};
	\end{scope}
	\begin{scope}[xshift=6.5cm]
	\draw [blue,thick,dir] (0.75,0.75) -- (0.35,-0.35);
	\draw [red,thick,dir] (0,0.75) -- (0,0);
	\draw [thick,red,wavy] (0,0.75) -- (-0.75,0.75);
	\filldraw [thick] (0,0) circle (1pt) node[below=0pt]{$l_2$} (0.35,-0.35) circle (1pt) node[right=0pt]{$j_2$} (0,0.75) circle (1pt) node[above=0pt]{$b_1$};
	\filldraw [thick]  (0.75,0.75) circle (1pt) node[above=0pt]{$i$} (-0.75,0.75) circle (1pt) node[above=0pt]{$a_1$};
	\draw [thick] (-0,-0.35) ++(-110:0.35) arc (-110:-430:0.35);
	\node at (0,-1.25) [align=center] {$b_1\in W_1$ \\ $j_2,l_2\in W_2$};
	\end{scope}
	\end{tikzpicture}\,\,,
\end{align}
in which we have used the reference order $\pmb{R}=i\prec W_1$. All the paths are directed towards the root, and different ones are illustrated by different colors. For each $T\in\mathbf{T}(W_{m+1})$, the map $\mathcal{C}$ is defined as
\begin{align}\label{Zmap}
&\begin{tikzpicture}[baseline={([yshift=0.5ex]current bounding box.center)},every node/.style={font=\footnotesize,},vertex/.style={inner sep=0,minimum size=3pt,circle,fill},wavy/.style={decorate,decoration={coil,aspect=0, segment length=2.2mm, amplitude=0.5mm}},dir/.style={decoration={markings, mark=at position \halfway with {\arrow{Latex[scale=0.9]}}},postaction={decorate}}]
\filldraw [thick] (0,0) circle (1pt) node[below=0]{$i$}  (1,0) circle (1pt) node[below=0]{$j$};
\draw [thick,dir] (0,0) -- (1,0);
\end{tikzpicture}\rightarrow \frac{1}{z_{ij}}\,,
& &\begin{tikzpicture}[baseline={([yshift=0.5ex]current bounding box.center)},every node/.style={font=\footnotesize,},vertex/.style={inner sep=0,minimum size=3pt,circle,fill},wavy/.style={decorate,decoration={coil,aspect=0, segment length=2.2mm, amplitude=0.5mm}}]
\filldraw (3.5,0) circle (1pt) node[below=0]{$a_i$} (4.5,0) circle (1pt) node[below=0]{$b_i$};
\draw [thick,wavy] (3.5,0) -- (4.5,0);
\end{tikzpicture}\rightarrow\,\text{PT}(W_i)z_{b_ia_i}\,.
\end{align}
This definition is compatible with the one introduced in section~\ref{sec:fusion}. On the other hand, for each path $p$ in the path set $\mathcal{P}[T]$, we can define a path factor $\varphi(p)$ obtained from the following rule:
\begin{equation}\label{Nmap}\renewcommand{\arraystretch}{1.2}
\begin{tabular}{|c|c|c|c|} \hline
	\multirow{2}{*}{node} & \multicolumn{3}{c|}{position in the path} \\ \cline{2-4}
	& start & middle & end \\ \hline
	gluon $i$ & $\epsilon^{\mu}_i$ & $f_i^{\mu\nu}$ & $k_i^\nu$ \\ \hline
	trace $W_i$ & $k_{b_i}^\mu$ & $k_{a_i}^{\mu}k_{b_i}^{\nu}$ & $k_{j_i}^\nu$ \\ \hline
\end{tabular}\,,
\end{equation}
where the Lorentz indices are contracted with their neighbors on the path. The map $N_{\pmb{R}}(T)$ is given by the product of all these path factors:
\begin{align}
N_{\pmb{R}}(T)=\prod_{p\in\mathcal{P}[T]}\varphi(p)\,.
\end{align}
The outcome of $N_{\pmb{R}}$ depends on the reference order $\pmb{R}$, since different $\pmb{R}$'s lead to different path sets for a given tree.

According to Eq.~\eqref{Zmap} and~\eqref{Nmap}, the labeled trees in Eq.~\eqref{labeledTreesExample} are evaluated as follows under the reference order $\pmb{R}=i\prec W_1$:
\begin{equation}
	\begin{tabular}{|c|c|c|c|} \hline
		 & $\vphantom{\Big[}\sum_{T}$ & $\displaystyle N_{\pmb{R}}(T)$ & $\mathcal{C}(T)$ \\ \hline
		\begin{tikzpicture}[baseline={([yshift=0ex]current bounding box.center)},every node/.style={font=\scriptsize},wavy/.style={decorate,decoration={coil,aspect=0, segment length=2mm, amplitude=0.5mm}},dir/.style={decoration={markings, mark=at position \halfway with {\arrow{Latex[scale=0.9]}}},postaction={decorate}},scale=0.8]
		\draw [thick,blue,wavy] (0,0.75) -- (-0.75,0.75);
		\draw [thick,blue,dir] (0.75,0.75) -- (0,0.75) ;
		\draw [thick,blue,dir] (-0.75,0.75) -- (-0.75,0);
		\filldraw [thick] (-0.75,0) circle (1pt) node[below=0pt]{$j_2$} (0,0.75) circle (1pt) node[above=0pt]{$a_1$} (-0.75,0.75) circle (1pt) node[above=0pt]{$b_1$};
		\filldraw [thick] (0.75,0.75) circle (1pt) node[above=0pt]{$i$};
		\draw [thick] (-0.75,-0.35) ++(-110:0.35) arc (-110:-430:0.35);
		\node at (0,-0.65) {};
		\end{tikzpicture} & $\displaystyle\sum_{a_1,b_1\in W_1}\sum_{j_2\in W_2}$ & $\displaystyle(\epsilon_i\!\cdot\!k_{a_1}) (k_{b_1}\!\cdot\!k_{j_2})$ & $\displaystyle\vphantom{\sum_T}\frac{\textrm{PT}(W_1)z_{b_1a_1}}{z_{ia_1}z_{b_1j_2}}$ \\ \hline
		\begin{tikzpicture}[baseline={([yshift=-0.ex]current bounding box.center)},every node/.style={font=\scriptsize},wavy/.style={decorate,decoration={coil,aspect=0, segment length=2mm, amplitude=0.5mm}},dir/.style={decoration={markings, mark=at position \halfway with {\arrow{Latex[scale=0.9]}}},postaction={decorate}},scale=0.8]
		\draw [thick,blue,dir] (0,0.75) -- (0,0);
		\draw [thick,red,dir] (0.75,0.75) -- (0,0.75);
		\draw [thick,red,wavy] (0.75,0.75) -- (1.5,0.75);
		\filldraw [thick] (0,0) circle (1pt) node[below=0pt]{$j_2$} (0,0.75) circle (1pt) node[above=0pt]{$i$};
		\filldraw [thick] (0.75,0.75) circle (1pt) node[above=0pt]{$b_1$} (1.5,0.75) circle (1pt) node[above=0pt]{$a_1$};
		\draw [thick] (-0,-0.35) ++(-110:0.35) arc (-110:-430:0.35);
		\node at (0,-0.65) {};
		\end{tikzpicture} & $\displaystyle\sum_{b_1\in W_1}\sum_{j_2\in W_2}$ & $\displaystyle(\epsilon_i\!\cdot\!k_{j_2})(k_{b_1}\!\cdot\!k_i)$ & $\displaystyle\vphantom{\sum_T}\frac{\textrm{PT}(W_1)z_{b_1a_1}}{z_{b_1i}z_{ij_2}}$ \\ \hline
		\begin{tikzpicture}[baseline={([yshift=-0.ex]current bounding box.center)},every node/.style={font=\scriptsize},wavy/.style={decorate,decoration={coil,aspect=0, segment length=2mm, amplitude=0.5mm}},dir/.style={decoration={markings, mark=at position \halfway with {\arrow{Latex[scale=0.9]}}},postaction={decorate}},scale=0.8] 
		\draw [blue,thick,dir] (0.75,0.75) -- (0.35,-0.35);
		\draw [red,thick,dir] (0,0.75) -- (0,0);
		\draw [thick,red,wavy] (0,0.75) -- (-0.75,0.75);
		\filldraw [thick] (0,0) circle (1pt) node[below=0pt]{$l_2$} (0.35,-0.35) circle (1pt) node[right=0pt]{$j_2$} (0,0.75) circle (1pt) node[above=0pt]{$b_1$};
		\filldraw [thick]  (0.75,0.75) circle (1pt) node[above=0pt]{$i$} (-0.75,0.75) circle (1pt) node[above=0pt]{$a_1$};
		\draw [thick] (-0,-0.35) ++(-110:0.35) arc (-110:-430:0.35);
		\node at (0,-0.65) {};
		\end{tikzpicture} & $\displaystyle\sum_{b_1\in W_1}\sum_{j_2,l_2\in W_2}$ & $\displaystyle(\epsilon_i\!\cdot\!k_{j_2})(k_{b_1}\!\cdot\!k_{l_2})$ & $\displaystyle\vphantom{\sum_T}\frac{\textrm{PT}(W_1)z_{b_1a_1}}{z_{b_1j_2}z_{il_2}}$ \\ \hline
	\end{tabular}\,\,,
\end{equation}
such that the logarithmic integrand $\mathcal{T}_{W_2}(i,W_1)$ is obtained simply by adding the three rows together. In particular, the $a_1\in W_1$ in the second and third row is the same, and not summed over. Different choice of $a_1$ leads to equivalent $\mathcal{T}_{W_2}$, and thus it exposes certain redundancy in both the string and CHY integrand. We note that if we choose $\pmb{R}=W_1\prec i$ instead, the first two classes of labeled trees in the above table are modified into
\begin{subequations}
\begin{align}
	& \begin{tikzpicture}[baseline={([yshift=0ex]current bounding box.center)},every node/.style={font=\scriptsize},wavy/.style={decorate,decoration={coil,aspect=0, segment length=2mm, amplitude=0.5mm}},dir/.style={decoration={markings, mark=at position \halfway with {\arrow{Latex[scale=0.9]}}},postaction={decorate}}]
	\draw [thick,blue,wavy] (0,0.75) -- (-0.75,0.75);
	\draw [thick,red,dir] (0.75,0.75) .. controls (0.375,0.45) and (0,0.45) ..  (-0.375,0.75);
	\draw [thick,blue,dir] (-0.75,0.75) -- (-0.75,0);
	\filldraw [thick] (-0.75,0) circle (1pt) node[below=0pt]{$j_2$} (0,0.75) circle (1pt) node[above=0pt]{$a_1$} (-0.75,0.75) circle (1pt) node[above=0pt]{$b_1$};
	\filldraw [thick] (0.75,0.75) circle (1pt) node[above=0pt]{$i$};
	\filldraw [thick] (-0.375,0.75) circle (1pt) node[above=0pt]{$j_1$};
	\draw [thick] (-0.75,-0.35) ++(-110:0.35) arc (-110:-430:0.35);
\end{tikzpicture} & &\longrightarrow & & \sum_{j_1,b_1\in W_1}\sum_{j_2\in W_2}(k_{b_1}\!\cdot\!k_{j_2})(\epsilon_i\!\cdot\!k_{j_1})\frac{\text{PT}(W_1)z_{b_1a_1}}{z_{ij_1}z_{b_1j_2}}\,, \\
	& \begin{tikzpicture}[baseline={([yshift=-0.ex]current bounding box.center)},every node/.style={font=\scriptsize},wavy/.style={decorate,decoration={coil,aspect=0, segment length=2mm, amplitude=0.5mm}},dir/.style={decoration={markings, mark=at position \halfway with {\arrow{Latex[scale=0.9]}}},postaction={decorate}}]
	\draw [thick,blue,dir] (0,0.75) -- (0,0);
	\draw [thick,blue,dir] (0.75,0.75) -- (0,0.75);
	\draw [thick,blue,wavy] (0.75,0.75) -- (1.5,0.75);
	\filldraw [thick] (0,0) circle (1pt) node[below=0pt]{$j_2$} (0,0.75) circle (1pt) node[above=0pt]{$i$};
	\filldraw [thick] (0.75,0.75) circle (1pt) node[above=0pt]{$b_1$} (1.5,0.75) circle (1pt) node[above=0pt]{$a_1$};
	\draw [thick] (-0,-0.35) ++(-110:0.35) arc (-110:-430:0.35);
\end{tikzpicture} & &\longrightarrow & & \sum_{b_1\in W_1}\sum_{j_2\in W_2}(k_{b_1}\!\cdot\!f_i\!\cdot\!k_{j_2})\frac{\text{PT}(W_1)z_{b_1a_1}}{z_{b_1i}{z_{ij_2}}}\,,
\end{align}
\end{subequations}
while the third class remains the same. The resultant integrand $\mathcal{T}_{W_2}(W_1,i)$ is of course equivalent to $\mathcal{T}_{W_2}(i,W_1)$ both as string and CHY integrand.

\section{IBP reduction of  multibranch graphs}\label{sec:multibranch}

In our previous letter~\cite{He:2018pol}, we have shown that generic multibranch graphs can be algebraically rearranged into tadpoles and then processed by using Eq.~\eqref{stibp}.
In this section, we introduce a new IBP reduction for multibranch graphs that naturally leads to our recursive expansion~\eqref{stringExpansion}.  

We may view a multibranch graph as a collection of subtrees planted on a subcycle consisting of a color trace and/or gluons. If we denote the subcycle as $W$, each node $i\in W$ is the root of a tree $\mathsf{B}_i$. Moreover, we use $\mathsf{s}_i$ to denote the immediate successors of $i$ in the tree $\mathsf{B}_i$. By definition, the set $\mathsf{s}_i$ can be empty while $\mathsf{B}_i$ at least contains one node, the root $i$. For generic multibranch graphs, we have
\begin{align}
\mathcal{C}\!\left[\begin{tikzpicture}[baseline={([yshift=-0.75ex]current bounding box.center)},dir/.style={decoration={markings, mark=at position \halfway with {\arrow{Latex[scale=0.8]}}},postaction={decorate}}]
\fill [rounded corners,fill=red!20] (-0.6,0.2) -- (-2.95,0.8) -- (-2.95,-0.8) -- (-0.6,-0.2) -- cycle;
\fill [rounded corners,fill=red!20] (0.6,0.2)  -- (1.9,0.6) -- (2.95,0.6) -- (2.95,-0.6) -- (1.9,-0.6) -- (0.6,-0.2) -- cycle;
\draw [thick] (0,0) circle (1cm);
\filldraw (1,0) circle (1pt) node[left=0pt,font=\footnotesize]{$j$} (-1,0) circle (1pt) node[right=0pt,font=\footnotesize]{$i$} (-2,0) circle (1pt) (-2.75,0.5) circle (1pt) (-2.75,-0.5) circle (1pt) (2,0.3) circle (1pt) (2.75,0.3) circle (1pt) (2,-0.3) circle (1pt);
\foreach \x in {75,90,105,-75,-90,-105}{
	\filldraw (\x:0.8) circle (0.75pt);
}
\draw [dashed] (-2,0) circle (0.35cm) node [below=0pt,font=\footnotesize]{$\mathsf{s}_i$};
\node at (0,0) {$W$};
\node at (-2.75,0) [font=\footnotesize]{$\mathsf{B}_i$};
\node at (2.75,0) [font=\footnotesize]{$\mathsf{B}_j$};
\node at (2,0) [font=\footnotesize]{$\mathsf{s}_j$};
\draw [dashed] (2,0) ellipse (0.2cm and 0.5cm);
\draw [thick,dir] (-1,0) -- (-2,0);
\draw [thick,dir] (-2,0) -- (-2.75,0.5);
\draw [thick,dir] (-2,0) -- (-2.75,-0.5);
\draw [thick,dir] (1,0) -- (2,0.3);
\draw [thick,dir] (2,0.3) -- (2.75,0.3);
\draw [thick,dir] (1,0) -- (2,-0.3);
\end{tikzpicture}\right]=\text{PT}(W)\prod_{i\in W}\mathcal{C}[\mathsf{B}_i]\,,
\end{align}
where $\mathcal{C}$ is defined in Appendix~\ref{sec:trees}. As special cases, single subcycles correspond to all $\mathsf{B}_i=\{i\}$ while tadpoles have exactly one nontrivial $\mathsf{B}_i$ that at the same time is a chain.

For any multibranch graph, we can absorb all the $\frac{1}{z_{bj}}$ factors with $b\in W$ and $j\in \mathsf{s}_b$ into the Koba-Nielsen factor. Then using directly Eq.~\eqref{stibp}, we get  
\begin{align}\label{stibp222}
\text{PT}(W)(\cdots)\!\prod_{i\in W}\!\mathcal{C}[\mathsf{B}_i]\overset{\text{IBP}}{\cong}\frac{(\cdots)}{1-s_W}\sum_{b\in W}\text{PT}(W)z_{ba}\Bigg[\sum_{j\notin W\cup\mathsf{s}_b}\frac{s_{bj}}{z_{bj}}+\sum_{j\in\mathsf{s}_b}\frac{s_{bj}-1}{z_{bj}}\Bigg]\prod_{i\in W}\!\mathcal{C}[\mathsf{B}_i]\,,
\end{align}
where as before $a\in W$ is arbitrary and $(\cdots)$ does not involve any punctures in $W$. Both terms on the right hand side of Eq.~\eqref{stibp222} contain \emph{induced subcycles}, which are not present in the original integrand but appear as a result of IBP. Here, they consists of nodes originally in the branches but only part of the nodes in $W$. To manifest the recursive pattern, we need further operations to make all the nodes in $W$ to appear in the induced subcycles. We first demonstrate this process by an example. 



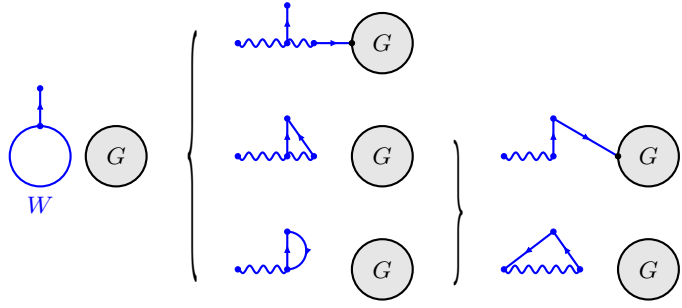
\begin{figure}[t]
\centering
\begin{tikzpicture}[every node/.style={font=\footnotesize,},vertex/.style={inner sep=0,minimum size=3pt,circle,fill},wavy/.style={decorate,decoration={coil,aspect=0, segment length=1.8mm, amplitude=0.5mm}},dir/.style={decoration={markings, mark=at position \halfwayb with {\arrow[scale=0.5]{Latex}}},postaction={decorate}},scale=1]
\begin{scope}[xshift=0cm,yshift=0cm]
\draw[thick,blue] (1,0) circle (0.4);
\draw[thick,blue,dir] (1,0.4) -- (1,0.9);
\filldraw [thick,fill=gray!20,draw=black] (2,0) circle (0.4) node [black]{$G$};
\filldraw [blue] (1,0.4) circle (1pt) (1,0.9) circle (1pt); 
\node at (1,-0.65) [blue] {$W$};
\node at (3,0) {$\stretchto[2000]{\{}{95pt}$};
\node at (6.5,-0.75) {$\stretchto[2000]{\}}{55pt}$};
\end{scope}
\begin{scope}[xshift=5.5cm,yshift=1.5cm]
\filldraw [thick,fill=gray!20,draw=black] (0,0) circle (0.4) node [black]{$G$};
\draw[thick,blue,dir] (-0.9,0) -- (-0.4,0);
\draw[thick,blue,wavy] (-0.9,0) -- ++(-1,0);
\draw[thick,blue,dir] (-0.9,0) ++(-0.35,0) -- ++(0,0.5);
\filldraw (-0.4,0) circle (1pt);
\filldraw [blue] (-0.9,0) circle (1pt) ++(-0.35,0) circle (1pt) ++(-0.65,0) circle (1pt);
\filldraw [blue] (-1.25,0.5) circle (1pt);
\end{scope}
\begin{scope}[xshift=5.5cm,yshift=0cm]
\filldraw [thick,fill=gray!20,draw=black] (0,0) circle (0.4) node [black]{$G$};
\draw[thick,blue,wavy] (-0.9,0) -- ++(-1,0);
\draw[thick,blue,dir] (-0.9,0) ++(-0.35,0) -- ++(0,0.5);
\filldraw [blue] (-0.9,0) circle (1pt) ++(-0.35,0) circle (1pt) ++(-0.65,0) circle (1pt);
\filldraw [blue] (-1.25,0.5) circle (1pt);
\draw[thick,blue,dir] (-0.9,0) -- (-1.25,0.5);
\end{scope}
\begin{scope}[xshift=5.5cm,yshift=-1.5cm]
\filldraw [thick,fill=gray!20,draw=black] (0,0) circle (0.4) node [black]{$G$};
\draw[thick,blue,wavy](-1.25,0) -- (-1.9,0);
\filldraw [blue] (-1.25,0.5) circle (1pt);
\filldraw [blue] (-1.25,0) circle (1pt) (-1.9,0) circle (1pt);
\draw[thick,blue,dir] (-1.25,0) -- ++(0,0.5);
\draw[thick,blue,dir] (-1.25,0.5) arc (90:-90:0.25);
\end{scope}
\begin{scope}[xshift=9cm,yshift=-1.5cm]
\filldraw [thick,fill=gray!20,draw=black] (0,0) circle (0.4) node [black]{$G$};
\draw[thick,blue,wavy] (-0.9,0) -- ++(-1,0);
\draw[thick,blue,dir] (-0.9,0) -- (-1.25,0.5);
\filldraw [blue] (-0.9,0) circle (1pt) (-1.9,0) circle (1pt);
\filldraw [blue] (-1.25,0.5) circle (1pt);
\draw[thick,blue,dir] (-1.25,0.5) -- (-1.9,0);
\end{scope}
\begin{scope}[xshift=9cm,yshift=0cm]
\filldraw [thick,fill=gray!20,draw=black] (0,0) circle (0.4) node [black]{$G$};
\draw[thick,blue,wavy](-1.25,0) -- (-1.9,0);
\filldraw [blue] (-1.25,0.5) circle (1pt);
\filldraw [blue] (-1.25,0) circle (1pt) (-1.9,0) circle (1pt);
\draw[thick,blue,dir] (-1.25,0) -- ++(0,0.5);
\draw[thick,blue,dir] (-1.25,0.5) -- (-0.4,0);
\filldraw (-0.4,0) circle (1pt);
\end{scope}
\end{tikzpicture}
\caption{The IBP reduction of $\text{PT}(W)\frac{1}{z_{pq}}$, where $G$ denotes other parts of the graph disconnected to the tadpole. }
\label{fig:tadpole}
\end{figure}

We can treat the simplest tadpole $\text{PT}(W)\frac{1}{z_{pq}}$ as multibranch and apply Eq.~\eqref{stibp222}. The generalization is that we can now break the subcycle at any point, not just the tail attach point $p$. The result is
\begin{align}\label{qfqq}
\frac{\text{PT}(W)(\cdots)}{z_{pq}}\overset{\text{IBP}}{\cong}\frac{\text{PT}(W)(\cdots)}{1-s_W}\Bigg[\sum_{\substack{b\in W \\ j\notin W\cup\{q\}}}\frac{z_{ba}s_{bj}}{z_{bj}z_{pq}}-\sum_{b\in W\backslash\{p\}}\frac{z_{ba}s_{bq}}{z_{bq}z_{qp}}+\frac{z_{pa}(1-s_{pq})}{z_{pq}z_{qp}}\Bigg]\,,
\end{align}
where $p\in W$ and $\mathsf{B}_p=\!\tikz[baseline={([yshift=-1ex]current bounding box.center)},every node/.style={font=\footnotesize},dir/.style={decoration={markings, mark=at position \halfwayb with {\arrow{Latex[scale=0.6]}}},postaction={decorate}}]{\filldraw (0,0) circle (1pt)  node[above=-1.5pt]{$p$} (0.5,0) circle (1pt)  node[above=-1.5pt]{$q$};\draw[thick,dir] (0,0) -- (0.5,0);}$. The last term is a tadpole, and the numerator cancels the tachyon pole introduced by the IBP relation~\eqref{stibp},
\begin{align}\label{fqewuhou}
\frac{\text{PT}(W)z_{pa}(1-s_{pq})(\cdots)}{z_{pq}z_{qp}}\overset{\text{IBP}}{\cong}\text{PT}(W)\Bigg[\sum_{b\in W\backslash\{p\}}\frac{z_{pa}s_{bq}}{z_{bq}z_{qp}}+\sum_{j\notin W\cup\{q\}}\frac{z_{pa}s_{qj}}{z_{pq}z_{qj}}\Bigg](\cdots)\,.
\end{align}
Noticing that $z_{ba}-z_{pa}=z_{bp}$, we can collapse the first term of~\eqref{fqewuhou} and the second term of~\eqref{qfqq} into a single subcycle. The final result is
\begin{align}\label{vpqoierf}
\frac{\text{PT}(W)(\cdots)}{z_{pq}}\overset{\text{IBP}}{\cong}\frac{\text{PT}(W)(\cdots)}{1-s_{W}}\Bigg[\sum_{\substack{b\in W \\ j\notin W\cup\{q\}}}\frac{z_{ba}s_{bj}}{z_{bj}z_{pq}}+\sum_{j\notin W\cup\{q\}}\frac{z_{pa}z_{qj}}{z_{pq}z_{qj}}+\sum_{b\in W}\frac{z_{pb}s_{bq}}{z_{bq}z_{qp}}\Bigg]\,.
\end{align}
The first two terms are trees planted on the remaining integrand. 
The third term features a ``fusion'' between the subcycle $W$ and the branch, and it can be further broken by using Eq.~\eqref{stibp}. Schematically, we can represent the above reduction process as in figure~\ref{fig:tadpole}.

The example gives some important ideas on the reduction of generic multibranch graphs. As discussed before, the goal is to include all nodes in the original subcycle $W$ (represented by the wavy line) into the induced subcycles. In Eq.~\eqref{stibp222}, we have generated a family of length-two subcycles $\text{PT}(b,j)$ featuring a numerator $(s_{bj}-1)$, where $b\in W$ and $j\in\mathsf{s}_j\subset\mathsf{B}_p$. Further IBP reduction on them will not lead to new tachyon poles. Next, we can absorb all the edges connecting $j$ and its immediate successors $\mathsf{s}_j$ into the Koba-Nielsen and break the subcycle at $b$. In this way, we can push the induced length-two subcycles with numerators $(s_{bj}-1)$ towards the end of the branches. This process ends at the leaves of each branch $\mathsf{B}_p$, where length-two tadpoles are generated and processed using relations like Eq.~\eqref{fqewuhou}. Then after using some algebraic identities, we arrive at

\begin{align}\label{stfuhwoefpe}
\text{PT}(W)(\cdots)\!\prod_{i\in W}\!\mathcal{C}[\mathsf{B}_i]\overset{\text{IBP}}{\cong}&\frac{\text{PT}(W)(\cdots)}{1-s_W}\nonumber\\
&\times\Bigg[\sum_{\substack{p\in W \\ j\notin\cup_{i}\mathsf{B}_i}}\sum_{\tilde{p}\in\mathsf{B}_p}\frac{z_{pa}s_{\tilde{p}j}}{z_{\tilde{p}j}}+\frac{1}{2}\!\sum_{\substack{p\in W \\ r\in W}}\sum_{\substack{\tilde{p}\in \mathsf{B}_p,\tilde{r}\in \mathsf{B}_r \\ (\tilde{p},\tilde{r})\neq (p,r)}}\!\!\frac{z_{pr}s_{\tilde{p}\tilde{r}}}{z_{\tilde{p}\tilde{r}}}\Bigg]\!\prod_{i\in W}\!\mathcal{C}[\mathsf{B}_i]\,,
\end{align}
where the $\frac{1}{2}$ in the second term cancels a double counting in the summation. The two terms in Eq.~\eqref{stfuhwoefpe} can be represented by two kinds of graphs
\begin{align}
\begin{tikzpicture}[baseline={([yshift=-0.75ex]current bounding box.center)},every node/.style={font=\footnotesize},dir/.style={decoration={markings, mark=at position \halfway with {\arrow{Latex[scale=0.8]}}},postaction={decorate}},wavy/.style={decorate,decoration={coil,aspect=0, segment length=1.8mm, amplitude=0.5mm}}]
\filldraw [fill=red!20,draw=black,thin,xshift=1cm] (0,0) ellipse (0.4cm and 0.2cm);
\filldraw [fill=red!20,draw=black,thin,xshift=-1cm] (0,0) ellipse (0.4cm and 0.2cm);
\filldraw [fill=red!20,draw=black,thin,xshift=-0.707107cm,yshift=0.707107cm,rotate=135] (0,0) ellipse (0.4cm and 0.2cm);
\draw[thick] (0,0) circle (0.75cm) node[font=\normalsize]{$W$};
\filldraw (0.75,0) circle (1pt) (-0.75,0) circle (1pt);
\filldraw (135:0.75) circle (1pt);
\foreach \x in {45,67.5,90}{
\filldraw (\x:0.65) circle (0.7pt);
}
\end{tikzpicture}
\;\xrightarrow{\text{Eq.~\eqref{stfuhwoefpe}}}\left\{\,
\begin{tikzpicture}[baseline={([yshift=-0.75ex]current bounding box.center)},every node/.style={font=\footnotesize},dir/.style={decoration={markings, mark=at position \halfway with {\arrow{Latex[scale=0.8]}}},postaction={decorate}},wavy/.style={decorate,decoration={coil,aspect=0, segment length=1.9mm, amplitude=0.5mm}}]
\foreach \x in {0,1,2}{
	\filldraw [fill=red!20,draw=black,thin,xshift=\x cm,yshift=0.25cm] (0,0) ellipse (0.2cm and 0.4cm);
	\filldraw (\x,0) circle (1pt);}
\draw [thick,wavy] (0,0) -- (2,0);
\draw [thick,dir] (2,0.5)  -- (3,0.5); 
\filldraw (2,0.5) circle (1pt) node [above=1.5pt]{$\tilde{p}$} (3,0.5) circle (1pt) node[above=1.5pt]{$j$};
\node at (2,0) [below=1.5pt]{$p$};
\node at (0,0) [below=1.5pt]{$a$};
\node at (3.5,0) {$,$};
\begin{scope}[xshift=4cm]
\foreach \x in {0,1,2}{
	\filldraw [fill=red!20,draw=black,thin,xshift=\x cm,yshift=0.25cm] (0,0) ellipse (0.2cm and 0.4cm);
	\filldraw (\x,0) circle (1pt);}
\draw [thick,wavy] (0,0) -- (2,0);
\draw [thick,dir] (2,0.5) .. controls (1.5,0.9) and (0.5,0.9) .. (0,0.5); 
\filldraw (2,0.5) circle (1pt) node [above=1.5pt]{$\tilde{p}$} (0,0.5) circle (1pt) node[above=1.5pt]{$\tilde{r}$};
\node at (2,0) [below=1.5pt]{$p$};
\node at (0,0) [below=1.5pt]{$r$};
\node at (1,1.6) [below=2pt]{$(\tilde{p},\tilde{r})\neq(p,r)$};
\end{scope}
\end{tikzpicture}\,\right\},
\end{align}
where we represent each branch $\mathsf{B}_i$ by a blob. Very nicely, the first term has a tree topology and there exists an arbitrary choice $a\in W$, while the second term has a multibranch topology and no arbitrary choice is involved.\footnote{In practice, we often choose $a$ to be the attach point of a certain branch, see Appendix~\ref{sec:appb}.} This resembles the structure of our recursive expansion~\eqref{stringExpansion}. The induced subcycles all contain the original subcycle $W$ as a whole (the wavy line) and at least one node from the branches, such that the total number of nodes in the branches is reduced.
We end our discussion with an example, 
\ba
\frac{\PT(W) (\cdots)}{z_{12}z_{23}z_{45}} \overset{\rm IBP}{\cong} & 
\frac{\PT(W) (\cdots)}{(1-s_W)z_{12}z_{23}z_{45}} \Bigg[\sum_{j\notin W\cup\{2,3,5\}} \!\Big( \sum_{ b\in W} z_{ba}\frac{s_{bj}}{z_{bj}}+  \frac{z_{1a}s_{2j}}{z_{2j}}+\frac{z_{1a}s_{3j}}{z_{3j}} +   \frac{z_{4a} s_{5j}}{z_{5j}} \Big)
\nl
&+ \frac{z_{14}s_{25}}{z_{25}}+\frac{z_{14}s_{35}}{z_{35}} + \sum_{b\in W} 
\Big( \frac{z_{b1} s_{b2}}{z_{b2}}+\frac{z_{b1} s_{b3}}{z_{b3}}+ \frac{z_{b4}s_{b5}}{z_{b5}} \Big)\Bigg]\,,
\ea
where $1,4\in W$ and  $2,3,5\notin W$.  The branches are $\mathsf{B}_1=\!\patha{1}{2}{3}$ and $\mathsf{B}_4=\!\pathb{4}{5}$. As before $a\in W$ is arbitrary and $(\cdots)$ does not involve any punctures in $W\cup\{2,3,5\}$.

\section{Towards the generic recursive expansion}\label{sec:appb}
In this section, we try to derive the generic recursive expansion~\eqref{stringExpansion} from the string integrand~\eqref{eq:strInt} with $r$ gluons and $m{+}1$ traces. The reduced gluon integrand $R(i_1,\ldots,i_r)$ gives rise to several new features compared with the pure-scalar case. Besides the traces $\text{PT}(W_i)$, there are new length-two gluon subcycles introduced by the $\mathcal{R}_{(ij)}=-\frac{\epsilon_i\cdot\epsilon_j}{\alpha'}\text{PT}(i,j)$ factors. There are additional gluon subcycles contributed by the product of $C_i$'s, which are of the form $C_{ij}C_{jk}C_{ki}$. To derive the generic recursive expansion, a very useful start point is to rewrite the gluon part \emph{algebraically} into the following form:
\ba\label{cqeihf}
R(i_1,\ldots,i_r)
=& 
-
\sum_{\substack{\mathsf{A}\in\mathbb{P}[i_1,\ldots,i_r]\\
		|\mathsf{A}|<r
}}\!\! (-1)^{|\mathsf{A}|+|\text{sg}(\mathsf{A})|} R\big(\text{sg}(\mathsf{A})\big)   \! \prod_{|\mathsf{A}_j|=2}\mathcal{S}_{0}(\mathsf{A}_j)
\! 
\prod_{|\mathsf{A}_j|\geqslant 3} \Big[\mathcal{S}_{0}(\mathsf{A}_j)_{\e\cdot \e\to 0} \Big]
\nl
&+
\sum_{\substack{\mathsf{A}\in\mathbb{P}[i_1,\ldots,i_r]\\
		\text{with all }|\mathsf{A}_j|\leqslant 2
}}\!\! \det(-{\bm C}_{\text{sg}(\mathsf{A})}) \!\prod_{|\mathsf{A}_j|=2} (1-s_{\mathsf{A}_j}){\cal R}_{(\mathsf{A}_j)}
\,,
\ea
where  
the summation in the first line is over all the partitions $\mathsf{A}=\{\mathsf{A}_1,\mathsf{A}_2,\ldots,\mathsf{A}_{|\mathsf{A}|}\}$ of the gluon set $\{i_1,\cdots,i_r\}$ except for the all-singleton partition  $\mathsf{A}=\{i_1,\cdots,i_r\}$. The set $\text{sg}(\mathsf{A}):=\big\{\mathsf{A}_j\in\mathsf{A}\,\big |\,|\mathsf{A}_j|=1\big\}$ is the collection of all the singleton blocks in $\mathsf{A}$.
Suppose there are exactly $s$ gluons  $\{i'_1,i'_2,\cdots, i'_s\}\subset \{i_1,i_2,\cdots,i_r\}$ which are singleton blocks in a certain partition $\mathsf{A}$, then we have $R\big(\text{sg}(\mathsf{A})\big) =  R(i'_1,\cdots, i'_u)$. The second factor $\prod_{|\mathsf{A}_j|=2}{\cal S}_0(\mathsf{A}_j)$ is a product of all length-two blocks in the certain partition $\mathsf{A}$. The third factor comes from the gluon subcycles of the form $C_{ij}C_{jk}C_{ki}$ mentioned before. For example,
\begin{align}\label{Cijk}
\mathcal{S}_{0}(i,j,k)_{\epsilon\cdot\epsilon\rightarrow 0}=\Big(\langle i,j,k\rangle+\langle i,k,j\rangle\Big)_{\epsilon\cdot\epsilon\rightarrow 0}=C_{ij}C_{jk}C_{ki}+C_{ik}C_{kj}C_{ji}\,.
\end{align}
The summation in the second line is over all the partitions with only singleton and length-two blocks.
The matrix $ {\bm C}_{\text{sg}(\mathsf{A})} = {\bm C}_{i'_1,\cdots, i'_s}$ is an $s\times s$ matrix whose off-diagonal entries are $C_{ij}$ and the diagonal ones are $-C_i$. According to the matrix-tree theorem~\cite{Stanley:2011}, $\det(-{\bm C}_{i'_1,\cdots, i'_s})$ is a combination of labeled trees rooted on the complement set of $\{i'_1,\cdots, i'_s\}$ in $\{i_1,\ldots,i_r,W_1,\ldots,W_{m+1}\}$.   



The benefit of this rewriting is the following. Comparing with the recursive expansion~\eqref{stringExpansion}, one can easily see that the first line of Eq.~\eqref{cqeihf} is already part of the $\mathcal{J}$ in Eq.~\eqref{stringExpansion}. More precisely, it is the part that only gluons are involved in the fusions. However, the $\epsilon\cdot\epsilon$ contributions in the fusions with length-three and beyond are missing. Therefore, to reach the recursive expansion, this line does not need any further manipulation and we only need to perform IBP reduction on the second line of Eq.~\eqref{cqeihf}. Interestingly, all the gluon subcycles there are length-two and dressed by a numerator $(1-s_{ij})$, which cancels the tachyon pole generated by the IBP. For convenience, we later refer the length-two gluon subcycles in the second line of Eq.~\eqref{cqeihf} as \emph{B-type}, and all the subcycles in the first line of Eq.~\eqref{cqeihf} as \emph{C-type}.






We are now ready to give the IBP algorithm that leads to the generic recursive expansion~\eqref{stringExpansion}. We need to first gauge fix a puncture in $W_{m+1}$ to infinity and then fix a reference order $\pmb{R}$ for gluons and the rest of traces as the priority list of being processed by IBP. It is convenient to put gluons before traces, for example, using the order~\eqref{reference}, although there are no restrictions in principle. Starting with the first element in $\pmb{R}$, 
\begin{itemize}
\item \emph{If it is a gluon, say $i$,} for each term in the string integrand, we do the following:
\begin{enumerate}[label=(g\arabic*)]
	\item If $i$ appears in the B-type subcycle $(1-s_{ij})\mathcal{R}_{(ij)}$, choose $a=i$ and break it using~\eqref{stfuhwoefpe}.\label{g1}
	\item If $i$ appears in a branch of a trace $W_j$ or B-type subcycle $(1-s_{jk})\mathcal{R}_{(jk)}$, we choose $a$ as the attach point of that branch and break the subcycle using~\eqref{stfuhwoefpe}.\label{g2}
	\item Repeat this process until in every term $i$ is connected to the root $W_{m+1}$ or an induced subcycle.\footnote{See the definition of induced subcycles below Eq.~\eqref{stibp222}. We note that $i$ can never connect to a C-type subcycle since they do not appear in the second line of Eq.~\eqref{cqeihf}.} Then proceed to the next element in~$\pmb{R}$.\label{g3}
\end{enumerate}
\item \emph{If it is a trace, say $W_i$,} we choose the same $a_i\in W_i$ for every term in the string integrand and do the following:
\begin{enumerate}[label=(t\arabic*')]
	\item If $W_i$ is in its original form and has not been processed, break it at $a_i$ using~\eqref{stfuhwoefpe}. This turns $W_i$, together with its branches, into a tree planted on the other part of the integrand.\label{step1'}
	\item If the tree generated in step~\ref{step1'} appears in a branch of another trace or B-type subcycle in its original form, choose $a$ as the attach point of the branch and break the subcycle using~\eqref{stfuhwoefpe}.\label{step2'}
	\item Repeat this process until in every term $W_i$ is connected to the root $W_{m+1}$ or an induced subcycle. Then proceed to the next element in $\pmb{R}$.\label{step3'}
\end{enumerate}
\end{itemize}
We note that the trace rules are the direct generalization of~\ref{step1}, \ref{step2} and~\ref{step3} for pure-scalar cases. The algorithm terminates when all the elements in $\pmb{R}$ are traversed. the outcome will provide the $\epsilon\cdot\epsilon$ terms in the labeled trees and in the gluon fusions together with the fusions involving traces. We will then arrive at the recursive expansion~\eqref{stringExpansion} after some straightforward algebra.


\subsection{Examples: two and three gluons with a single trace}\label{sec:examples}
We will demonstrate by two examples that the above algorithm indeed leads to the correct expansion~\eqref{stringExpansion}.

The single-trace two-gluon string integrand is  $R(1,2) \PT(W_1)=( C_1C_2+{\cal R}_{(12)})\PT(W_1)$. We can expand the $C_1 C_2\PT(W_1)$ part as
\ba
C_1C_2\PT(W_1)
= \Bigg[\sum_{\ell_1,\ell_2\in W_1}\!\!C_{1\ell_1}C_{2\ell_2} + \!\sum_{j\in W_1} 
\underbrace{\frac{ \e_1\C k_2 \, \e_2 \C k_{j}}{z_{12}z_{2,j}}}_{
C_{12} C_{2j}} + C_{21}  \!\sum_{j\in W_1}\!C_{1j} + 
\underbrace{ \frac{\e_1\C k_2\, \e_2\C k_1}{z_{12}z_{21}}}_
 {C_{12}C_{21}} \Bigg]  \PT(W_1),
\ea
where the first three terms of the right hand side are combinations in labeled trees and the last term is a C-type subcycle.  We rewrite the subcycle ${\cal R}_{(12)}$ as ${\cal R}_{(12)}(1-s_{12})+ {\cal R}_{(12)}s_{12}$ and perform IBP on the the first part,
\ba\la{cqepfae}
{\cal R}_{(12)}\PT(W_1)\overset{\rm IBP}{\cong}
\left( \sum_{j\in W_1} \frac{\e_1\C \e_2}{z_{12}} \frac{k_2\C k_j}{z_{2j}} - \frac{\e_1\C \e_2\, k_2\C k_1}{z_{12}z_{21}}
\right)\PT(W_1)\,.
\ea
Combining these two equations, we have
\ba\la{iqpofe}
R(1,2)\PT(W_1)
\overset{\rm IBP}{\cong}\Big(\underbrace{\sum_{\ell_1,\ell_2\in W_1}C_{1\ell_1}C_{2\ell_2} + \sum_{j\in W_1} 
\frac{ \e_1\C f_2 \C k_{j}}{z_{12}z_{2,j}}+C_{21}  \sum_{j\in W_1}C_{1j}}_{\mathcal{T}_{W_1}(1,2)} + 
\<1,2\>\Big)  \PT(W_1)\,,
\ea
which agrees with the general formula~\eqref{stringExpansion}. The result corresponds to the reference order $1\prec 2$ because we choose to break the subcycle at gluon $1$ in Eq.~\eqref{cqepfae}, \emph{cf.} the rule~\ref{g1}.

We then try to derive the recursive expansion for the single-trace three-gluon string integrand. According to Eq.~\eqref{stringExpansion}, it is
\ba\la{fqeof;}
R(1,2,3)\PT(W_1) \overset{\text{IBP}}{\cong} & \Big[{\cal T}_{W_1}(1,2,3)+ \<1,2,3\> +\<1,3,2\> +\Big(\<1,2\> C_3+{\rm cyclic}\Big)
\Big] \PT(W_1)\,.
\ea
As described in Appendix~\ref{sec:trees},  there are 16 relevant spanning trees  rooted on $W_{1}$.   Each such spanning tree can be decomposed into a collection of paths according to the reference ordering $1\prec 2 \prec 3$. In the following,  we will use a set of paths to denote a spanning tree. For examples, 
\begin{align}
\begin{tikzpicture}[baseline={([yshift=-0.75ex]current bounding box.center)},every node/.style={font=\footnotesize},dir/.style={decoration={markings, mark=at position \halfway with {\arrow{Latex[scale=0.8]}}},postaction={decorate}},wavy/.style={decorate,decoration={coil,aspect=0, segment length=1.8mm, amplitude=0.5mm}},scale=0.9]
\draw[thick,blue,dir] (-1,1) -- (0,0);
\draw[thick,red,dir] (0,1) -- (0,0);
\draw[thick,dir] (1,1) -- (0,0);
\filldraw (0,0) circle (1pt) node[below=0pt]{$W_1$} (-1,1) circle (1pt) node[above=0pt]{$1$} (0,1) circle (1pt) node[above=0pt]{$2$} (1,1) circle (1pt) node[above=0pt]{$3$};
\end{tikzpicture}=\{\!\pathb{1}{W_1},\pathb{2}{W_1},\pathb{3}{W_1}\!\}\,, & &
\begin{tikzpicture}[baseline={([yshift=-0.75ex]current bounding box.center)},every node/.style={font=\footnotesize},dir/.style={decoration={markings, mark=at position \halfway with {\arrow{Latex[scale=0.8]}}},postaction={decorate}},wavy/.style={decorate,decoration={coil,aspect=0, segment length=1.8mm, amplitude=0.5mm}},scale=0.9]
\draw[thick,blue,dir] (-1,1) -- (0,1);
\draw[thick,blue,dir] (0,1) -- (0,0);
\draw[thick,red,dir] (1,1) -- (0,1);
\filldraw (0,0) circle (1pt) node[below=0pt]{$W_1$} (-1,1) circle (1pt) node[above=0pt]{$1$} (0,1) circle (1pt) node[above=0pt]{$2$} (1,1) circle (1pt) node[above=0pt]{$3$};
\end{tikzpicture}=\{\!\patha{1}{2}{W_1},\pathb{3}{2}\!\}\,,
\end{align} 
where different paths are drawn with different colors.

We rewrite the string integrand this way,
\ba\la{fqwo;f}
R(1,2,3)\PT(W_1)=& \Big[ \det(-{\bm C}_{123}) + \Big(\<1,2\>C_3+  {\rm cyclic} \Big)+C_{12}C_{23}C_{31}+C_{13}C_{32}C_{21} 
\nl &  + \Big( (1- s_{12})  {\cal R}_{(12)}  C_3 +  {\rm cyclic} \Big)  \Big] \PT(W_1)\,,
\ea
where ${\bm C}_{123}$ is the $3\times 3$ matrix where the off-diagonal elements are $C_{ij}$ and the diagonal ones are $-C_i$.  According to matrix tree theorem~\cite{Stanley:2011},   $\det(-{\bm C}_{123})$ is a combination of labeled trees rooted on $W_1$.  Actually, $\det(-{\bm C}_{123}) $ is the part of ${\cal T}_{W_1}(1,2,3)$ with $\e\cdot \e$ absent,
\ba\la{aifojew}
\det(-{\bm C}_{123})= {\cal T}_{W_1}(1,2,3) \Big|_{\e\cdot \e\to 0}\,.
\ea
Equivalently, they are the reference order independent part of $\mathcal{T}_{W_1}$. Therefore, they are already part of the final recursive expansion. Similarly, $\<1,2\>C_3$ and its cyclic are the ingredient of the recursive expansion, while $C_{12}C_{23}C_{31}+C_{13}C_{32}C_{21}$ is the part of  $\<1,2,3\>+\<1,3,2\>$ with $\e\cdot \e$ absent according to Eq.~\eqref{Cijk}.

All that left to be done is to perform IBP on $ (1- s_{12})  {\cal R}_{(12)}  C_3  $ and its cyclic to get the missing $\e\cdot \e$ pieces. For this simple example, Eq.~\eqref{vpqoierf} is adequate since only tadpoles appear after expanding $C_i$. We proceed with the reference order $1\prec 2\prec 3$.


For  $ (1- s_{12})  {\cal R}_{(12)}  C_3  $,  we set $a=1$ according to~\ref{g1} and break the subcycle $\PT(1,2)$ using Eq.~\eqref{vpqoierf}. Note that $C_3=C_{31}+C_{32}+ \sum_{j\in W_1} C_{3j}$. For $C_{31}$,  we have
 \ba\la{fiqopruifa0}
  (1- s_{12})  {\cal R}_{(12)}  C_{31}  \PT(W_1)
\overset{\rm IBP}{\cong} &-\frac{\e_1\C\e_2}{z_{12}} 
  C_{31}  \Bigg(\frac{ k_2\C k_3}{z_{23}} +  \sum_{j\in W_1}  \frac{  k_2\C k_j}{z_{2j}}   \Bigg) \PT(W_1)\,,
  \ea 
where the first  term contains a subcycle $\PT(1,2,3)$, which contributes to $\<1,2,3\>+\<1,3,2\>$.  The second term contributes to the tree $\{\!\patha{1}{2}{W_1},\pathb{3}{1}\!\}$ in  ${\cal T}_{W_1}(1,2,3)$. According to the rule~\eqref{Nmap}, the path $\patha{1}{2}{W_1}$ has the kinematic factor $\e_1\cdot f_2 \cdot k_j= \e_{1} \cdot k_2 \e_2 \cdot k_j- \e_{1} \cdot \e_2 k_2 \cdot k_j$ with $j\in W_1$. Indeed, the first part $ \e_{1} \C k_2\e_2 \C k_j$ is given by Eq.~\eqref{aifojew}, while the second part is supplemented by Eq.~\eqref{fiqopruifa0}. 
Similarly, for $C_{32}$,  we have
    \ba\la{fiqofqprui0}
  (1- s_{12})  {\cal R}_{(12)}  C_{32}  \PT(W_1)
\overset{\rm IBP}{\cong} \!-\frac{\e_1\C\e_2}{z_{12}} C_{32}  \Bigg[\frac{ k_3\C k_1}{z_{31}} +  \sum_{j\in W_1}\!\! \Big( \frac{  k_2\C k_j}{z_{2j}} + \frac{  k_3\C k_j}{z_{3j}} \Big)  \Bigg] \PT(W_1),
  \ea 
where the first  term contributes to  $\<1,2,3\>+\<1,3,2\>$, and second term contributes to the labeled tree $\{\!\patha{1}{2}{W_1},\pathb{3}{2}\!\}$  and  $\{\!\pathc{1}{2}{3}{W_1}\!\}$. Finally, for $\sum_{j\in W_1} C_{3,j}$, we have
    \ba\la{fiqofq;rfprui0}
  (1- s_{1,2})  {\cal R}_{(12)} \sum_{j\in W_1} C_{3,j}  \PT(W_1)
\overset{\rm IBP}{\cong} &-\frac{\e_1\C\e_2}{z_{12}}  \sum_{j\in W_1} C_{3,j}  \sum_{\ell\in W_1\cup \{3\}}  \frac{  k_2\C k_\ell}{z_{2\ell}} \PT(W_1)\,,
  \ea 
which contributes to the labeled tree $\{\!\patha{1}{2}{W_1}, \pathb{3}{W_1}\!\}$ and $\{\!\pathc{1}{2}{3}{W_1}\!\}$ .

Next, $ (1- s_{1,3})  {\cal R}_{(13)}  C_2  $ can be processed similarly by Eq.~\eqref{vpqoierf} with $a=1$,
  \ba\la{quhrq}
 (1- s_{1,3})  {\cal R}_{(13)}  C_2  \PT(W_1)
\overset{\rm IBP}{\cong} &-\frac{\e_1\C\e_3}{z_{13}}  \PT(W_1)  \Bigg[    C_{21} \frac{  k_3\C k_2}{z_{32}} + C_{23}  \frac{ k_2\C k_1}{z_{21}} 
\nonumber\\
&+ C_2  \!  \sum_{j\in W_1} \! \frac{  k_3\C k_j}{z_{3j}}+ C_{23}\! \sum_{j\in W_1} \! \frac{  k_2\C k_j}{z_{2j}} +  \frac{k_3\c k_2}{z_{32}} \! \sum_{j\in W_1}\! C_{2,j}
  \Bigg]\,,
  \ea 
where the first line contributes to $\<1,2,3\>+\<1,3,2\>$ and the second line contributes to the labeled tree $\{\!\patha{1}{3}{W_1},\pathb{2}{1}\!\}$, $\{\!\patha{1}{3}{W_1},\pathb{2}{3}\!\}$, $\{\!\patha{1}{3}{W_1},\pathb{2}{W_1}\!\}$ and $\{\!\pathc{1}{3}{2}{W_1}\!\}$.

    
Finally, for  $ (1- s_{23})  {\cal R}_{(23)}  C_1 = (1- s_{23})  {\cal R}_{(23)} \big( C_{12}+ C_{13}+   \sum_{j \in W_1}  C_{1j}\big)  $,  we invoke~\ref{g2} to set $a=2$ and $3$ for the first two terms. In the last term, since $1$ is connected to the root, we move on to the next particle in the reference order, which is $2$, according to~\ref{g3}. For the first two terms, using~\eqref{vpqoierf} with the proper $a$, we get
\begin{align}\la{fqupo}
(1- s_{23})  {\cal R}_{(23)}  C_{13} \PT(W_1)&\overset{\rm IBP}{\cong} -C_{13} \frac{\e_2\C\e_3}{z_{23}}  \Bigg(\frac{  k_2\C k_1}{z_{21}}+    \sum_{j\in W_1}  \frac{  k_2\C k_j}{z_{2j}}    \Bigg) \PT(W_1)\,, \nonumber\\
   (1- s_{23})  {\cal R}_{(23)}  C_{12} \PT(W_1)& \overset{\rm IBP}{\cong} -C_{12} \frac{\e_2\C\e_3}{z_{23}}  \Bigg(\frac{  k_3\C k_1}{z_{31}}+    \sum_{j\in W_1}  \frac{  k_3\C k_j}{z_{3j}}    \Bigg)\PT(W_1)\,.
\end{align}
In both results, the first term contributes to $\<1,2,3\>+\<1,3,2\>$, while the second term contributes to the labeled tree $\{\!\pathc{1}{3}{2}{W_1}\!\}$ and $\{\!\pathc{1}{2}{3}{W_1}\!\}$ respectively. For the $\sum_{j}C_{1j}$ part, we set $a=2$  to break the subcycle $\PT(2,3)$ because of the rule~\ref{g1}, 
  \ba
  (1- s_{23})  {\cal R}_{(23)} \sum_{j \in W_1}  C_{1j} \PT(W_1)\overset{\rm IBP}{\cong}-   \frac{\e_2\C\e_3}{z_{23}}    \sum_{j \in W_1}  C_{1j}   \sum_{\ell\in W_1\cup \{1\}}  \frac{  k_3\C k_\ell}{z_{3\ell}}    \PT(W_1)\,.
  \ea 
  We see that this choice of $a$,  the result contributes to  the labeled tree $\{\!\patha{2}{3}{W_1},\pathb{1}{W_1}\!\}$ and  $\{\!\pathc{2}{3}{1}{W_1}\!\}$.
  

  
  The six terms containing subcycle $\PT(1,2,3)$ in Eq.~\eqref{fiqopruifa0}, \eqref{fiqofqprui0},   \eqref{quhrq} and \eqref{fqupo} are the exact the remaining components to make up $\<1,2,3\>+\<1,3,2\>$ together with $C_{12}C_{23}C_{31}+C_{13}C_{32}C_{21}$. According to the definition of fusion~\eqref{fqoe}, we have
  \ba\la{fqpfu}
    \<1,2,3\>+\<1,3,2\>=&\,C_{12}C_{23}C_{31}+C_{13}C_{32}C_{21} - \frac{\e_1\C\e_2}{z_{12}}  \Big[    C_{31} \frac{  k_2\C k_3}{z_{23}} + C_{32}  \frac{  k_3\C k_1}{z_{31}} \Big] 
\nl 
&- \frac{\e_1\C\e_3}{z_{13}}  \Big[    C_{21} \frac{  k_3\C k_2}{z_{32}} + C_{23}  \frac{ k_2\C k_1}{z_{21}}\Big] -   \frac{\e_2\C\e_3}{z_{23}}  \Big[C_{13} \frac{  k_2\C k_1}{z_{21}} + C_{12}   \frac{  k_3\C k_1}{z_{31}}\Big].
\ea
Meanwhile, summing over all the aforementioned labeled trees, we exactly reproduce $\mathcal{T}_{W_1}(1,2,3)$ constructed from the rules in Appendix~\ref{sec:trees}. The explicit expression is 
\ba\la{aifojeafdfw}
  {\cal T}_{W_1}(1,2,3)= &\det(-{\bm C}_{123})-  \frac{\e_1\C\e_2}{z_{12}}   \Bigg[ C_3      \sum_{j\in W_1}  \frac{  k_2\C k_j}{z_{2j}} 
 + C_{32} \sum_{j\in W_1}  \frac{  k_3\C k_j}{z_{3j}} +  \frac{k_2\c k_3}{z_{23}}  \sum_{j\in W_1} C_{3,j}
\Bigg]
 \nl
 &-  \frac{\e_1\C\e_3}{z_{13}} \Bigg[  C_2     \sum_{j\in W_1}  \frac{  k_3\C k_j}{z_{3j}}   + C_{23} \sum_{j\in W_1}  \frac{  k_2\C k_j}{z_{2j}} +  \frac{k_3\c k_2}{z_{32}}  \sum_{j\in W_1} C_{2,j}\Bigg]
\nl
& -  \frac{\e_2\C\e_3}{z_{23}}       \Bigg[ \sum_{\ell_1 \in W_1} \!\!\! C_{1\ell_1}  \sum_{j\in W_1\cup\{1\}} \!\! \frac{  k_3\C k_j}{z_{3j}} +
 C_{13} \! \sum_{j\in W_1} \! \frac{  k_2\C k_j}{z_{2j}}  
 +C_{12} \!  \sum_{j\in W_1} \! \frac{  k_3\C k_j}{z_{3j}} \Bigg].
\ea
This finishes the derivation of the recursive expansion~\eqref{fqeof;} for three gluons and one trace from the the string integrand \eqref{eq:strInt}\,.



\bibliographystyle{JHEP}
\bibliography{reference}

\providecommand{\href}[2]{#2}\begingroup\raggedright\begin{thebibliography}{10}

\bibitem{He:2018pol}
S.~He, F.~Teng and Y.~Zhang, \emph{{String amplitudes from field-theory
  amplitudes and vice versa}},
  \href{http://dx.doi.org/10.1103/PhysRevLett.122.211603}{\emph{Phys. Rev.
  Lett.} {\bfseries 122} (2019) 211603},
  [\href{https://arxiv.org/abs/1812.03369}{{\ttfamily 1812.03369}}].

\bibitem{ArkaniHamed:2012nw}
N.~Arkani-Hamed, J.~L. Bourjaily, F.~Cachazo, A.~B. Goncharov, A.~Postnikov and
  J.~Trnka, \emph{{Grassmannian Geometry of Scattering Amplitudes}}.
\newblock Cambridge University Press, 2016,
  \href{http://dx.doi.org/10.1017/CBO9781316091548}{10.1017/CBO9781316091548}.

\bibitem{Henn:2014yza}
J.~M. Henn and J.~C. Plefka, \emph{{Scattering Amplitudes in Gauge Theories}},
  \href{http://dx.doi.org/10.1007/978-3-642-54022-6}{\emph{Lect. Notes Phys.}
  {\bfseries 883} (2014) pp.1--195}.

\bibitem{Elvang:2015rqa}
H.~Elvang and Y.-t. Huang, \emph{{Scattering Amplitudes in Gauge Theory and
  Gravity}}.
\newblock Cambridge University Press, 2015.

\bibitem{Kawai:1985xq}
H.~Kawai, D.~C. Lewellen and S.~H.~H. Tye, \emph{{A Relation Between Tree
  Amplitudes of Closed and Open Strings}},
  \href{http://dx.doi.org/10.1016/0550-3213(86)90362-7}{\emph{Nucl. Phys.}
  {\bfseries B269} (1986) 1--23}.

\bibitem{Bern:1998sv}
Z.~Bern, L.~J. Dixon, M.~Perelstein and J.~S. Rozowsky, \emph{{Multileg one
  loop gravity amplitudes from gauge theory}},
  \href{http://dx.doi.org/10.1016/S0550-3213(99)00029-2}{\emph{Nucl. Phys.}
  {\bfseries B546} (1999) 423--479},
  [\href{https://arxiv.org/abs/hep-th/9811140}{{\ttfamily hep-th/9811140}}].

\bibitem{Bern:2008qj}
Z.~Bern, J.~J.~M. Carrasco and H.~Johansson, \emph{{New Relations for
  Gauge-Theory Amplitudes}},
  \href{http://dx.doi.org/10.1103/PhysRevD.78.085011}{\emph{Phys. Rev.}
  {\bfseries D78} (2008) 085011},
  [\href{https://arxiv.org/abs/0805.3993}{{\ttfamily 0805.3993}}].

\bibitem{Bern:2010ue}
Z.~Bern, J.~J.~M. Carrasco and H.~Johansson, \emph{{Perturbative Quantum
  Gravity as a Double Copy of Gauge Theory}},
  \href{http://dx.doi.org/10.1103/PhysRevLett.105.061602}{\emph{Phys. Rev.
  Lett.} {\bfseries 105} (2010) 061602},
  [\href{https://arxiv.org/abs/1004.0476}{{\ttfamily 1004.0476}}].

\bibitem{Bern:2013uka}
Z.~Bern, S.~Davies, T.~Dennen, A.~V. Smirnov and V.~A. Smirnov,
  \emph{{Ultraviolet Properties of N=4 Supergravity at Four Loops}},
  \href{http://dx.doi.org/10.1103/PhysRevLett.111.231302}{\emph{Phys. Rev.
  Lett.} {\bfseries 111} (2013) 231302},
  [\href{https://arxiv.org/abs/1309.2498}{{\ttfamily 1309.2498}}].

\bibitem{Bern:2014sna}
Z.~Bern, S.~Davies and T.~Dennen, \emph{{Enhanced ultraviolet cancellations in
  $\mathcal N=5$ supergravity at four loops}},
  \href{http://dx.doi.org/10.1103/PhysRevD.90.105011}{\emph{Phys. Rev.}
  {\bfseries D90} (2014) 105011},
  [\href{https://arxiv.org/abs/1409.3089}{{\ttfamily 1409.3089}}].

\bibitem{Johansson:2017bfl}
{Johansson, Henrik and K\"{a}lin, Gregor and Mogull, Gustav}, \emph{{Two-loop
  supersymmetric QCD and half-maximal supergravity amplitudes}},
  \href{http://dx.doi.org/10.1007/JHEP09(2017)019}{\emph{JHEP} {\bfseries 09}
  (2017) 019}, [\href{https://arxiv.org/abs/1706.09381}{{\ttfamily
  1706.09381}}].

\bibitem{Bern:2018jmv}
Z.~Bern, J.~J. Carrasco, W.-M. Chen, A.~Edison, H.~Johansson, J.~Parra-Martinez
  et~al., \emph{{Ultraviolet Properties of $\mathcal N = 8$ Supergravity at
  Five Loops}}, \href{http://dx.doi.org/10.1103/PhysRevD.98.086021}{\emph{Phys.
  Rev.} {\bfseries D98} (2018) 086021},
  [\href{https://arxiv.org/abs/1804.09311}{{\ttfamily 1804.09311}}].

\bibitem{Cachazo:2013hca}
F.~Cachazo, S.~He and E.~Y. Yuan, \emph{{Scattering of Massless Particles in
  Arbitrary Dimensions}},
  \href{http://dx.doi.org/10.1103/PhysRevLett.113.171601}{\emph{Phys. Rev.
  Lett.} {\bfseries 113} (2014) 171601},
  [\href{https://arxiv.org/abs/1307.2199}{{\ttfamily 1307.2199}}].

\bibitem{Cachazo:2013iea}
F.~Cachazo, S.~He and E.~Y. Yuan, \emph{{Scattering of Massless Particles:
  Scalars, Gluons and Gravitons}},
  \href{http://dx.doi.org/10.1007/JHEP07(2014)033}{\emph{JHEP} {\bfseries 07}
  (2014) 033}, [\href{https://arxiv.org/abs/1309.0885}{{\ttfamily 1309.0885}}].

\bibitem{Cachazo:2013gna}
F.~Cachazo, S.~He and E.~Y. Yuan, \emph{{Scattering equations and
  Kawai-Lewellen-Tye orthogonality}},
  \href{http://dx.doi.org/10.1103/PhysRevD.90.065001}{\emph{Phys. Rev.}
  {\bfseries D90} (2014) 065001},
  [\href{https://arxiv.org/abs/1306.6575}{{\ttfamily 1306.6575}}].

\bibitem{Adamo:2013tsa}
T.~Adamo, E.~Casali and D.~Skinner, \emph{{Ambitwistor strings and the
  scattering equations at one loop}},
  \href{http://dx.doi.org/10.1007/JHEP04(2014)104}{\emph{JHEP} {\bfseries 04}
  (2014) 104}, [\href{https://arxiv.org/abs/1312.3828}{{\ttfamily 1312.3828}}].

\bibitem{Geyer:2015bja}
Y.~Geyer, L.~Mason, R.~Monteiro and P.~Tourkine, \emph{{Loop Integrands for
  Scattering Amplitudes from the Riemann Sphere}},
  \href{http://dx.doi.org/10.1103/PhysRevLett.115.121603}{\emph{Phys. Rev.
  Lett.} {\bfseries 115} (2015) 121603},
  [\href{https://arxiv.org/abs/1507.00321}{{\ttfamily 1507.00321}}].

\bibitem{Cachazo:2015aol}
F.~Cachazo, S.~He and E.~Y. Yuan, \emph{{One-Loop Corrections from Higher
  Dimensional Tree Amplitudes}},
  \href{http://dx.doi.org/10.1007/JHEP08(2016)008}{\emph{JHEP} {\bfseries 08}
  (2016) 008}, [\href{https://arxiv.org/abs/1512.05001}{{\ttfamily
  1512.05001}}].

\bibitem{Geyer:2016wjx}
Y.~Geyer, L.~Mason, R.~Monteiro and P.~Tourkine, \emph{{Two-Loop Scattering
  Amplitudes from the Riemann Sphere}},
  \href{http://dx.doi.org/10.1103/PhysRevD.94.125029}{\emph{Phys. Rev.}
  {\bfseries D94} (2016) 125029},
  [\href{https://arxiv.org/abs/1607.08887}{{\ttfamily 1607.08887}}].

\bibitem{Geyer:2018xwu}
Y.~Geyer and R.~Monteiro, \emph{{Two-Loop Scattering Amplitudes from
  Ambitwistor Strings: from Genus Two to the Nodal Riemann Sphere}},
  \href{http://dx.doi.org/10.1007/JHEP11(2018)008}{\emph{JHEP} {\bfseries 11}
  (2018) 008}, [\href{https://arxiv.org/abs/1805.05344}{{\ttfamily
  1805.05344}}].

\bibitem{Cachazo:2014xea}
F.~Cachazo, S.~He and E.~Y. Yuan, \emph{{Scattering Equations and Matrices:
  From Einstein To Yang-Mills, DBI and NLSM}},
  \href{http://dx.doi.org/10.1007/JHEP07(2015)149}{\emph{JHEP} {\bfseries 07}
  (2015) 149}, [\href{https://arxiv.org/abs/1412.3479}{{\ttfamily 1412.3479}}].

\bibitem{He:2016mzd}
S.~He and O.~Schlotterer, \emph{{New Relations for Gauge-Theory and Gravity
  Amplitudes at Loop Level}},
  \href{http://dx.doi.org/10.1103/PhysRevLett.118.161601}{\emph{Phys. Rev.
  Lett.} {\bfseries 118} (2017) 161601},
  [\href{https://arxiv.org/abs/1612.00417}{{\ttfamily 1612.00417}}].

\bibitem{He:2017spx}
S.~He, O.~Schlotterer and Y.~Zhang, \emph{{New BCJ representations for one-loop
  amplitudes in gauge theories and gravity}},
  \href{http://dx.doi.org/10.1016/j.nuclphysb.2018.03.003}{\emph{Nucl. Phys.}
  {\bfseries B930} (2018) 328--383},
  [\href{https://arxiv.org/abs/1706.00640}{{\ttfamily 1706.00640}}].

\bibitem{Mason:2013sva}
L.~Mason and D.~Skinner, \emph{{Ambitwistor strings and the scattering
  equations}}, \href{http://dx.doi.org/10.1007/JHEP07(2014)048}{\emph{JHEP}
  {\bfseries 07} (2014) 048},
  [\href{https://arxiv.org/abs/1311.2564}{{\ttfamily 1311.2564}}].

\bibitem{Casali:2015vta}
E.~Casali, Y.~Geyer, L.~Mason, R.~Monteiro and K.~A. Roehrig, \emph{{New
  Ambitwistor String Theories}},
  \href{http://dx.doi.org/10.1007/JHEP11(2015)038}{\emph{JHEP} {\bfseries 11}
  (2015) 038}, [\href{https://arxiv.org/abs/1506.08771}{{\ttfamily
  1506.08771}}].

\bibitem{Siegel:2015axg}
W.~Siegel, \emph{{Amplitudes for left-handed strings}},
  \href{https://arxiv.org/abs/1512.02569}{{\ttfamily 1512.02569}}.

\bibitem{Casali:2016atr}
E.~Casali and P.~Tourkine, \emph{{On the null origin of the ambitwistor
  string}}, \href{http://dx.doi.org/10.1007/JHEP11(2016)036}{\emph{JHEP}
  {\bfseries 11} (2016) 036},
  [\href{https://arxiv.org/abs/1606.05636}{{\ttfamily 1606.05636}}].

\bibitem{Azevedo:2017yjy}
T.~Azevedo and R.~L. Jusinskas, \emph{{Connecting the ambitwistor and the
  sectorized heterotic strings}},
  \href{http://dx.doi.org/10.1007/JHEP10(2017)216}{\emph{JHEP} {\bfseries 10}
  (2017) 216}, [\href{https://arxiv.org/abs/1707.08840}{{\ttfamily
  1707.08840}}].

\bibitem{Mizera:2017rqa}
S.~Mizera, \emph{{Scattering Amplitudes from Intersection Theory}},
  \href{http://dx.doi.org/10.1103/PhysRevLett.120.141602}{\emph{Phys. Rev.
  Lett.} {\bfseries 120} (2018) 141602},
  [\href{https://arxiv.org/abs/1711.00469}{{\ttfamily 1711.00469}}].

\bibitem{Bjerrum-Bohr:2014qwa}
N.~E.~J. Bjerrum-Bohr, P.~H. Damgaard, P.~Tourkine and P.~Vanhove,
  \emph{{Scattering Equations and String Theory Amplitudes}},
  \href{http://dx.doi.org/10.1103/PhysRevD.90.106002}{\emph{Phys. Rev.}
  {\bfseries D90} (2014) 106002},
  [\href{https://arxiv.org/abs/1403.4553}{{\ttfamily 1403.4553}}].

\bibitem{Mafra:2011kj}
C.~R. Mafra, O.~Schlotterer and S.~Stieberger, \emph{{Explicit BCJ Numerators
  from Pure Spinors}},
  \href{http://dx.doi.org/10.1007/JHEP07(2011)092}{\emph{JHEP} {\bfseries 07}
  (2011) 092}, [\href{https://arxiv.org/abs/1104.5224}{{\ttfamily 1104.5224}}].

\bibitem{Mafra:2011nv}
C.~R. Mafra, O.~Schlotterer and S.~Stieberger, \emph{{Complete N-Point
  Superstring Disk Amplitude I. Pure Spinor Computation}},
  \href{http://dx.doi.org/10.1016/j.nuclphysb.2013.04.023}{\emph{Nucl. Phys.}
  {\bfseries B873} (2013) 419--460},
  [\href{https://arxiv.org/abs/1106.2645}{{\ttfamily 1106.2645}}].

\bibitem{Mafra:2011nw}
C.~R. Mafra, O.~Schlotterer and S.~Stieberger, \emph{{Complete N-Point
  Superstring Disk Amplitude II. Amplitude and Hypergeometric Function
  Structure}},
  \href{http://dx.doi.org/10.1016/j.nuclphysb.2013.04.022}{\emph{Nucl. Phys.}
  {\bfseries B873} (2013) 461--513},
  [\href{https://arxiv.org/abs/1106.2646}{{\ttfamily 1106.2646}}].

\bibitem{Mafra:2014gja}
C.~R. Mafra and O.~Schlotterer, \emph{{Towards one-loop SYM amplitudes from the
  pure spinor BRST cohomology}},
  \href{http://dx.doi.org/10.1002/prop.201400076}{\emph{Fortsch. Phys.}
  {\bfseries 63} (2015) 105--131},
  [\href{https://arxiv.org/abs/1410.0668}{{\ttfamily 1410.0668}}].

\bibitem{He:2015wgf}
S.~He, R.~Monteiro and O.~Schlotterer, \emph{{String-inspired BCJ numerators
  for one-loop MHV amplitudes}},
  \href{http://dx.doi.org/10.1007/JHEP01(2016)171}{\emph{JHEP} {\bfseries 01}
  (2016) 171}, [\href{https://arxiv.org/abs/1507.06288}{{\ttfamily
  1507.06288}}].

\bibitem{Chiodaroli:2014xia}
M.~Chiodaroli, M.~G\"{u}naydin, H.~Johansson and R.~Roiban, \emph{{Scattering
  amplitudes in $ \mathcal{N}=2 $ Maxwell-Einstein and Yang-Mills/Einstein
  supergravity}}, \href{http://dx.doi.org/10.1007/JHEP01(2015)081}{\emph{JHEP}
  {\bfseries 01} (2015) 081},
  [\href{https://arxiv.org/abs/1408.0764}{{\ttfamily 1408.0764}}].

\bibitem{BjerrumBohr:2009rd}
N.~E.~J. Bjerrum-Bohr, P.~H. Damgaard and P.~Vanhove, \emph{{Minimal Basis for
  Gauge Theory Amplitudes}},
  \href{http://dx.doi.org/10.1103/PhysRevLett.103.161602}{\emph{Phys. Rev.
  Lett.} {\bfseries 103} (2009) 161602},
  [\href{https://arxiv.org/abs/0907.1425}{{\ttfamily 0907.1425}}].

\bibitem{Stieberger:2009hq}
S.~Stieberger, \emph{{Open \& Closed vs. Pure Open String Disk Amplitudes}},
  \href{https://arxiv.org/abs/0907.2211}{{\ttfamily 0907.2211}}.

\bibitem{Stieberger:2016lng}
S.~Stieberger and T.~R. Taylor, \emph{{New relations for
  Einstein–Yang–Mills amplitudes}},
  \href{http://dx.doi.org/10.1016/j.nuclphysb.2016.09.014}{\emph{Nucl. Phys.}
  {\bfseries B913} (2016) 151--162},
  [\href{https://arxiv.org/abs/1606.09616}{{\ttfamily 1606.09616}}].

\bibitem{Schlotterer:2016cxa}
O.~Schlotterer, \emph{{Amplitude relations in heterotic string theory and
  Einstein-Yang-Mills}},
  \href{http://dx.doi.org/10.1007/JHEP11(2016)074}{\emph{JHEP} {\bfseries 11}
  (2016) 074}, [\href{https://arxiv.org/abs/1608.00130}{{\ttfamily
  1608.00130}}].

\bibitem{Broedel:2013tta}
J.~Broedel, O.~Schlotterer and S.~Stieberger, \emph{{Polylogarithms, Multiple
  Zeta Values and Superstring Amplitudes}},
  \href{http://dx.doi.org/10.1002/prop.201300019}{\emph{Fortsch. Phys.}
  {\bfseries 61} (2013) 812--870},
  [\href{https://arxiv.org/abs/1304.7267}{{\ttfamily 1304.7267}}].

\bibitem{Carrasco:2016ldy}
J.~J.~M. Carrasco, C.~R. Mafra and O.~Schlotterer, \emph{{Abelian Z-theory:
  NLSM amplitudes and $\alpha$'-corrections from the open string}},
  \href{http://dx.doi.org/10.1007/JHEP06(2017)093}{\emph{JHEP} {\bfseries 06}
  (2017) 093}, [\href{https://arxiv.org/abs/1608.02569}{{\ttfamily
  1608.02569}}].

\bibitem{Mafra:2016mcc}
C.~R. Mafra and O.~Schlotterer, \emph{{Non-abelian $Z$-theory: Berends-Giele
  recursion for the $\alpha'$-expansion of disk integrals}},
  \href{http://dx.doi.org/10.1007/JHEP01(2017)031}{\emph{JHEP} {\bfseries 01}
  (2017) 031}, [\href{https://arxiv.org/abs/1609.07078}{{\ttfamily
  1609.07078}}].

\bibitem{Carrasco:2016ygv}
J.~J.~M. Carrasco, C.~R. Mafra and O.~Schlotterer, \emph{{Semi-abelian
  Z-theory: NLSM$+\phi^{3}$ from the open string}},
  \href{http://dx.doi.org/10.1007/JHEP08(2017)135}{\emph{JHEP} {\bfseries 08}
  (2017) 135}, [\href{https://arxiv.org/abs/1612.06446}{{\ttfamily
  1612.06446}}].

\bibitem{Huang:2016tag}
Y.-t. Huang, O.~Schlotterer and C.~Wen, \emph{{Universality in string
  interactions}}, \href{http://dx.doi.org/10.1007/JHEP09(2016)155}{\emph{JHEP}
  {\bfseries 09} (2016) 155},
  [\href{https://arxiv.org/abs/1602.01674}{{\ttfamily 1602.01674}}].

\bibitem{Azevedo:2018dgo}
T.~Azevedo, M.~Chiodaroli, H.~Johansson and O.~Schlotterer, \emph{{Heterotic
  and bosonic string amplitudes via field theory}},
  \href{http://dx.doi.org/10.1007/JHEP10(2018)012}{\emph{JHEP} {\bfseries 10}
  (2018) 012}, [\href{https://arxiv.org/abs/1803.05452}{{\ttfamily
  1803.05452}}].

\bibitem{Johansson:2017srf}
H.~Johansson and J.~Nohle, \emph{{Conformal Gravity from Gauge Theory}},
  \href{https://arxiv.org/abs/1707.02965}{{\ttfamily 1707.02965}}.

\bibitem{Schlotterer:2012ny}
O.~Schlotterer and S.~Stieberger, \emph{{Motivic Multiple Zeta Values and
  Superstring Amplitudes}},
  \href{http://dx.doi.org/10.1088/1751-8113/46/47/475401}{\emph{J. Phys.}
  {\bfseries A46} (2013) 475401},
  [\href{https://arxiv.org/abs/1205.1516}{{\ttfamily 1205.1516}}].

\bibitem{Stieberger:2013wea}
S.~Stieberger, \emph{{Closed superstring amplitudes, single-valued multiple
  zeta values and the Deligne associator}},
  \href{http://dx.doi.org/10.1088/1751-8113/47/15/155401}{\emph{J. Phys.}
  {\bfseries A47} (2014) 155401},
  [\href{https://arxiv.org/abs/1310.3259}{{\ttfamily 1310.3259}}].

\bibitem{Stieberger:2014hba}
S.~Stieberger and T.~R. Taylor, \emph{{Closed String Amplitudes as
  Single-Valued Open String Amplitudes}},
  \href{http://dx.doi.org/10.1016/j.nuclphysb.2014.02.005}{\emph{Nucl. Phys.}
  {\bfseries B881} (2014) 269--287},
  [\href{https://arxiv.org/abs/1401.1218}{{\ttfamily 1401.1218}}].

\bibitem{Schlotterer:2018zce}
O.~Schlotterer and O.~Schnetz, \emph{{Closed strings as single-valued open
  strings: A genus-zero derivation}},
  \href{http://dx.doi.org/10.1088/1751-8121/aaea14}{\emph{J. Phys.} {\bfseries
  A52} (2019) 045401}, [\href{https://arxiv.org/abs/1808.00713}{{\ttfamily
  1808.00713}}].

\bibitem{Brown:2018omk}
F.~Brown and C.~Dupont, \emph{{Single-valued integration and superstring
  amplitudes in genus zero}},
  \href{https://arxiv.org/abs/1810.07682}{{\ttfamily 1810.07682}}.

\bibitem{Schnetz:2013hqa}
O.~Schnetz, \emph{{Graphical functions and single-valued multiple
  polylogarithms}},
  \href{http://dx.doi.org/10.4310/CNTP.2014.v8.n4.a1}{\emph{Commun. Num. Theor.
  Phys.} {\bfseries 08} (2014) 589--675},
  [\href{https://arxiv.org/abs/1302.6445}{{\ttfamily 1302.6445}}].

\bibitem{Brown:2013gia}
F.~Brown, \emph{{Single-valued Motivic Periods and Multiple Zeta Values}},
  \href{http://dx.doi.org/10.1017/fms.2014.18}{\emph{SIGMA} {\bfseries 2}
  (2014) e25}, [\href{https://arxiv.org/abs/1309.5309}{{\ttfamily 1309.5309}}].

\bibitem{BjerrumBohr:2010yc}
N.~E.~J. Bjerrum-Bohr, P.~H. Damgaard, B.~Feng and T.~Sondergaard, \emph{{Proof
  of Gravity and Yang-Mills Amplitude Relations}},
  \href{http://dx.doi.org/10.1007/JHEP09(2010)067}{\emph{JHEP} {\bfseries 09}
  (2010) 067}, [\href{https://arxiv.org/abs/1007.3111}{{\ttfamily 1007.3111}}].

\bibitem{Mizera:2017sen}
S.~Mizera and G.~Zhang, \emph{{A String Deformation of the Parke-Taylor
  Factor}}, \href{http://dx.doi.org/10.1103/PhysRevD.96.066016}{\emph{Phys.
  Rev.} {\bfseries D96} (2017) 066016},
  [\href{https://arxiv.org/abs/1705.10323}{{\ttfamily 1705.10323}}].

\bibitem{Mizera:2017cqs}
S.~Mizera, \emph{{Combinatorics and Topology of Kawai-Lewellen-Tye Relations}},
  \href{http://dx.doi.org/10.1007/JHEP08(2017)097}{\emph{JHEP} {\bfseries 08}
  (2017) 097}, [\href{https://arxiv.org/abs/1706.08527}{{\ttfamily
  1706.08527}}].

\bibitem{Arkani-Hamed:2017tmz}
N.~Arkani-Hamed, Y.~Bai and T.~Lam, \emph{{Positive Geometries and Canonical
  Forms}}, \href{http://dx.doi.org/10.1007/JHEP11(2017)039}{\emph{JHEP}
  {\bfseries 11} (2017) 039},
  [\href{https://arxiv.org/abs/1703.04541}{{\ttfamily 1703.04541}}].

\bibitem{Arkani-Hamed:2017mur}
N.~Arkani-Hamed, Y.~Bai, S.~He and G.~Yan, \emph{{Scattering Forms and the
  Positive Geometry of Kinematics, Color and the Worldsheet}},
  \href{http://dx.doi.org/10.1007/JHEP05(2018)096}{\emph{JHEP} {\bfseries 05}
  (2018) 096}, [\href{https://arxiv.org/abs/1711.09102}{{\ttfamily
  1711.09102}}].

\bibitem{Teng:2017tbo}
F.~Teng and B.~Feng, \emph{{Expanding Einstein-Yang-Mills by Yang-Mills in CHY
  frame}}, \href{http://dx.doi.org/10.1007/JHEP05(2017)075}{\emph{JHEP}
  {\bfseries 05} (2017) 075},
  [\href{https://arxiv.org/abs/1703.01269}{{\ttfamily 1703.01269}}].

\bibitem{Du:2017gnh}
Y.-J. Du, B.~Feng and F.~Teng, \emph{{Expansion of All Multitrace Tree Level
  EYM Amplitudes}},
  \href{http://dx.doi.org/10.1007/JHEP12(2017)038}{\emph{JHEP} {\bfseries 12}
  (2017) 038}, [\href{https://arxiv.org/abs/1708.04514}{{\ttfamily
  1708.04514}}].

\bibitem{Lam:2016tlk}
C.~S. Lam and Y.-P. Yao, \emph{{Evaluation of the Cachazo-He-Yuan gauge
  amplitude}}, \href{http://dx.doi.org/10.1103/PhysRevD.93.105008}{\emph{Phys.
  Rev.} {\bfseries D93} (2016) 105008},
  [\href{https://arxiv.org/abs/1602.06419}{{\ttfamily 1602.06419}}].

\bibitem{Stanley:2011}
R.~P. Stanley, \emph{Enumerative Combinatorics: Volume 2}.
\newblock Cambridge University Press, New York, NY, USA, 1st~ed., 2001.

\bibitem{DelDuca:1999rs}
V.~Del~Duca, L.~J. Dixon and F.~Maltoni, \emph{{New color decompositions for
  gauge amplitudes at tree and loop level}},
  \href{http://dx.doi.org/10.1016/S0550-3213(99)00809-3}{\emph{Nucl. Phys.}
  {\bfseries B571} (2000) 51--70},
  [\href{https://arxiv.org/abs/hep-ph/9910563}{{\ttfamily hep-ph/9910563}}].

\bibitem{Johansson:2018ues}
H.~Johansson, G.~Mogull and F.~Teng, \emph{{Unraveling conformal gravity
  amplitudes}}, \href{http://dx.doi.org/10.1007/JHEP09(2018)080}{\emph{JHEP}
  {\bfseries 09} (2018) 080},
  [\href{https://arxiv.org/abs/1806.05124}{{\ttfamily 1806.05124}}].

\bibitem{He:2016iqi}
S.~He and Y.~Zhang, \emph{{New Formulas for Amplitudes from Higher-Dimensional
  Operators}}, \href{http://dx.doi.org/10.1007/JHEP02(2017)019}{\emph{JHEP}
  {\bfseries 02} (2017) 019},
  [\href{https://arxiv.org/abs/1608.08448}{{\ttfamily 1608.08448}}].

\bibitem{Garozzo:2018uzj}
L.~M. Garozzo, L.~Queimada and O.~Schlotterer, \emph{{Berends-Giele currents in
  Bern-Carrasco-Johansson gauge for $F^3$- and $F^4$-deformed Yang-Mills
  amplitudes}}, \href{http://dx.doi.org/10.1007/JHEP02(2019)078}{\emph{JHEP}
  {\bfseries 02} (2019) 078},
  [\href{https://arxiv.org/abs/1809.08103}{{\ttfamily 1809.08103}}].

\bibitem{Du:2017kpo}
Y.-J. Du and F.~Teng, \emph{{BCJ numerators from reduced Pfaffian}},
  \href{http://dx.doi.org/10.1007/JHEP04(2017)033}{\emph{JHEP} {\bfseries 04}
  (2017) 033}, [\href{https://arxiv.org/abs/1703.05717}{{\ttfamily
  1703.05717}}].

\bibitem{Du:2018khm}
Y.-J. Du and Y.~Zhang, \emph{{Gauge invariance induced relations and the
  equivalence between distinct approaches to NLSM amplitudes}},
  \href{http://dx.doi.org/10.1007/JHEP07(2018)177}{\emph{JHEP} {\bfseries 07}
  (2018) 177}, [\href{https://arxiv.org/abs/1803.01701}{{\ttfamily
  1803.01701}}].

\bibitem{Hou:2018bwm}
L.~Hou and Y.-J. Du, \emph{{A graphic approach to gauge invariance induced
  identity}}, \href{http://dx.doi.org/10.1007/JHEP05(2019)012}{\emph{JHEP}
  {\bfseries 05} (2019) 012},
  [\href{https://arxiv.org/abs/1811.12653}{{\ttfamily 1811.12653}}].

\bibitem{Mizera:2019gea}
S.~Mizera, \emph{{Aspects of Scattering Amplitudes and Moduli Space
  Localization}},  \href{https://arxiv.org/abs/1906.02099}{{\ttfamily
  1906.02099}}.

\bibitem{Vanhove:2018elu}
P.~Vanhove and F.~Zerbini, \emph{{Closed string amplitudes from single-valued
  correlation functions}},  \href{https://arxiv.org/abs/1812.03018}{{\ttfamily
  1812.03018}}.

\bibitem{He:2018pue}
S.~He, G.~Yan, C.~Zhang and Y.~Zhang, \emph{{Scattering Forms, Worldsheet Forms
  and Amplitudes from Subspaces}},
  \href{http://dx.doi.org/10.1007/JHEP08(2018)040}{\emph{JHEP} {\bfseries 08}
  (2018) 040}, [\href{https://arxiv.org/abs/1803.11302}{{\ttfamily
  1803.11302}}].

\bibitem{Arkani-Hamed2}
N.~Arkani-Hamed,
  \emph{\href{https://indico.cern.ch/event/750565/contributions/3439541/attachments/1873668/3084360/Arkani-Hamed.pdf}{Talk
  at Amplitudes 2019}}, July 2019 at Trinity College, Dublin.

\bibitem{Baadsgaard:2016fel}
C.~Baadsgaard, N.~E.~J. Bjerrum-Bohr, J.~L. Bourjaily and P.~H. Damgaard,
  \emph{{String-Like Dual Models for Scalar Theories}},
  \href{http://dx.doi.org/10.1007/JHEP12(2016)019}{\emph{JHEP} {\bfseries 12}
  (2016) 019}, [\href{https://arxiv.org/abs/1610.04228}{{\ttfamily
  1610.04228}}].

\bibitem{Tsuchiya:1988va}
A.~Tsuchiya, \emph{{More on One Loop Massless Amplitudes of Superstring
  Theories}}, \href{http://dx.doi.org/10.1103/PhysRevD.39.1626}{\emph{Phys.
  Rev.} {\bfseries D39} (1989) 1626}.

\bibitem{Dolan:2007eh}
L.~Dolan and P.~Goddard, \emph{{Current Algebra on the Torus}},
  \href{http://dx.doi.org/10.1007/s00220-008-0542-1}{\emph{Commun. Math. Phys.}
  {\bfseries 285} (2009) 219--264},
  [\href{https://arxiv.org/abs/0710.3743}{{\ttfamily 0710.3743}}].

\bibitem{Mafra:2017ioj}
C.~R. Mafra and O.~Schlotterer, \emph{{Double-Copy Structure of One-Loop
  Open-String Amplitudes}},
  \href{http://dx.doi.org/10.1103/PhysRevLett.121.011601}{\emph{Phys. Rev.
  Lett.} {\bfseries 121} (2018) 011601},
  [\href{https://arxiv.org/abs/1711.09104}{{\ttfamily 1711.09104}}].

\bibitem{Mafra:2018nla}
C.~R. Mafra and O.~Schlotterer, \emph{{Towards the n-point one-loop superstring
  amplitude I: Pure spinors and superfield kinematics}},
  \href{https://arxiv.org/abs/1812.10969}{{\ttfamily 1812.10969}}.

\bibitem{Mafra:2018pll}
C.~R. Mafra and O.~Schlotterer, \emph{{Towards the n-point one-loop superstring
  amplitude II: Worldsheet functions and their duality to kinematics}},
  \href{https://arxiv.org/abs/1812.10970}{{\ttfamily 1812.10970}}.

\bibitem{Mafra:2018qqe}
C.~R. Mafra and O.~Schlotterer, \emph{{Towards the n-point one-loop superstring
  amplitude III: One-loop correlators and their double-copy structure}},
  \href{https://arxiv.org/abs/1812.10971}{{\ttfamily 1812.10971}}.

\bibitem{Gerken:2018jrq}
J.~E. Gerken, A.~Kleinschmidt and O.~Schlotterer, \emph{{Heterotic-string
  amplitudes at one loop: modular graph forms and relations to open strings}},
  \href{http://dx.doi.org/10.1007/JHEP01(2019)052}{\emph{JHEP} {\bfseries 01}
  (2019) 052}, [\href{https://arxiv.org/abs/1811.02548}{{\ttfamily
  1811.02548}}].

\bibitem{Gao:2017dek}
X.~Gao, S.~He and Y.~Zhang, \emph{{Labelled tree graphs, Feynman diagrams and
  disk integrals}},
  \href{http://dx.doi.org/10.1007/JHEP11(2017)144}{\emph{JHEP} {\bfseries 11}
  (2017) 144}, [\href{https://arxiv.org/abs/1708.08701}{{\ttfamily
  1708.08701}}].

\end{thebibliography}\endgroup

\end{document}